%% file: msrSHORTABS_finalcorrections.tex
\documentclass[12pt,preprint]{aastex}

\shorttitle{Water and Ice}
\shortauthors{D. Hollenbach}

\newcommand{\be}{\begin{equation}}
\newcommand{\ee}{\end{equation}}

\newcommand{\ep}{\varepsilon}

\def\ltsimeq{\,\raise 0.3 ex\hbox{$ < $}\kern -0.75 em
 \lower 0.7 ex\hbox{$\sim$}\,}
\def\gtsimeq{\,\raise 0.3 ex\hbox{$ > $}\kern -0.75 em
 \lower 0.7 ex\hbox{$\sim$}\,}

\let\gta=\gtsimeq
\let\lta=\ltsimeq
\begin{document}

\title{WATER, O$_2$, AND ICE IN MOLECULAR CLOUDS}

\author{David Hollenbach\altaffilmark{1}, Michael J. Kaufman\altaffilmark{1,2}, Edwin A. Bergin\altaffilmark{3}, and Gary J. Melnick\altaffilmark{4}}

\altaffiltext{1}{NASA Ames Research Center, Moffett Field, CA 94035}

\altaffiltext{2}{Department of Physics \& Astronomy, San Jose State University, San Jose, CA 95192-0106}

\altaffiltext{3}{Astronomy Department, University of Michigan, Ann Arbor, MI 48109}

\altaffiltext{4}{Smithsonian Astrophysical Observatory, Cambridge, MA 02138}

\begin{abstract}
We model the temperature and chemical structure of molecular clouds as a function
of depth into the cloud, assuming a cloud of constant density $n$ illuminated
by an external FUV (6 eV $< h\nu < 13.6$ eV) flux $G_0$ (scaling factor in multiples of the local
interstellar field). Extending previous photodissociation region models, we
include the freezing of species, simple grain surface
chemistry, and desorption (including FUV photodesorption)
 of ices. We also treat the opaque cloud interior
with time-dependent chemistry.  Here, under certain
conditions, gas phase elemental oxygen freezes out as water ice and the
elemental C/O abundance ratio can exceed unity, leading to complex carbon
chemistry.  Gas phase
H$_2$O and O$_2$ peak in abundance at intermediate depth into the
cloud, roughly $A_V \sim 3-8$ from the surface, the depth proportional
to $\ln (G_0/n)$.  Closer to the surface, molecules are photodissociated.
Deeper into the cloud,  molecules freeze to grain surfaces.  At
intermediate depths photodissociation rates are attenuated by dust
extinction, but photodesorption prevents total freezeout.
For $G_0 < 500$,  abundances of H$_2$O and O$_2$ peak at values
$\sim 10^{-7}$, producing columns 
$\sim 10^{15}$ cm$^{-2}$, independent of $G_0$ and $n$. The peak abundances
depend primarily on the product of the photodesorption yield of water
ice and the grain surface area per H nucleus.  At higher
values of $G_0$, thermal desorption of O atoms from grains enhances
the gas phase H$_2$O peak abundance and column slightly, whereas the
gas phase O$_2$ peak abundance rises to $\sim 10^{-5}$ and the column
to $\sim 2 \times 10^{16}$ cm$^{-2}$. We present simple analytic
equations for the abundances as a function of depth which clarify
the dependence on parameters.  The models are applied to observations
of H$_2$O, O$_2$, and water ice in a number of sources, including
B68, NGC 2024, and $\rho$ Oph.

\end{abstract}


\section{INTRODUCTION}

Oxygen is the third most abundant element in the universe, after hydrogen
and helium, so that a basic knowledge of oxygen chemistry in molecular
clouds is essential in order to understand 
 the chemical structure, thermal balance,
and diagnostic line emission from star-forming molecular gas in galaxies.
Early gas-phase chemical models (e.g.,  Langer \& Graedel 1989, Millar 1990,
Bergin et al 1998) predicted large abundances of H$_2$O ($\sim 10^{-6}$)
and O$_2$ ($\sim 10^{-5}-10^{-4}$) relative to hydrogen nuclei in molecular gas
well shielded from far ultraviolet (FUV, 6 eV $< h\nu < 13.6$ eV) photons.
If gas phase H$_2$O and O$_2$ were that abundant, they would be important
coolants for dense gas (Goldsmith \& Langer 1978, Hollenbach 1988, Neufeld,
Lepp, \& Melnick 1995).  However, the {\it Submillimeter Wave Astronomy
Satellite} (SWAS) made pointed observations in low energy transitions
of ortho-H$_2$O (the 1$_{10}-1_{01}$ transition at 557 GHz) and O$_2$
(the 3$_3-1_2$ transition at 487 GHz) toward numerous dense (but unshocked)
molecular cores and  determined that the {\it line-of-sight-averaged}
and beam-averaged (SWAS beam $\sim 4'$)
abundance of H$_2$O is of order $\lta 3\times 10^{-8}$ (Snell et al 2000) and that
of O$_2$ is $\lta 10^{-6}$ (Goldsmith et al 2000).  More recent observations by the
{\it Odin} mission set 
more stringent upper limits on O$_2$, $\lta 10^{-7}$ (Pagani et al 2003), 
with a reported detection 
at the $\sim 2.5\times 10^{-8}$ level in $\rho$ Oph (Larsson et al 2007).  
Although the water abundance derived from the observed water emission
depends inversely on the gas density, and therefore is somewhat uncertain,
understanding the two order of magnitude
discrepancy between the gas phase chemical models and the observations is
essential to astrochemistry and to the basic understanding of the
physics of molecular clouds.

Previous attempts to explain the low abundance of H$_2$O and O$_2$ observed
by SWAS showed that time dependent gas-phase chemistry by itself would not
be sufficient (Bergin et al 1998, 2000).  Starting from atomic gas, a dense
($n(H_2) \sim 10^5$ cm$^{-3}$) cloud only took $\gta 10^4$ years to 
reach H$_2$O abundances $\sim 10^{-6}$,
close to the final steady-state values and much greater than observed.

The best previous explanation involved time-dependent chemistry linked 
with the freezeout of oxygen species on grain surfaces
and the formation of substantial water ice mantles on grains (Bergin et
al 2000).  In these models, all the oxygen in the molecular gas which 
is {\it not} tied up in CO adsorbs quickly (in $\sim 10^5$ years for a gas density
of $\sim 10^4$ cm$^{-3}$) to grain
surfaces, forms water ice and remains stuck on the grain as an
ice mantle. As a result, the gas phase H$_2$O abundance drops
from  $\sim 10^{-6}$  to $\sim 10^{-8}$.  The grains are
assumed too warm for CO to freeze as a CO ice mantle, so that for a
period from about $10^5$ years to about $3\times 10^6$ years, the gas phase 
H$_2$O
abundance remains at the $\sim 10^{-8}$ level, fed by the slow dissociation
of CO into O and the subsequent reaction of some of this O with H$_3^+$, which 
then ultimately forms gas-phase H$_2$O. 
 The slow dissociation
of CO is driven by cosmic ray ionization of He; the resultant He$^+$ reacts
with CO to form O and C$^+$. After  $\sim 3\times 10^6$ years, 
even the gas phase CO abundance drops significantly as the dissociation
process bleeds away the oxygen which ends up as water ice on grain
surfaces.  Therefore, after about $3\times 10^6$ years, the gas
phase H$_2$O also drops from $\sim 10^{-8}$ to $ < 10^{-9}$.  This 
previous scenario  explained the observed H$_2$O emission as arising
from the central, opaque regions of the cloud, where the abundance has
dropped to the observed values but have not had time to reach the
extremely low steady state values.  The model relied, then, on a ``tuning''
of molecular cloud timescales so that they are long enough for the freeze-out
of existing gas phase oxygen not in CO  onto grains, but not so long that the CO is 
broken down and the resultant O converted to water ice, which would cause the 
gas phase H$_2$O abundances to drop below the observed values.\footnote{For 
another similar model which relies on time-dependent effects, coupled
with grain freezeout and grain surface processes, to explain the
low O$_2$ and H$_2$O abundances observed by SWAS, see Roberts \& Herbst (2002).}
The model also relied on CO {\it not} freezing out on grains in the
opaque cloud centers.

To add to the problems of modeling the SWAS data, the SWAS team observed a 
number
of strong submillimeter continuum sources such as SgrB, W49 and W51
and found the 557 GHz line of H$_2$O in {\it absorption}, as the continuum
passed through translucent clouds ($A_V \sim 1-5$) along the line of
sight (Neufeld et al 2002, Neufeld et al 2003, Plume et al 2004). The
absorption measurement provided an even better estimate of the H$_2$O
column, $N$H$_2$O, through these clouds because the absorption line strengths
are only proportional to $N(H_2O)$, whereas
the emission line strengths are proportional to $n(H_2)N(H_2O)
e^{-27 {\rm K}/T}$ since the line is subthermal, effectively optically
thin, and lies $\Delta E/k = 27$ K above the ground state.  Therefore,
to obtain columns from emission line observations requires the
separate knowledge of both the gas density and the gas temperature.
The absorption measurements showed column-averaged abundances of
H$_2$O of $\sim 10^{-7}-10^{-6}$ with respect to H$_2$,  using observed 
CO columns and multiplying by $10^4$ to obtain H$_2$ columns. 
 In the context of the time-dependent models
with freezeout, the difference between the abundances measured in absorption
compared with those measured in emission was attributed to the presumed
lower gas densities in the absorbing clouds compared to the emitting
clouds.  Because the freezeout time depends on $n^{-1}$, it was assumed
that the lower density clouds had not had time to freeze as much oxygen
from the gas phase, so that gas phase H$_2$O abundances were higher.   

A variation of this model is that of Spaans \& van Dishoeck (2001)
and Bergin et al (2003), where it was noticed that the water emission seemed to
trace the photodissociation region (PDR, see Hollenbach \& Tielens 1999),
which lies on the surface ($A_V \lta 5$) of the molecular cloud.  A
two component model was invoked in which the water froze out as ice
in the dense clumps, but remained relatively undepleted in the low density
interclump gas because of the longer timescales for freezeout.  Thus, the 
average gas phase water abundance was derived from a mix of heavily 
depleted and undepleted gas.

We propose in this paper a new model of the H$_2$O and O$_2$ chemistry
in a cloud.  In this model, we assume that the molecular cloud lifetime
is sufficiently long to allow freezeout\footnote{Indeed, the freezeout
of volatile molecules such as CO is now commonly observed in the dense, cold
regions of molecular clouds (e.g., Bergin \& Tafalla 2007)} and reduce
 the gas phase H$_2$O and O$_2$ abundances to
very low values in the central, opaque regions of the cloud.   
The key to understanding the H$_2$O emission and the O$_2$ upper limits
observed by SWAS and {\it Odin} is to model the {\it spatially-dependent} H$_2$O
and O$_2$ abundances through each cloud.  At the surface of the cloud
the gas-phase H$_2$O and O$_2$ abundances are very low because of the 
photodissociation
by the ISRF or by the FUV flux from nearby OB stars.  Near the cloud
surface the dust grains have little water ice because of photodesorption
by the FUV field.  Deeper into the cloud, the attenuation of the FUV
field leads to a rapid rise of the gas-phase H$_2$O and O$_2$ abundances, which
peak at a hydrogen nucleus column $N_{f} \sim 10^{21.5} - 10^{22}$ cm$^{-2}$
(or depths $A_{Vf} \sim $ several into the cloud) and plateau at this value for $\Delta A_V
\sim$ several beyond $A_{Vf}$, insensitive to
the gas density $n$ and  the incident FUV flux $G_0$ (scaling factor in multiples of the 
avearage local
interstellar radiation field).  At these 
intermediate depths, photodesorption of some of the water ice by FUV photons keeps
the gas phase water abundance
high (the ``$f$'' in $N_f$ and $A_{Vf}$ 
signifies the onset of water ice freezeout, as will
be discussed in \S 2.5 and \S3.3).  The FUV is strong enough to keep some of
the ice off the grains, but, due to efficient water ice formation followed by
photodesorption, it is not strong enough to dissociate all gas-phase water.  At
greater depths, gas phase reactions other than photodissociation
begin to dominate gas phase H$_2$O destruction, and the steady state
gas phase H$_2$O and O$_2$ abundances plummet as all the gas phase elemental O
is converted to water ice.  Thus, the overall structure of a molecular cloud
has three subregions: at the surface is a highly  ``photodissociated region'',
at intermediate depths is the ``photodesorbed region'', and deep in the cloud
is the ``freezeout region''.

In this new model, the H$_2$O and O$_2$ emission mainly arise
from the layer of high abundance gas in the plateau which starts at $N_{f}$
and extends somewhat beyond, so that the
emission is a (deep) ``surface'' process rather than a ``volume'' process.
For clouds with columns $N>N_{f}$, the H$_2$O and O$_2$ emission
becomes independent of cloud column, assuming the incident FUV field
and gas density are fixed. The model gives {\it column-averaged} abundances
 through the dense cores of $\lta  10^{-8}$ for H$_2$O
and O$_2$ for low values of $G_0 < 500$, but the local abundances peak at values which
are at least an order of magnitude larger than these values.  For cloud
columns $N>N_{f}$, the average H$_2$O abundance scales as $N^{-1}$.

The implications of this scenario for molecular clouds
may be much broader than this particular model of H$_2$O and O$_2$ chemistry
in clouds.  Other molecules such as CO, CS, CN, and HCN require
a spatial model of their distribution in a cloud, with photodissociation
and photodesorption the dominant processes near the cloud surface, and
the freezing out of the molecules the dominant process deeper into
the cloud.  In addition, the adsorption process and creation of abundant
ice mantles changes the relative gas phase abundances of the elements.
In the case considered in this paper, the C/O ratio in the gas may
go from 0.5 at the cloud surface to unity or greater in the H$_2$O freezeout
region.
Such changes in the gas phase C/O ratio have major implications on {\it
all} gas phase chemistry (e.g., Langer et al 1984, Bergin et al 1997).

The interesting and important caveat to this relatively simple 
steady state model is the effect of time-dependence in raising abundances
of, for example, gas phase H$_2$O, O$_2$, and CO above steady state values
in the opaque ($A_V > 5-10$) interiors of clouds.  One such effect is 
that at low densities, $n\lta 10^3$ cm$^{-3}$, species do not have time to freeze
out within cloud lifetimes and therefore have much higher gas phase abundances.
We have also uncovered a new {\it time-dependent} process
that may elevate the H$_2$O and O$_2$ abundances for times $t\lta 10^7$ years
in the freezeout region at very high $A_V$ {\it even at high cloud densities}
$n\sim 10^5$ cm$^{-3}$. If the grains are sufficiently cold
to freeze out CO ($T_{gr} \lta 20$ K or $G_0 \lta 500$) at high $A_V$, a
CO/H$_2$O ice mix rapidly ($t \lta 10^{5-6}$ years) forms. The steady
state solution has very little CO ice with most of the O in H$_2$O ice.
However, the time to convert CO ice to H$_2$O ice is very long, and the
bottleneck is the cosmic ray desorption of CO from the CO/H$_2$O ice mixture.
This desorption provides gas phase CO which then acts as a reservoir to
produce gas phase H$_2$O and O$_2$, until all the oxygen eventually freezes
out as water ice (i.e., the same mechanism as described in Bergin et al (2000)
except the gas phase CO in this case comes from the CO ice). Depending on
the assumptions regarding the CO desorption process, this cosmic ray CO
desorption timescale may range from $10^5$ to $>>10^7$ years.  If the
timescale is roughly 0.1 to 1 times  the cloud age, a maximum gas phase H$_2$O
and O$_2$ abundance is produced at that time.  
Although not as large as
the peak abundances produced at $A_{Vf}$, these abundances can be significant
and can contribute to the total column of these species if the cloud
has a high total column ({\it but only if the CO desorption time is
0.1 to 1 times the cloud age}).  

This paper is organized as follows.  In \S 2 we describe the new
chemical/thermal model of an opaque molecular cloud illuminated
by FUV radiation.  The major changes implemented in our older
photodissociation region (PDR) models (Kaufman et al 1999) include
the adsorption of gas species onto grain surfaces, chemical reactions on grain 
surfaces,
and the desorption of molecules and atoms from grain surfaces.
In \S 3 we show the results of the numerical code
as functions of the cloud gas density $n$, the incident FUV flux $G_0$,
and the grain properties.  We  present a simple analytical model
that explains the numerical results and how the abundances scale with
depth and with the other model parameters. We discuss time dependent models
for the opaque cloud center.  In \S 4
we apply the numerical model to both diffuse clouds and dense clouds. 
 Finally, we summarize our results and conclusions in \S 5.  For the
convenience of the reader, Table 3 in the Appendix lists the symbols 
used in the paper.

\section{THE CHEMICAL AND THERMAL MODEL OF AN OPAQUE MOLECULAR CLOUD}

\subsection{Summary of Prior PDR Model and Modifications}

The numerical code we have developed to model the chemical
and thermal structure of an opaque cloud externally illuminated by FUV
flux is based on our previous models of photodissociation regions (PDRs),
which are described in Tielens \& Hollenbach (1985), Hollenbach et al (1991), and
Kaufman et al (1999,2006). This 1D code modeled a 
constant density slab of
gas, illuminated from one side by an FUV flux of $1.6\times 10^{-3}G_0$ erg 
cm$^{-2}$ s$^{-1}$. The unitless parameter $G_0$ is defined above in
such a way that $G_0 \sim 1$ corresponds to the average local interstellar radiation
field in the FUV band (Habing 1968).  The subscript ``0'' indicates that the flux is
the incident FUV flux, as opposed to the attenuated FUV flux deeper into
the slab.  The code calculated the steady state chemical abundances and
the gas temperature from thermal balance as a function of depth into the
cloud.  It incorporated $\sim 45$ chemical species, $\sim 300$ chemical 
reactions, and a 
large number of heating mechanisms and cooling processes.  The chemical
reactions included FUV photoionization and photodissociation; cosmic ray
ionization; neutral-neutral, ion-neutral, and electronic recombination
reactions, including reactions with charged dust grains; and the 
formation of H$_2$ on grain surfaces.  No other grain
surface chemistry was included in these prior models.  The  main route
to gas phase H$_2$O and O$_2$  through these gas-phase reactions is schematically 
shown in Fig. 1. It starts with
the ionization of H$_2$ by cosmic rays or X-rays, which eventually leads
to H$_3$O$^+$ that recombines to form gas phase water.  It also recombines
to form OH which reacts with O to form gas phase O$_2$. The code contained a
relatively limited carbon chemistry; 
we included CH and CH$_2$,  as well as their associated ions, but 
did not include CH$_3$, CH$_4$, C$_2$H or longer carbon chains, 
nor any carbon molecules  that include S or N.  

The code modeled regions which lie at hydrogen column densities 
$N \lta 2 \times 10^{22}$ cm$^{-2}$ (or $A_V \lta 10$) from the surface
of a cloud.  Therefore, it applied not only to the                   
photodissociated  surface region, where gas phase hydrogen and oxygen
are nearly entirely atomic and where gas phase carbon is mostly C$^+$,
but also to regions deeper into the molecular cloud where hydrogen is in
H$_2$ and carbon is in CO molecules.  Even in these molecular regions,
the attenuated FUV field can play a significant role in photodissociating
H$_2$O and O$_2$, and in heating the gas.

The dust opacity is normalized to give the observed visual extinction per unit H
nucleus column density that is observed in the diffuse ISM (e.g., Savage \& Mathis
1979, Mathis 1990).  The extinction in the FUV, assumed to go as $e^{-b_\lambda A_V}$,
is taken from PDR models described most recently in Kaufman et al (1999, 2006),
and mainly derived from the work of Roberge et al (1991).  Here $b_\lambda$ is the
scale factor which expresses the increase in extinction as one moves from the visual
to shorter wavelengths.  However, the FUV extinction law in molecular clouds is not
well known.  We discuss in \S3.2 how our results would vary if $b_\lambda$ varies.

In the following subsections (\S 2.2 - \S 2.7) we describe the modifications 
to the basic
PDR code which we have made to properly model the gas phase H$_2$O and O$_2$
abundances from the surfaces of molecular clouds to the deep interiors,
where the gas phase elemental oxygen  freezes out as water 
ice mantles on grain surfaces.  The modifications include the freezeout of gas
phase species onto grains, the formation of OH and H$_2$O as well
as CH, CH$_2$, CH$_3$ and CH$_4$ on grain 
surfaces,
and various desorption processes of species from grain surfaces.  Fig. 
\ref{fig:grainsurface} 
shows the grain surface formation process for water ice.

One key modification is the application of time dependent chemistry to the
deep, opaque interior.  Here, certain chemical timescales are comparable
to cloud lifetimes, and steady state chemical networks such as our PDR code
do not apply. The time dependent code also increases the number of reactions
and species we follow, and this particularly improves the treatment of the carbon
chemistry in the cloud interior.  Further discussion is found in \S 2.3, 2.4 and  3.7.

\subsection{Dust Properties}

The basic PDR code implicitly assumed a ``MRN'' (Mathis, Rumpl \& Nordsieck 
1977) grain size distribution
$n_{gr}(a) \propto a^{-3.5}$, where $a$ is the grain radius, extending
from very small ($\lta 10$ \AA) polycyclic aromatic hydrocarbons (PAHs)
to large ($\sim 0.25$ $\mu$m) dust grains.  With this distribution the
largest grains provide the bulk of the grain mass, and the intermediate-sized
grains ($\sim 100$ \AA) provide the bulk of the FUV extinction.  The smallest
particles and PAHs provide the bulk of the surface area, and dominate
the grain photoelectric heating process and the formation of H$_2$ on
grain surfaces.  The code implicitly assumes an MRN distribution in the 
sense that the equations for attenuation of incident photons by dust, for
grain photoelectric heating, for H$_2$ formation, and for gas-grain collisions
and the charge transfer that may occur in such collisions all assume
such a distribution.

For the process of the freezeout of oxygen species onto grain surfaces, we also 
implicitly assume an MRN distribution, in the sense that we adopt a cross sectional
grain area 
$\sigma _H$ appropriate for an MRN distribution. As will be shown in \S3, $\sigma _H$,
which is one fourth the surface area available for freezeout, 
is an important parameter for determining the abundance of gas-phase $\rm 
H_2O$ in our models. Below in \S 2.5, we show that $\sigma _H$ is 
roughly the total cross sectional area of grains (per H) that are larger than about 20 \AA.  
Smaller grains experience single photon heating events which clear them of 
ice mantles (Leger, Jura, \& Omont 1985).  In addition, cosmic rays may
clear small grains of ices (Herbst \& Cuppen 2006). 
In an MRN distribution, the cross sectional area  of grains with radii larger 
than about 20 \AA \ is $\sigma _H \sim 2\times 10^{-21}$ cm$^{2}$ per H.

We note, however, that the cross sectional area is somewhat uncertain.  It could 
be smaller if grains coagulate inside of dense molecular clouds.  It could 
be larger due to ice buildup. 
We note that once all oxygen-based ices freeze out onto such an MRN distribution, it forms a 
constant  mantle thickness of $\Delta a \sim 50$ \AA \ regardless of grain size.  
While this is 
insignificant to the large ($\sim 0.1$ $\mu$m) grains which contain most of the 
mass, it represents a very large change in radius and area of the smallest 
($\sim 20$ \AA) grains which provide most of the grain surface area.  
The increased surface area of grains at the end 
of this
freezeout will proportionally increase subsequent gas-grain interaction rates. 
Thus, neutralization of ions by grains, formation
of H$_2$ on grain surfaces, the freezeout of species, and the cooling or heating 
of the gas by gas-grain  collisions may all be
enhanced by significant amounts in regions where water ice has totally frozen 
out on grains. 
However, because of the possibility of grain coagulation which reduces grain 
surface area, and because of the uncertainly in the lower size cutoff of 
effective grains, we have chosen to fix $\sigma _H$
and  treat it as a (constrained) variable with values ranging
from $\sigma _H = 6\times 10^{-22}$ to 
6 $\times 10^{-21}$ cm$^{2}$ per H. 

\subsection{Adsorption and Time Dependent Chemistry} 

Adsorption is the process by which a gas phase atom or molecule hits a grain
and sticks to the surface due to van der Waals or stronger surface bonding 
forces.
Typically, at the low gas temperatures we will consider, $T_{gas} \lta 50$ K, 
the
sticking probability is of order unity for most species (Burke \& Hollenbach 
1983).
We assume unit sticking coefficients here for all species heavier than helium.
Thus, the timescale
to adsorb (freeze) is the timescale for a species to strike a grain surface: 

\begin{equation}
t_{f,i} \sim [n_{gr} \sigma _{gr} v_i]^{-1}, 
\end{equation}
where $v_i$ is the 
thermal speed of species $i$ and where the grain number density $n_{gr}$
times the grain cross section $\sigma _{gr}$ is the average over the size 
distribution.
As discussed above, in an MRN distribution with a lower cutoff size of $a \sim 
20$ \AA,
$n_{gr} \sigma _{gr} \simeq 2 \times 10^{-21} n$ cm$^{-1}$, where $n$ is the gas phase
hydrogen nucleus density in units of cm$^{-3}$.  Therefore, 
$t_{f,i} \sim 8 \times 10^4 (m_O/m_i)^{1/2} (10^4 {\rm cm^{-3}}/n) (30 {\rm K}/T)^{1/2}$ 
years, where $m_O$ is the mass of atomic O and $m_i$ is the mass of the species.
Since molecular cloud lifetimes
are thought to be $2\times 10^6$ to $2\times 10^7$ years, and the molecules reside in regions 
with $n \gta 10^3$ cm$^{-3}$, this short timescale suggests that molecules should 
mostly freeze out in molecular clouds unless some process desorbs them from grains.  
Dust vacuums the condensibles quickly!

Consider the 
depletion of CO in regions where the grain
temperatures are too high to directly freeze out CO.  CO freezes out when grain 
temperatures
are $T_{gr} \lta 20$ K (see \S 2.4 below).  However, H$_2$O freezes out when 
grain temperatures
are $T_{gr} \lta 100$ K at the densities in molecular clouds.  Suppose that 20 
K$< T_{gr} <
100$ K, so that gas phase elemental oxygen not in CO 
freezes out as water ice, but CO does not directly 
freeze.
CO is continuously destroyed by He$^+$ ions created by cosmic rays, producing 
C$^+$ and O.
The resultant O has two basic paths: it can reform CO, or it can freeze out on 
grain surfaces as water ice, never to return to the gas phase in the absence of 
desorption
mechanisms. The latter process therefore slowly depletes the CO gas phase 
abundance, by
lowering the available gas phase elemental O.  We see this effect in the late 
time
evolution of CO in Bergin et al (2000).
One effect, not included in Bergin et al, that slows this process is that for 
grain temperatures $\gta 20$ K adsorbed atomic O may be rapidly thermally
desorbed before reacting with H and therefore the production of H$_2$O ice by
grain surface reactions may be substantially curtailed (this will
be discussed further in \S 2.5 and \S 3.2).  Nevertheless, gas phase 
H$_2$O will form and will freeze out, and therefore, given enough time, the
elemental O not incorporated in refractory (e.g., silicate) grain material
is converted to H$_2$O ice.  If elemental C does not evolve to a similarly
tightly bound carbonaceous ice, there is the potential for the gas phase C/O
number ratio to exceed unity.
This will have consequences for astrochemistry 
(e.g., Langer et al 1984, Bergin et al 1997, Nilsson et al 2000), 
and we show some exemplary results in \S 3.5.

Consider as well the case where $T_{gr}\lta 20$ K deep in the cloud
(i.e., at $A_V > 5$; this low $T_{gr}$  corresponds to $G_0 \lta 500$).  
In this case, the gas phase
CO freezes out with the H$_2$O, forming a CO/H$_2$O ice mix.
In order for this CO ice to be converted
to H$_2$O ice, the CO needs to be desorbed by cosmic rays.  It then reacts
in the gas phase with He$^+$ as described above and eventually the freed O
goes on to form mostly water ice.  The timescale for this CO cosmic ray
desorption can be quite long, and may dominate the slow evolution of the
chemistry toward steady state. 

Unfortunately, this timescale is not well known.  First of all, cosmic ray
desorption rates are not well constrained.  But secondly, the timescales
depend on the structure of the CO/H$_2$O ice mix, and this structure  is not certain.
If the molecules are immobile, the CO and H$_2$O ice molecules are initally mixed
and each monolayer contains this mix.  However, there is some evidence that
either pure CO ice pockets form, or the CO accretes on top of an already formed
layer of water ice (Tielens et al 1991, Lee et al 2005, Pontoppidan 
et al 2004, 2008).  Even if the CO and H$_2$O
are initially mixed and immobile, 
cosmic rays desorb CO much more efficiently than H$_2$O due to the much smaller
binding energy of adsorbed CO compared with adsorbed H$_2$O (Leger, Jura, \& Omont 1985).  
This could
lead to surface monolayers with very little CO compared to the protected layers
underneath the surface.  The cosmic ray desorption rate is directly proportional
to the fraction of {\it surface} sites occupied by CO.  In the examples above, this
fraction could range from zero to unity.  Thus, the desorption timescale could
range from large values ($>>10^7$ years) to small values ($\sim 10^5$ years).
We compute these timescales in \S 2.4.

Because of the moderately long timescale for the depletion of either gas phase
or ice phase CO via 
this freezing out
of the elemental O in the form of water ice, we apply time dependent chemical 
models (see \S 3.7)
to the cloud interiors which are run for the presumed lifetimes ($\sim 10^7$ 
years) of molecular clouds.  Time dependent models are also warranted for
clouds of relatively low density, $n \lta 10^3$ cm$^{-3}$, where $t_f \gta
2\times 10^6$ years (see Eq. 1).

\subsection{Desorption}

To desorb a species from a grain surface requires overcoming the binding 
energy
which holds the species to the surface.  Table 1 lists the binding energies of
the key species we treat.  Note that water has a very high binding energy and is 
therefore
more resistant to desorption.  

{\it Thermal Desorption}.  The rate $R_{td,i}$  per atom or molecule of thermal 
desorption from
a surface can be written:

\begin{equation}
R_{td,i} \simeq \nu _i e^{-E_i/kT_{gr}},
\end{equation}
where $E_i$ is the adsorption binding energy of species $i$ and $\nu_i= 
1.6\times 10^{11}\sqrt{(E_i/k)/(m_i/m_{\rm H})}$ s$^{-1}$ is the vibrational
frequency of the species in the surface potential well, $k$ is Boltzmann's 
constant  and $m_i$ and $m_H$ are the mass of species $i$ and 
hydrogen, respectively.  One can find the temperature at
which a species freezes by equating the flux $F_{td,i}$ of desorbing molecules 
from the ice surface
to the flux of adsorbing molecules from the gas, 

\begin{equation}
F_{td,i}\equiv N_{s,i} R_{td,i}f_{s,i} = 0.25 n_i v_i,
\end{equation}
where $N_{s,i}$ is the number of adsorption sites per cm$^2$ ($N_{s,i} \sim 
10^{15}$ sites cm$^{-2}$), $f_{s,i}$ is the fraction of the 
surface adsorption sites that are occupied by species $i$,
$n_i$ is the gas phase number density of species $i$ and $v_i$ is its thermal 
speed.  We have
assumed sticking probability of unity.  From these equations, one can calculate 
the
freezing temperature $T_{f,i}$ of a species by solving for $T_{gr}$.

\begin{equation}
T_{f,i} \simeq (E_i/k)\left[ \ln \left({{4 N_{s,i}f_{s,i} \nu _i} \over {n_i v_i}}\right) 
\right]^{-1}
\end{equation}
\begin{equation}
T_{f,i} \simeq (E_i/k)\left[ 57 + \ln \left[ \left({N_{s,i} \over {10^{15}\ {\rm 
cm^{-2}}}}\right)
\left({\nu _i \over {10^{13}\ {\rm s^{-1}}}}\right)\left({{1\ {\rm cm^{-
3}}}\over n_i}\right)
\left({{10^4\ {\rm cm\ s^{-1}}}\over v_i}\right)\right]\right]^{-1}.
\end{equation}
This equation shows that at molecular cloud conditions the freezing temperature 
$T_{f,i}$
is about $0.02E_i/k$, or about 100 K for H$_2$O but only $\sim 20$ K for CO (see 
Table 1)
We include thermal desorption rates from our surfaces as given by these 
equations. We compute
the number of layers of an adsorbed species on the grain, and only thermally 
desorb from the surface layer. 

{\it Photodesorption}.  The flux $F_{pd,i}$ of adsorbed particles leaving a 
surface due to photodesorption is given by

\begin{equation}
F_{pd,i}= Y_i F_{FUV}f_{s,i},
\end{equation}
where $Y_i$ is the photodesorption yield for species $i$, averaged over the 
FUV wavelength band, and $F_{FUV}$ is the
incident FUV photon flux on the grain surface.  We write 
the FUV photon flux at a depth $A_V$ into the cloud as (see Tielens \& Hollenbach 1985)

\begin{equation}
 F_{FUV}= G_0 F_0 e^{-1.8 A_V},
\end{equation}
where $F_0\simeq 10^8$ photons 
cm$^{-2}$ s$^{-1}$ is approximately the local interstellar flux of 6eV--13.6eV 
photons and $G_0$ is the commonly used scaling factor (see \S 2.1).

Reliable photodesorption yields are often hard to find in the literature, but 
for
the case of H$_2$O a laboratory study has been done for the yield created by
Lyman $\alpha$ photons (Westley et al 1995a, b).  The mass loss at large photon 
doses is
proportional to the photon flux, showing that desorption is not due to 
sublimation
caused by absorbed heat.  The yield is negligible when infrared or optical 
photons
are incident.  As an approximation, we will assume the average yield in
the FUV (912-2000 \AA) is the same as the Lyman $\alpha$ (1216 \AA) yield.  Most
of the photodesorption occurs when the incident photon is absorbed in the first
two surface monolayers of the water ice (Andersson et al 2006). The 
value
of the yield ($\sim 1-8 \times 10^{-3}$) is perhaps 10 times smaller than
 the probability that the
incident photon is absorbed in the first two surface monolayers, so the 
photodesorption efficiency
from these layers is of order 10\%.  We note that this yield from Westley et al
is similar to the values that \"Oberg (private communication) has found in 
preliminary experiments with a broad band FUV source.

 One peculiar result from the Westley et al experiments
 is that the yield first increases with the dose of photons,
until it hits a saturation value for that flux. The saturation dose is of order
3 $\times 10^{18}$ photons cm$^{-2}$. In the fields we are considering, $G_0 
\sim 1-10^5$,
this dose is achieved in less than 1000 years.  Thus, our grains may well be 
saturated.
The interpretation of the increase of the photodesorption yield with dose
is that photolysed radicals like O, H and OH build up on the surface to a 
saturation value.  Chemical reactions of a newly-formed radical with a 
pre-existing radical may lead to the desorption of the product (usually H$_2$O)
and explain the dose dependence of the yield.
Adding further weight to this interpretation is that the saturated yields 
decline with decreasing
surface temperature ($Y_{H_2O}$ is 0.008 at $T_{gr}=100$ K, 0.004 at 75 K, and 
0.0035
at 35 K).  As temperatures decrease the radicals cannot diffuse across the
surface as well to react with other radicals, nor can activation bonds be as 
easily
overcome.  Since presumably radicals like O, H  and OH are being created, and it 
is
the process of reforming more stable molecules that may help initiate 
desorption--and may kick off neighboring molecules, 
other molecules may desorb such as OH, H, O$_2$, H$_2$. These are indeed observed,  
but Westley et al report that most of
the desorbing molecules are water molecules. We note, however, a recent 
theoretical calculation by Andersson et al (2006)
which suggests that one might 
expect OH and H to be major desorption products, but confirming that the yield of H$_2$O 
photodesorption may still be of order 10$^{-3}$. We also find (see \S 4.1
below) that adopting a yield for OH+H photodesorption from water ice
which is twice that of H$_2$O photodesorption provides better agreement
to the OH/H$_2$O abundance ratio measured in translucent clouds.  We define
$Y_{OH,w}$ as the yield of OH photodesorbing from water ice; $Y_{OH,w}=2Y_{H_2O}$.
This distinguishes $Y_{OH,w}$ from $Y_{OH}$, which is the yield of OH photodesorbing
from a surface covered with adsorbed OH molecules.

The interpretation that radical formation in the surface layer is important in 
creating
photodesorption leads one to re-examine the issue of whether grains near the 
surfaces of
molecular clouds will indeed be saturated with radicals on their surfaces.  
While it is
true that the photon dose is sufficient, this dose was delivered over a long 
timescale.
Westley et al (1995b) point out that it is possible in this case that radicals 
may
recombine in the time between photon impacts within an area of molecular dimensions
($\sim 10^{-15}$ cm$^2$), which might result in smaller yields than the 
saturated ones.

In light of the uncertainties in the wavelength dependence of the yield over the 
FUV range, the photodesorption products, and
the level of saturation, we adopt $Y_{H_2O} = 10^{-3}$ and $Y_{OH,w}= 2\times 10^{-3}$
as our  ``standard'' values,
but we will show models in which these yields are  varied by a factor of 10 from 
their standard values, maintaining their ratio of 0.5.
We will also  use the observations of $A_V$ thresholds for water ice formation 
to constrain
the absolute values of these yields (\S 4.2.1), and the observations of OH and
H$_2$O in diffuse clouds to help determine their ratio (\S 4.1).  

Recent laboratory measurements of CO photodesorption imply a yield similar
to that for H$_2$O (\"Oberg et al 2007). Therefore, we adopt $Y_{CO}=10^{-3}$
as our standard value for CO.
Unless $T_{gr} < 20$ K, adsorbed CO is anticipated
to be rare
because it thermally desorbs so rapidly. However, in cases with low $G_0$, CO 
may occupy $\sim 1/3$ of the surface sites at the H$_2$O freezeout column.  

For the other species there is less or no reliable photodesorption data in the literature.
Since our grain surface is largely H$_2$O ice, and the photodesorption is via the production
of radicals (and possibly the reformation of H$_2$O from radicals), we assume that the yield 
for O is $Y_O=10^{-4}$ (a smaller yield since no reformation possible). 
The yields for O, OH and O$_2$ are less important, because in our models they 
rarely occupy
many of the surface sites. The first two are rare because they rapidly react 
with H to form H$_2$O.  Table 1 lists the photodesorption yields we adopted.

A complication arises if grains move {\it rapidly} from surface regions (higher 
temperature and possible lower density) where H$_2$O ice can exist to deeper
(lower temperature and higher density) regions where CO ice forms and then
back again to the surface region.  The fraction of the surface now covered
by H$_2$O ice, $f_{s,H_2O}$, is now nearly zero, since CO covers the surface.
Therefore, H$_2$O and OH photodesorption rates are very low.  We have ignored this
effect, assuming a mix of CO and H$_2$O ice in each adsorbed monolayer.\footnote{Except,
as discussed later in this section, when we test the sensitivity of our results to
CO cosmic ray desorption.}  In this paper the region where H$_2$O and OH photodesorption
are important is the intermediate or ``plateau'' region where gas phase H$_2$O
and OH abundances peak.  Here, the photodesorption and/or thermal desorption timescales
of CO are much shorter than the dynamical timescales to bring CO-coated ice grains
from the deep interior to the plateau region.

{\it Desorption by Chemical Reaction}.  There have been some attempts to model 
the desorption
of molecules due to chemical reactions on grain surfaces.  D'Hendecourt et al (1982)
and Tielens \& Allamandola (1987) discuss an exploding
grain hypothesis in which the UV photons produce radicals in the grains, which 
are relatively
immobile until the grain temperature is increased.  At this point, they begin 
to diffuse
and react, and the chemical  heat causes a chain reaction exploding the grain and 
``desorbing'' species
in large bursts.  Our photodesorption is a sort of steady state version of this.  
O'Neill,
Viti, \& Williams (2002) examine the effects of hydrogenation on grain surfaces, 
and in
some of their models they assume that the reaction of a newly adsorbed H atom 
with radicals
on the surface such as OH will release enough energy to desorb the resultant 
hydrogenated molecule.
In this case, it is the chemical energy carried in by the H atom, 
and not
the photodissociation of the ice itself, which leads to desorption.  In our standard case
we assume that H$_2$ desorbs on formation because of its low adsorption binding energy
and its low mass, but that other forming molecules do not.  However, we ran one 
case (see \S 3.5) in
which we include this desorption process for all forming molecules
in our grain surface chemistry.

{\it Desorption by Cosmic Rays and X-rays}.  Deep in the cloud, cosmic ray desorption of ices 
becomes important for maintaining trace amounts of O-based and C-based species 
in the gas phase. Cosmic ray desorption rates per surface molecule (i.e., molecule in
the top monolayer of ice) for various
species are listed
in Table 1.
As discussed above, the  desorption rates depend
on the abundance of a particular adsorbed species in the surface monolayer.
In our standard steady state models, we assume that if there are a number
of species (e.g., CO, CH$_4$, H$_2$O) that are adsorbed, that they are well
mixed and each monolayer of ice contains the same mix.  However, as discussed
above, for CO the fraction of the surface layer containing CO may range from
zero to unity.  The timescale for the cosmic ray desorption of a species $i$
is given:

\begin{equation}
t_{cr,i} =   {{x(i)}\over{ 4 R_{cr,i} N_{s,i} f_{s,i} \sigma _H}},
\end{equation}
where $x(i)$ is the abundance of the adsorbed ice species and $R_{cr,i}$
is the cosmic ray desorption rate per surface atom or molecule and is given in
Table 1.  The cosmic ray desorption time for H$_2$O$_{ice}$ is long 
($\sim 10^{10}$ years) and is not significant in our models.  However,
the timescale for CO desorption under certain circumstances may be
comparable to the age of the cloud, and can therefore lead to a source of
gas phase CO which can then serve as a source of gas phase H$_2$O and
O$_2$, as discussed above.  Substituting from Table 1 for $R_{cr,CO}$,
assuming an initially high abundance of CO ice, $x(CO_{ice}) = 10^{-4}$,
and taking $f_{s,CO}=0.3$, we obtain $t_{cr,CO} \sim 2.3 \times 10^6$
years.  However, both $R_{cr,CO}$ and $f_{s,CO}$ are uncertain.  In
the case of the fractional surface coverage $f_{s,CO}$, the cosmic
ray desorption process itself, acting on the top monolayer, could 
result in a very low value of $f_{s,CO}$ by essentially purging the
surface of CO and leaving a fairly pure monolayer of H$_2$O$_{ice}$ (see \S 2.3).
On the other hand, there may be evidence for a non-polar CO ice surface
on top of a polar H$_2$O ice mantle resulting in $f_{s,CO}=1$ (see \S 2.3).
Because of the uncertainly in
the product $f_{s,CO} R_{cr,CO}$ we will examine in \S 3.7 a range
of values of this parameter extending from $10^{-15}$ to $10^{-11}$
s$^{-1}$.  
The corresponding timescale to remove all
the ice is about  $3 \times 10^5$
to $3\times 10^9$ years, ranging from less than to much greater than cloud ages. 

Cosmic rays also lead to the production of FUV 
photons (Gredel et al. 1989) at a level equivalent to $G_0\sim 10^{-3}$ 
throughout the 
cloud.     We include this flux in determining both the photodissociation of 
gas-phase
molecules and the photodesorption of ices.  

Following Leger, Jura, \& Omont (1985) we assume that, in general, cosmic
ray desorption dominates over X-ray desorption and we therefore ignore this process.
Of course, in regions close to X-ray sources and where the UV is shielded by dust,
 X-ray desorption will become more significant.

\subsection{Timescales Related to Freezeout}

Several timescales are needed to 
understand the freezing process fully. Assuming that the grain size distribution 
extends to very small grains (or large molecules) whose size is of order a few 
Angstr\"oms, we first compute timescales relevant for single photons to clear 
grains of ice due to transient heating by the photon.
%
The peak temperature that a silicate grain of radius
$a$ achieves after absorbing a typical $\sim 10$ eV FUV photon is
given (Drapatz \& Michel 1977, Leger, Jura \& Omont 1985)

\begin{equation}
T_{max}\sim 100 \left({{20\ {\rm\AA}}\over a}\right)^{1.3} \ {\rm K}.
\end{equation}
The timescale for this grain to radiate away the energy of this photon is

\begin{equation}
t_r \sim 100\left({{100\ {\rm K}}\over T_{max}}\right)^5 \left({{20\ {\rm \AA}}\over 
a} \right)^3\ {\rm seconds}.
\end{equation}
On the other hand, the timescale for a single H$_2$O molecule to thermally 
evaporate from
an ice surface on the grain is given
\begin{equation}
t_{evap,H_2O} \sim 4 \times 10^{-13} e^{{4800\ {\rm K}}\over T_{max}} \ {\rm seconds},
\end{equation}
where $\Delta E/k=4800$ K is the binding energy of the H$_2$O molecule (Fraser et al 2001)
 and  $4\times 10^{-13}$
seconds is the period of one vibration of the molecule in the ice lattice (see 
Eqn 2. and following text).  For
comparison with the above equations, $t_{evap,H_2O}\simeq 10^{29}$ seconds at 
$T_{max}=50$ K,
$3 \times 10^8$ seconds at $T_{max}=100$ K, and 30 seconds at $T_{max}=150$ K.  Comparison
with $t_r$ indicates that the ice mantle will evaporate before cooling when 
$T_{max}
\sim 150$ K.  Comparison with the equation for $T_{max}$ shows that $a
\gta 15-20$
\AA \ is required for grains to be large enough such that single photon events 
do not clear ice mantles.  Thermal evaporation of H$_2$O will also cool the 
grain.  Since
the binding energy of an H$_2$O molecule is $\sim 0.5$ eV, $\sim 20$ H$_2$O 
molecules must evaporate to offset the heating due to the absorption of a single
UV photon. However,  we show below that
10 eV photons are absorbed by 20 \AA \ grains more rapidly than O atoms
hit the grains to
form ice.  Therefore, thermal evaporation cannot significantly cool the 
transiently heated
grains compared with radiative cooling, and an entire ice mantle never builds up 
on the grain surface. 
The conclusion holds that $a \lta 20$ \AA \  grains
lose their ice in the presence of single photon heating events. In summary,
for the purposes of modeling oxygen freezeout and ice formation, we assume 
that the effective surface area for ice freezeout is given by the area of 
grains that are larger than $\sim 20 $ \AA.

Several more timescales reveal the fate of an oxygen atom once it sticks to a 
grain.  The
timescale $t_{gr,O}$ for a given grain of 
radius $a$ to be hit by  a gas phase oxygen atom
is given

\begin{equation}
t_{gr,O} \sim 2.5\times 10^4 \left({10^{-4}\over x(O)}\right)\left({{10\ {\rm 
K}}\over T_{gas}}
\right)^{1/2}\left({{0.1\mu\rm m}\over a}\right)^2\left({{10^5\ {\rm cm}^{-
3}}\over
n}\right)\ {\rm seconds},
\end{equation}
where $x(O)$ is the gas phase abundance of atomic oxygen relative to hydrogen.  
Similarly,
the timescale $t_{gr,H}$ for the same grain to be hit by a hydrogen atom from the gas 
is given

\begin{equation}
t_{gr,H} \sim 6.0\times 10^{-1} x(H)^{-1}\left({{10\ {\rm K}}\over T_{gas}}
\right)^{1/2}\left({{0.1\mu \rm m}\over a}\right)^2\left({{10^5\ {\rm 
cm}^{-3}}\over
n}\right)\ {\rm seconds},
\end{equation}
where $x(H)$ is the abundance of atomic hydrogen.  Note that at the column or 
$A_V$
into the cloud where water begins to freeze out on grain surfaces, $x(H)$ is likely
less than unity because most H is in H$_2$.  However, if $x(H) \gta  0.25x(O)$,
then H atoms stick more frequently than O atoms. Provided 
photodesorption or thermal desorption does not
intervene, and assuming that the H atoms are quite mobile on the surface 
and find the adsorbed O atom, O atoms will react with H atoms on the surface to 
form OH and then H$_2$O.\footnote{In our runs we find only a few cases
where $x(H)< x(O)$; this occurs at high $A_V$ for cases with high $G_0$, where
grains are so warm that O thermally desorbs before forming OH on the grain surface.
In these cases, there is no buildup of adsorbed O atoms on the surface and
therefore no resultant formation of O$_2$ or CO$_2$ on grain surfaces, as might otherwise be
expected
(Tielens \& Hagen 1982).}

These timescales need to be
compared to the timescale for UV photons to photodesorb the O atom and the 
timescale for O atoms to thermally desorb from grains. The timescale for a grain 
of size $a$ $\gta 100$ \AA \ to absorb an FUV photon is given by:

\begin{equation}
t_{\gamma} \sim 30 G_0^{-1} e^{1.8A_V} \left({{0.1\mu\rm m}\over 
a}\right)^2
\ {\rm seconds}.
\end{equation}

The timescale to photodesorb an O atom from a surface covered with a 
covering fraction $f_{s,O}$ of O atoms is then,

\begin{equation}
t_{\gamma -des} \sim 3\times 10^5 G_0^{-1} e^{1.8A_V} f_{s,O}^{-1} \left({{0.1\mu \rm 
m}\over a}\right)^2
\left({Y_O \over 10^{-4}}\right)^{-1}
\ {\rm seconds},
\end{equation}
where $Y_O$ is the photodesorption yield for atomic O.
The timescale for 
thermal desorption of an O atom on a grain is given

\begin{equation}
t_{evap,O} \sim 9\times 10^{-13} e^{{800\ {\rm K}}\over T_{max}} \ {\rm seconds}.
\end{equation}
Therefore, the timescale for thermal evaporation of an O atom is 
$10^{22}$  seconds for a 10 K grain, $2\times 10^5$ seconds for a 20 K grain, and 
0.3 seconds for a 30 K grain.
From Eqs (12-16) one concludes that if grains are 
cooler than about 20-25 K and if $ x(H) > 0.25x(O)$, an adsorbed O atom will react 
with H to form OH and then H$_2$O before it is thermally desorbed.  
Photodesorption of O never appears to be important. 

In this paper we are 
mainly interested in the abundance of gas phase H$_2$O and O$_2$ at the 
point where oxygen species start to freeze out in the form of water ice.  This 
``freezeout depth''  $A_V \equiv A_{Vf}$ is where the gas phase
oxygen abundance plummets and most oxygen is incorporated as water ice on
grain surfaces.  This occurs when the rate at which relatively undepleted gas phase 
oxygen atoms
hit and stick to a grain equals the photodesorption rate of water molecules from
grain surfaces completely covered in ice, or $t_{\gamma -des}(A_{Vf}) \sim t_{gr,O}(A_{Vf})$. 
Using
Eqs.(12 and 15) and taking $Y\equiv Y_{OH,w} + Y_{H_2O}$ as the total yield of
photodesorbing water ice, we find

\begin{equation}
A_{Vf} \simeq 0.56 \ln \left[ 0.8 G_0 \left({Y \over 10^{-3}}\right)\left({{10\ 
{\rm K}}\over T_{gas}}\right)^{1/2}\left({{10^5\ {\rm cm}^{-3}}\over
n}\right)\right],
\end{equation}
if the term in brackets exceeds unity.  If it is less than unity, there is 
freezeout
at the surface of the cloud. As can be seen in Eq. (17), freezeout at the surface 
only
occurs for very low values of $G_0/n \lta 10^{-5}$ cm$^3$. 

At $A_{Vf}$ the 
timescale $t_{\gamma f}$ for a grain of size $a$ to absorb an FUV
photon can be written, substituting
Eq. (17) for $A_{Vf}$ into Eq. (14),

\begin{equation}
t_{\gamma f} \sim 24 \left({{10\ {\rm K}}\over T_{gas}}\right)
^{1/2}\left({{0.1\mu \rm m}\over a}\right)^2\left({{10^5\ {\rm cm}^{-
3}}\over
n}\right)\left({Y \over 10^{-3}}\right)\ {\rm seconds}.
\end{equation} 
As discussed above, $t_{\gamma f} << t_{gr,O}$, so that, at this critical 
depth,
no ice forms on small ($\lta 20$ \AA)
grains  because single photons can thermally evaporate adsorbed O, OH or H$_2$O
before thermal radiative emission cools the grain and before another O atom
sticks to the grain.
Although ice may not form on $a \lta 20$ \AA\ grains, they may still contribute 
to the
formation of gas phase OH or H$_2$O. Note that once an O atom sticks to a given 
grain,
the timescale $t_{gr,H}$ for an H atom to stick (Eq. 13) may be shorter than the
timescales $t_{\gamma f}$ for a photon to be absorbed.  Thus, the O may be 
transformed into
H$_2$O on the grain surface before a photon transiently heats the small grain 
and thermally evaporates the H$_2$O to the gas phase.  Because of these 
complexities, the possibility of coagulation, and the uncertain nature of small 
grains, we have chosen to fix the cross sectional grain  area per H nucleus $\sigma _H$
as constant with depth $A_V$ in a given model, but to then run models with a 
range of assumed $\sigma _H$.

\subsection{Grain Surface Chemistry}

Figure 2 shows  the new grain surface oxygen chemistry that we adopt.  We 
maintain the old
method of treating H$_2$ formation on grain surfaces (Kaufman et al 1999). The 
main new 
grain surface chemistry reactions are the surface reactions H + O $\rightarrow$  OH and H 
+ OH $\rightarrow$
 H$_2$O. Miyauchi et al (2008) discuss and review recent laboratory experiments
which suggest rapid formation of water ice on grain surfaces. 
In addition, we include simple grain surface carbon chemistry.
Similar to Figure 2, we allow the sticking of C to grains to form C$_{ice}$, followed
by reactions of surface H atoms to produce CH$_{ice}$, CH$_{2ice}$, CH$_{3ice}$,
and CH$_{4ice}$, as well as the desorption of all these species.   We do not form
CO, CO$_2$, or methanol on grain surfaces in this paper. 

In the model we compare the timescale for O and OH to desorb with the timescale for an H atom
to hit and stick to the grain surface.  If the desorption timescale is shorter, 
then we desorb the
radical and the grain surface chemistry does not proceed.  However, if the 
desorption time is longer,
then we assume that the reaction proceeds.  We generally assume that the newly 
formed surface
molecule does not desorb from the heat of formation, but must wait for another 
desorption process, or, for a further chemical reaction with H, if possible.  However, as in
O'Neill et al (2002), we also run a case where the heat of formation desorbs the 
newly formed molecule.
Implicit in this simple grain chemical code is the assumption that the H atom 
migrates and finds
the O or OH before it desorbs or before another H atom adsorbs to possibly form 
H$_2$.  We note that, in general, it is rare to find even 
a single O or OH  on the grain surface, because  
desorption or the
reaction with H removes them before another O or OH sticks to the surface. 

\subsection{Ion-Dipole Reactions}
A final change to our chemical code is the inclusion of enhanced rate 
coefficients for the
reactions between atomic or molecular ions with molecules of large dipole 
moment, such as OH
and H$_2$O. We have used the UMIST rates (e.g., Woodall et al 2007) and the rates
quoted in Adams et al (1985) and in  Herbst \& Leung (1986).  These
rates  are larger than
the rates without the dipole interactions by a factor $\sim 10$ at $T=300\, \ K$ 
and, given their
 $T^{-1/2}$ dependence, increase further  at low temperatures.

\section{MODEL RESULTS}

In this section, we present computational results of our new model. 
In order to explore the sensitivity of the freeze-out processes to various 
input parameters, we first define a standard case. We then present a 
number of models in which the gas density, FUV field, grain properties 
and photoelectric yield are varied. For each of these runs, we use the 
steady-state model of \cite{kwhl99} modified to include grain surface 
reactions, ion-dipole reactions, and desorption mechanisms as detailed in \S2. 
Unlike our previous PDR models which were run to a depth $A_V=10$, we often
run our new models to $A_V=20$ to ensure that we are well beyond the regions
where photodesorption plays a role.
At the end of this section, we discuss the 
effects of time-dependence.

\subsection{Standard Case}

Most ortho-H$_2$O detections in ambient (non-shocked) gas are located in Giant
Molecular Clouds (GMCs), and, in particular, in the massive cores of these GMCs
which are illuminated by nearby OB stars.
Ortho-H$_2$O has not yet been detected toward quiescent ``dark clouds" such as Taurus.
Therefore, for our standard case, we choose a gas density $n=10^4\,\rm cm^{-3}$ and an
FUV field strength $G_0=10^2$.
We adopt the same gas 
phase abundances as in the PDR models of \cite{kwhl99} and \cite{kwh06}. Details
are given in Tables 1 and  2.

In the standard case, the gas (grain) temperature varies from $T= 120$ K ($T_{gr}=31$ K) at
the surface, to $T=22$ K ($T_{gr}=15$ K) at the water peak plateau, 
 to $T=16$ K ($T_{gr}=15$ K) for $A_V=10$.
The abundances of O-bearing species in the standard model are shown in 
Fig. \ref{fig:std}, while those of C-bearing species are shown in Fig. 
\ref{fig:stdC}. 
Here, the basic features of the results of freeze-out and desorption are 
clearly demonstrated. 

{\it The Photodissociation Layer.}  At the surface, the dominant species are precisely 
those found in gas-phase PDR models. Atomic oxygen is the most abundant 
O-bearing species, while the transition from C$^+$  to CO 
occurs
once significant shielding occurs at $A_V\sim 1.5$. Near the surface, all of the 
oxygen not in CO is atomic, and $x(O)\ge x(CO)$. 
At the low dust temperature at the surface, thermal desorption of H$_2$O is insignificant and the 
beginnings of an ice layer (less than a monolayer)
form despite rapid photodesorption by the FUV.  
In steady-state, photodesorption of this ice as well as gas phase 
chemistry
maintains a small abundance of gas-phase H$_2$O.

{\it The Photodesorbed Layer.} 
By $A_V\sim 2.8$, enough water has been deposited on grain surfaces to form a 
monolayer,
that is $x({\rm H_2O_{ice})}\approx 4 n_{gr}\sigma _{gr} N_{s,H_2O}/n \sim 8\times 10^{-6}
(\sigma _H/2\times 10^{-21}$ cm$^2$).
At this point, the water ice abundance increases rapidly with depth.  Here, 
the photodesorption rate decreases with depth due to attenuation of the FUV flux by
grains, but the rate of freezeout only decreases when the gas phase O abundance
decreases. For the gas phase O abundance to decrease significantly
requires {\it many}  monolayers of ice.
This is seen in Fig. 3  
as the increase in x(H$_2$O$_{ice})$ by two orders of magnitude at $A_V\sim 
3$.  Note the agreement of this freezeout depth with $A_{Vf}$ given in Eq. (16).
As the gas phase abundance of O decreases, the growth of water ice slows and
the ice abundance saturates (i.e., all O nuclei locked in water ice) near
$A_V\sim 3.5$.  Fed by photodesorption, the gas phase abundance of H$_2$O
maintains a broad plateau $x$(H$_2$O)$ \sim 10^{-7}$ from $A_V \sim A_{Vf}\sim 3$ to 
$A_V \sim 6$, beyond which (i.e., deeper in the cloud)
 other gas phase H$_2$O destruction mechanisms besides 
photodissociation become dominant.  
The abundance of gas-phase H$_2$O near the peak is determined by the balance of 
photodesorption from the surface layer of ice on the grains with 
photodissociation 
of H$_2$O by FUV photons. Both of these processes are proportional to the local
attenuated FUV field; this leads to a plateau in the H$_2$O abundance independent
of $G_0$ (see \S 3.3). 

The abundances of other O-bearing molecules closely track the abundance 
of H$_2$O. OH 
is mainly a product of photodissociation of gas-phase water, while it is 
destroyed by photodissociation and by reaction with S$^+$, both of 
which depend on the attenuated FUV field. 
O$_2$ is produced primarily through the reaction O + OH $\rightarrow$ O$_2$ + H 
and is
destroyed by photodissociation. Thus the gas-phase H$_2$O, OH and O$_2$ 
abundances track 
one another from the beginning of freezeout through the H$_2$O plateau.  We 
discuss this in more detail in \S 3.3 below.

{\it The Freezeout Layer}.   Beyond $A_V > 6$, photodesorption of water ice ceases
to produce enough gas phase H$_2$O or OH to maintain the H$_2$O and OH plateaus (and
therefore the O$_2$ plateau). Other destruction mechanisms than photodissociation
of H$_2$O take over, and gas phase abundances of all O-bearing species drop as
H$_2$O ice incorporates essentially all available elemental O.

The destruction of CO by He$^+$, followed by freezeout of the liberated O 
atom, as described in \S2.3, is responsible for the 
rapid drop in the steady-state gas phase CO abundance at $A_V \sim 6$ (see Fig. 4)  as all 
of the oxygen becomes locked in
water ice. In our rather limited steady state carbon chemistry, C atoms end up incorporated 
into  CH$_{4ice}$ once O is liberated from CO. In our time-dependent runs, which
have more extensive chemistry, the standard case at $A_V = 10$ has $\sim 0.6$ of the elemental
carbon in CH$_{4ice}$ and $\sim 0.4$ in CO$_{ice}$ after $2\times 10^7$ years.\footnote{The 
drop in the
{\it total} CO gas + CO$_{ice}$ abundance by a factor $\sim 2$ in the opaque interior compared
to the intermediate shielded regions would be hard to measure without a precise (better than
factor of 2) independent measure of the column along various lines of sight through the cloud.}
  The ratio of the CH$_{4ice}$
to the H$_2$O$_{ice}$ abundance is $\sim 0.15-0.25$ at this time.
\"Oberg et al (2008) discuss recent observational results on the ice content
of grains in molecular clouds with emphasis on CH$_{4ice}$; they conclude
that CH$_{4ice}$ does indeed form by hydrogenation on grain surfaces, as we have assumed
in our models.   However, they observe abundance ratios of CH$_{4ice}$ to H$_2$O$_{ice}$
that are typically 0.05, nearly 3-5 times smaller than in our standard model.
This likely arises because our time dependent code starts with carbon in atomic
or singly ionized form, so that there is more opportunity for C or C$^+$ to transform
to CH$_{4ice}$ than if the carbon started in the form of CO.  Aikawa et al (2005) provide
theoretical models that result in the observed ratios of CH$_{4ice}$/H$_2$O$_{ice}$ by initiating
carbon in the form of CO.  The purpose of this paper is to follow H$_2$O and O$_2$
chemistry, and not carbon chemistry.  We have therefore also neglected grain surface
reactions that could lead to CO$_{2ice}$ or methanol ice.\footnote{We note that under optimum
conditions
CO$_{2ice}$ and  methanol ice may carry nearly as much oxygen as water ice.  However, generally
they do not; therefore, their neglect should not appreciably affect our 
model of basic oxygen chemistry.} Our CO$_{2ice}$ is formed by first
forming gas phase CO$_2$
by the reaction of gas phase CO with OH, and then subsequently freezing the CO$_2$
onto the grain surfaces. We plan to study the possible fractionation
signature ($^{12}$CO$_{2ice}$/$^{13}$CO$_{2ice}$) of our gas phase route to forming 
CO$_{2ice}$ in our subsequent paper which examines carbon chemistry in more detail
(in preparation). 
In summary, our carbon 
chemistry here is simplified, and the abundances of carbon ices in the models are very crude.


\subsection{Dependence on $G_0$, $n$, $Y_{H_2O}$, $\sigma _H$, PAHs, and $b_\lambda$ }

{\it Dependence on Incident FUV Flux, $G_0$}. 
Fig. \ref{fig:varyg}  
shows the effect of changing the FUV field strength, $G_0$,
on the depth of the gas-phase water peak and the abundance at the peak, while 
keeping the
density fixed. Here $G_0$ is varied from 1 to 10$^3$.
The main effect of changing $G_0$ is to move the position of the water
peak  inwards with increasing $G_0$. Varying $G_0$ from  
1 to 10$^3$  changes the depth of freezeout from $ A_V\sim 1$ to $A_V \sim 7$. 
Gas phase H$_2$O peaks once ice has formed at least a monolayer on the
grain surfaces, so the location of the 
water peak also moves inwards (logarithmically) with increasing $G_0$. 
The peak gas-phase water abundance ($\sim 10^{-7}$) is insensitive to $G_0$
and insensitive to depth (i.e., $A_V$) in the water plateau.  We also see that the depth
where the gas phase water starts to decline from its peak (plateau) value
is also weakly (logarithmically) dependent on $G_0$, increasing with increasing
$G_0$.  Thus, the column of gas phase water remains fairly constant and
independent of $G_0$. 

However, some differences arise in the case with $G_0=10^3$. Here the assumption that every 
O sticking to a grain forms water ice breaks down. The high FUV field absorbed
at the cloud surface leads to
a high IR field that heats the 
grains to 
$T_{gr}\sim 25$ K even at $A_V\sim 8$, so that a significant fraction of 
O atoms are thermally desorbed from grains faster than they can form water ice,
resulting in a higher gas phase abundance of atomic O.  
At these large
depths, the formation of gas phase water is not mainly by photodesorption of
water ice, but by cosmic ray initiated ion-molecule reactions
(see Fig. 1).
The high gas phase elemental abundance of O leads to a high rate of formation of H$_2$O
and O$_2$, whereas the high $A_V$ leads to low photodestruction rates of these
two molecules.  In addition, although the FUV field is severely attenuated, the
residual FUV is sufficient to prevent the complete freezeout of gas phase H$_2$O.
The combined effect is  a deeper and higher peak in
the gas phase H$_2$O and O$_2$ abundances.  At very high $A_V$ ($A_V > 8$)
the freezeout of water ice finally drains the gas phase oxygen, and gas phase
H$_2$O and O$_2$ decrease with increasing $A_V$.  Our time dependent results
at $t= 2\times 10^7$ years find most of the oxygen in H$_2$O-ice ($\sim 70$\%)
and CO$_2$-ice ($\sim 30$\%) at $A_V > 8$, and the carbon is mostly in C$_3$H-ice
($\sim 60$\%) and CO$_2$-ice ($\sim 40$\%) since the warm grain temperatures lead to
thermal desorption of CO-ice and CH$_4$-ice.

{\it Dependence on Gas Density, $n$}.
  In Fig. 6
we vary the gas density 
from $10^3$ to $10^5\,\rm cm^{-3}$, keeping the rest of the parameters fixed
at their standard values.  Comparing this figure with Fig. 5 we see that
for $G_0< 10^3$ the plateau water abundance is
independent of 
gas density and $G_0$, depending only on the photoelectric yield $Y_{H_2O}$ and the grain 
cross sectional area
per H nucleus $\sigma _H$ as shown below.  The peak abundance is independent of $n$
because both the formation rate per unit volume (photodesorption) and the destruction
rate (photodissociation) are proportional to $n$. For a fixed value of $G_0$, lowering the
gas density moves the water peak further in;  the deposition of water ice
 depends on the product $n_{gr}n_O\propto n^2$, while the photodesorption 
rate goes only as the grain density $n_{gr} \propto n$  times the local FUV 
field. Therefore, the water peak moves into the cloud with decreasing density. 
However, the column or $A_V$ where the water starts to decline from its peak (plateau) 
value is insensitive to $n$. Thus, the plateau starts to become narrower and
narrower (in $\Delta A_V$) as $n$ decreases.  As a result, The column of gas phase water
increases with $n$ somewhat.  We explain all these trends in more detail
below in \S 3.3, where we present a simple analytic analysis.

{\it Dependence of Gas Temperature on $G_0$ and $n$}.
Figure 7 presents the gas temperature as a function of $A_V$ for a range of our $G_0$, $n$
parameter space.  This figure reveals the temperature at which most of the H$_2$O and O$_2$
is radiating, as we have marked the peak in the gas phase water abundance (which coincides
with the peak in the O$_2$ abundance) as solid portions of these curves.  We see
that over the parameter space $G_0= 1 - 10^5$ the emission from H$_2$O and O$_2$ arises
from regions of $T \sim 7-50$ K, with $T$ generally rising gradually with $G_0$. 
  For $G_0 < 500$, there is a true ``water
plateau'' and the solid line corresponds to the plateau region.  
 $T\sim 7-30$ K in the plateau, and the gas temperature is often higher than
the dust temperature because of grain photoelectric heating of the gas.  
For higher $G_0$ there is no true plateau but
the gas phase H$_2$O and O$_2$ peak at higher $A_V$, where the gas and dust temperatures
are nearly the same.   We will show
in \S 3.4 that at high $A_V$ the dust temperature scales roughly as $G_0^{0.24}$.
Here we have made solid lines where
the H$_2$O abundance is more than 0.33 of the peak abundance. 

{\it Dependence on Photodesorption Yield, $Y_{H_2O}$}.
Fig. 8
presents the profiles of the gas phase H$_2$O and O$_2$
abundances, and the H$_2$O ice abundance as the photodesorption yield $Y_{H_2O}$ of
water ice is varied, maintaining $Y_{OH,w}=2Y_{H_2O}$, but
keeping all other parameters standard.  In contrast
to the lack of dependence of the plateau value of the gas phase water
abundance $x_{pl}(H_2O)$ on $n$ and $G_0$, $x_{pl}(H_2O)$ scales roughly linearly
with $Y_{H_2O}$ for low $Y_{H_2O} < 10^{-3}$. This linear dependence can be easily understood: 
the formation rate (photodesorption)
of gas phase H$_2$O scales linearly with $Y_{H_2O}$ whereas the destruction rate
(photodissociation) at a given $A_V$  is independent of
$Y_{H_2O}$.\footnote{Where
thermal desorption of atomic O can compete with photodesorption, i.e. high grain temperatures
from high FUV fields, this is not the case.  In addition, at high values of $Y_{H_2O}
>10^{-3}$, the surface can become depleted in water ice compared with, for example,
CO ice, which then lowers the desorption rate of H$_2$O into the gas.  This
leads to a smaller increase in gas phase H$_2$O with increasing $Y_{H_2O}$ than
would otherwise be the case.}
On the other hand, at low values of $Y_{H_2O}< 10^{-3}$,
the abundance of O$_2$ rises  with $Y_{H_2O}^2$, or, in other words,
with  [$x(H_2O)$]$^2$.

{\it Dependence on Effective Grain Area, $\sigma _H$}. 
In Figure 9
we show the effect of varying $\sigma _H$ with
all other parameters at their standard values.  The abundance of gas phase
H$_2$O in the plateau $x_{pl}(H_2O)$ increases nearly linearly with $\sigma _H$.
The depth $A_V \sim A_{Vf}$ where the H$_2$O abundance first reaches its plateau
value is insensitive to  $\sigma _H$, as predicted by
Eq. (17).  Similarly, the depth where H$_2$O finally begins to drop from its
plateau value also is insensitive to $\sigma _H$.   The abundance
of O$_2$, like the abundance of H$_2$O,  scales with
$ \sigma _H$.

{\it Dependence on PAHs}.
We are most interested in the chemistry at $A_V \sim 3-6$ where the H$_2$O and
O$_2$ abundances peak.  Because it is not clear if PAHs exist at such depths
(Boulanger et al 1990, Bernard et al 1993, Ristorcelli et al 2003), we have not included
PAHs in our standard case
but have assumed a grain surface area per H given by $\sigma _H$ which is appropriate
to an MRN distribution extending down to 20 \AA \ grains  when computing the neutralization
of the gas by collisions with small dust particles.  If PAHs are present, the
effect of neutralization is enhanced, which changes the gas phase chemistry somewhat.
We have found that the inclusion of PAHs into our standard case does not affect
the results appreciably.  There is a somewhat greater effect at lower density
such as $n \sim 100$ cm$^{-3}$, where the inclusion of PAHs does increase the total
water column to high $A_V$ by a factor of about 10, both by broadening the peak
plateau and by increasing the peak abundance  (for the case $G_0=1$).

{\it Dependence on $b_\lambda$}.
We have run our standard case but have assumed that the small grains responsible
for FUV extinction are depleted, for example, by coagulation.  In our test run we have
decreased all our $b_\lambda$ by a factor of 2, reduced $\sigma _H$ by 2, and reduced
grain photoelectric heating accordingly.  The main effect is to increase the characteristic
$A_V$ where the H$_2$O and O$_2$ plateau by a factor of 2.  However, the abundances of
H$_2$O and O$_2$ in the plateau decrease because of the decrease in $\sigma _H$.
Therefore, the columns of
gas phase H$_2$O and O$_2$ only increase by a factor of about 1.5 from the standard case. 
The gas temperature does not appreciably change.   In summary, the change
in the intensities and average abundances of $H_2$O and O$_2$ is not large, and is
expected to be less than a factor of 2. 

{\it Dependence of H$_2$O and O$_2$ Columns on $G_0$ and $n$}.
In Fig. \ref{fig:nave}, we show the total columns of H$_2$O and O$_2$ for a cloud
with high $A_V > 10$ as a function of the incident FUV flux $G_0$ and the cloud density $n$.
For low $G_0 < 500$, the H$_2$O and O$_2$ columns are roughly independent of $G_0$
and $n$, and are of order $10^{15}$ cm$^{-2}$ for our standard values of photodesorption
yield.  These columns arise in the H$_2$O and O$_2$
plateaus, at intermediate cloud columns.  In the steady state models, there is so little gas
phase water in the interior of the cloud that the interior does not contribute significantly
to the total column.  Therefore, once the total column through the cloud $N$
is large enough to include the plateau ($N_H\gta 10^{22}$ cm$^{-2}$), the average abundance
will scale as  N$_H$$^{-1}$. Thus, part of the reason for the low
average abundances observed by SWAS and {\it Odin} is the dilution caused by the
lack of H$_2$O and O$_2$ at the surface and deep in the interior of very opaque,
high column, clouds.   At higher $G_0 > 500$ we see the effect of the thermal
evaporation
of the O atoms from the grain surfaces.   The effect is most dramatic for the gas
phase O$_2$ column, which rises to values $N(O_2) \sim 2 \times 10^{16}$ cm$^{-2}$.  This
is close to what is needed for a detection by {\it Herschel} and therefore it is in
this type of environment that {\it Herschel} may potentially detect O$_2$.
We note, however, that this result depends on the binding energy of O atoms to water
ice surfaces. The value we have adopted (see Table 1) is standard in the literature
but it refers to van der Waal binding to a chemically saturated surface.  It is 
quite possible that the O binding energy is larger, and this would then increase
the temperature (and $G_0$) where this effect would be initiated.

\subsection{Simple Analytic Analysis of the Results}

The results in \S 3.1 and \S 3.2 
can be understood by a simple analytic chemical model that
incorporates the main physics.
 Such a model, though approximate, has the
advantage of allowing one to determine and understand the sensitivity to
various model parameters, and serves to validate the numerical model.
 In this subsection the chemical rates are
all taken from the UMIST database, except for  the ion dipole rates
 and the photodesorption rates discussed in \S 2.7.  In the following,
the reader is referred to Table 3 in the Appendix for quick
reference to symbols. 

The abundance $x(H_2O)$ of gas phase H$_2$O can be roughly traced and understood
by simple analytic formulae.
 Photodissociation dominates the destruction of gas phase
H$_2$O from the surface to the depth $A_{Vd}$ where gas phase destruction 
with H$_3^+$ generally takes over.  From $0 < A_V < A_{Vd}$ the simple model
assumes that H$_2$O is formed by photodesorption of H$_2$O ice. Therefore, by equating
formation to destruction

\begin{equation}
G_0 F_0 e^{-1.8A_V} Y_{H_2O}f_{s, H_2O} n_{gr}\sigma_{gr} = G_0 R_{H_2O}e^{-1.7A_V} n(H_2O),
\end{equation}
or

\begin{equation}
x(H_2O) = F_0 Y_{H_2O} f_{s, H_2O} \sigma _H R_{H_2O}^{-1}e^{-0.1A_V},
\end{equation}
where $R_{H_2O} = 5.9 \times 10^{-10}$ s$^{-1}$ is the unshielded photodissociation
rate of H$_2$O in an FUV field of $G_0=1$.  The differences in the $b_\lambda$ factors of
1.7 and 1.8 is because the two processes are dominated by somewhat different FUV wavelengths.
The fractional coverage of the surface
by water ice $f_{s,H_2O}$ is given by equating the sticking of O atoms to grain surfaces
to the photodesorption rate of H$_2$O (including both desorption of H$_2$O and of OH + H).
Therefore, recalling that $Y= Y_{H_2O} + Y_{OH,w}$,

\begin{equation}
n(O)v_On_{gr}\sigma _{gr} = G_0F_0e^{-1.8A_V}Yf_{s,H_2O}n_{gr}\sigma _{gr}
\end{equation}
or
\begin{equation}
f_{s,H_2O} = {{n(O)v_Oe^{1.8A_V}}\over {G_0 F_0 Y}}
\end{equation}
Once a monolayer forms, $f_{s,H_2O} \sim 1$.  We have found in our numerical
runs that sometimes  $f_{s,H_2O}$ saturates at  $\sim 0.5$ rather than 1.0, because of the presence of
other ices such as CO ice or CH$_4$ ice. However, assuming a saturation value of unity,
the critical depth $A_{Vf}$ at which a monolayer
forms is then given (this is equivalent to Eq. 17 in \S 2.5):

\begin{equation}
A_{Vf}= 0.56 \ln \left({{G_0 F_0 Y}\over {n(O)v_O}}\right),
\end{equation}
or a critical H nucleus column (Spitzer 1978) from the surface of

\begin{equation}
N_f = 2\times 10^{21} A_{Vf} \ {\rm cm}^{-2}.
\end{equation}
We therefore find, using equations (20 and 22),

\begin{equation}
x(H_2O) = {{n(O) v_O Y_{H_2O}\sigma _H e^{1.7A_V}} \over {YG_0 R_{H_2O}}}\ \ {\rm for} \ 
A_V < A_{Vf}.
\end{equation}
\begin{equation}
x_{pl}(H_2O)= {{F_0 Y_{H_2O} \sigma _H e^{-0.1A_V}} \over {R_{H_2O}}}\simeq
2 \times 10^{-7} \left({Y_{H_2O} \over {10^{-3}}}\right)\left({\sigma _H
\over {2\times 10^{-21} \ \rm cm^2}}\right)e^{-0.1A_V}\
\ {\rm for}\   A_{Vf} < A_V < A_{Vd}.
\end{equation}
The abundance of gas phase H$_2$O rises exponentially with depth (as $e^{1.7A_V}$)
until a monolayer of water ice has formed on the grains.  It then forms a plateau
with value $x_{pl}(H_2O)$ that is independent of $G_0$ and $n$ and weakly dependent
on $A_V$.  The plateau abundance goes linearly with $Y$ and $\sigma _H$.
The lack of dependence on $G_0$ can be understood since both the formation and
destruction of gas phase H$_2$O depend linearly on $G_0$.  Higher fluxes do decrease
the surface abundance of gas phase water, but the freezeout depth is deeper, so 
there is more
dust attenuation, and the gas phase water abundance rises  and peaks at the 
same constant value.
 More precisely, regardless of the incident FUV flux, the local attenuated
FUV flux at the freezeout point is always the same, because it must be the flux 
which
photodesorbs an ice covered grain at the rate that oxygen from the gas 
resupplies the surface
with ice.  Therefore, the gas phase water abundance is the same as well, 
regardless of $G_0$, only depending on the the assumed values of the grain 
parameters and the photodesorption
yield. The lack of dependence
on $n$ can be understood similarly since both the formation rate per unit volume
and the destruction rate per unit volume depend linearly on $n$. On the other hand,
$x(H_2O) \propto Y\sigma _H$, since the formation rate by photodesorption
goes as this product whereas the destruction rate is independent of them.

In order to determine $A_{Vd}$ and to provide an approximation for $x(H_2O)$
that applies for $A_V > A_{Vd}$, we need to include other destruction routes
for gas phase H$_2$O that begin to dominate at high $A_V$. We find that generally the
route which takes over from photodissociation is reaction with 
H$_3^+$ to form H$_3$O$^+$.   H$_3$O$^+$ recombines
one fourth of the time to reform water (not a net destruction) and three fourths
of the time to form OH (Jensen et al 2000, Neufeld et al 2002).  The net rate coefficient for the
destruction of gas phase H$_2$O by H$_3^+$ is then

\begin{equation}
\gamma _{H_3^+} = 0.75(4.5 \times 10^{-9} T_{300}^{-1/2}) \ {\rm cm^3\ s^{-1}}.
\end{equation}
where $T_{300}= T/300$ K.
Equating the formation of gas phase H$_2$O by photodesorption to the destruction
of H$_2$O by both photodissociation and  H$_3^+$, we obtain a more general
solution for $x(H_2O)$ for $A_V > A_{Vf}$ that applies beyond $A_{Vd}$

\begin{equation}
x(H_2O) \simeq {{x_{pl}(H_2O)} \over {1 + \left[{{\gamma _{H_3^+}  n(H_3^+)}\over {G_0
R_{H_2O}}}\right]e^{1.7A_V}}} \ {\rm for} \ A_V > A_{Vf}.
\end{equation}
Note that the second term in the denominator, which represents destruction
by H$_3^+$, begins to dominate when its value reaches and exceeds unity.
At this point, $x(H_2O)$ begins to drop from its plateau value.  

H$_3^+$ is formed by cosmic rays ionizing H$_2$ followed by rapid reaction
of H$_2^+$ with H$_2$.  When gas phase CO is depleted, it is destroyed by dissociative
recombination with
electrons.  The electrons are provided by Si$^+$ and S$^+$ at moderate $A_V$
and by H$_3^+$, He$^+$, and H$^+$ at high $A_V$,
so the electron density is complicated.  At high $A_V$ the code results suggest
that the electron density is roughly $n(e) \simeq 2 n(H_3^+)$.  We then
obtain

\begin{equation}
n(H_3^+)_{highA_V} \simeq 10^{-5}T_{300}^{1/4} n^{1/2}\ {\rm cm^{-3}},
\end{equation}
with H nucleus density $n$ in cm$^{-3}$.  In Eq. (29)  we
have assumed a total (including secondary ionizations)
cosmic ray ionization rate of 5$\times 10^{-17}$ s$^{-1}$
for H$_2$ (Dalgarno 2006).  
At moderate $A_V$ Si$^+$ usually provides most of the electrons.
Gas phase Si$^+$ is produced by photoionization of gas phase Si
and destroyed by recombination with electrons. The Si$^+$ density, and therefore
the electrons provided by Si$^+$, exponentially drops with $A_V$ as the
FUV photons that photoionize Si drop with increasing dust attenuation.
As long as most of the gas phase silicon is Si$^+$, the electron density
is high and this suppresses the abundance of H$_3^+$, prolonging the water
plateau to higher $A_V$ (see Eq. 28). However, we find that once the gas phase
Si$^+$ density drops to $2n(H_3^+)_{highA_V}$, so that the electrons provided
by Si$^+$ are comparable to those provided by H$_3^+$, He$^+$, and H$^+$, 
$n(H_3^+)$ has risen to $\sim n(H_3^+)_{highA_V}$ and the H$_3^+$ term in
the denominator of Eq. (28) dominates.  Here the gas phase water abundance
starts to fall.  This occurs when

\begin{equation}
A_{Vd} = 0.59 \ln \left[ 4.5 \times 10^5 G_0 n^{-1/2} T_{300}^{-0.2}
\left({n_{SiT} \over n_{H_2O-gr}}\right)^{1/2} \right],
\end{equation}
where $n_{SiT} = 1.7 \times 10^{-6}n$ is the density of silicon both in the gas phase
and as ice mantles on grains, and $n_{H_2O-gr} \sim 3 \times 10^{-4}n$ is the
density of water molecules incorporated in ice mantles.

We have checked our analytic expressions for $A_{Vf}$ (Eq. 23), $A_{Vd}$ (Eq. 30),
and $x_{pl}(H_2O)$ (Eq. 26) and find good agreement ($A_{Vf}$
and $A_{Vd}$ better than $ \pm 25\%$ and $x_{pl}(H_2O)$ good to a factor of 2
over our ($G_0$,$n$) parameter space for $G_0 \lta 500$). For $G_0 > 500$ the grains are
so warm that O atoms adsorbed to grains thermally evaporate before reacting with
H atoms to form OH on the grain surface.  As discussed in \S 3.2, this results in
a change in the behaviour of gas phase H$_2$O. 

The abundance of O, OH, and O$_2$ in the H$_2$O plateau region $A_{Vf} < A_V <
A_{Vd}$ can also be analytically estimated from the above expression for
$x_{pl}(H_2O)$.  We first find $x_{pl}(OH)$ by equating formation
by photodissociation of H$_2$O and photodesorption of OH from water ice
with destruction by photodissociation of OH and reaction with S$^+$ and He$^+$.
The photodesorption rate of OH is 2 times the rate of H$_2$O photodesorption,
and therefore 2 times the rate of H$_2$O photodissociation.  On the other hand,
the destruction of OH by S$^+$ and He$+$ is roughly 2 times that OH photodissociation.
Therefore,

\begin{equation}
x_{pl}(OH) \sim \left({{R_{H_2O}} \over {R_{OH}}}\right) x_{pl}(H_2O) \sim  
1.7x_{pl}(H_2O), \ \ A_{Vf} < A_V < A_{Vd}
\end{equation}
where $R_{OH}=3.5 \times 10^{-10}$ s$^{-1}$ is the unshielded 
photodissociation rate of OH when $G_0=1$.
Thus, OH tracks H$_2$O with somewhat greater abundance in the plateau, as observed
in Figure 3.

The gas phase O abundance is determined by  balancing the formation of O by
photodissociation of OH with the removal of O by adsorption onto grains:

\begin{equation}
x_{pl}(O) \simeq \left[{{G_0 R_{OH} e^{-1.7 A_V}}\over {\sigma _H n v_O}}\right]
x_{pl}(OH) \simeq 4\times 10^{-5} \left({{G_0/n}\over 10^{-2}}\right) \left(
{e^{-1.7A_V}\over 10^{-3}}\right) \left({x_{pl}(H_2O) \over 10^{-7}}\right).
\end{equation}
Therefore, $x_{pl}(O)$ exponentially declines with $e^{-1.7A_V}$ due to the
declining photodissociation of OH, but the constant (with $A_V$) collision rate
of O with grains, as seen in Figure 3.  It also increases with $G_0/n$ for fixed
$A_V$. This formula breaks down when the RHS exceeds $\sim 10^{-4}$, where
$x_{pl}(O)$ saturates since all elemental gas phase O is now in atomic form.

The O$_2$ abundance follows from the abundances of O and OH since O$_2$ is formed
by the neutral reaction of OH with O (rate coefficient  $\gamma _{O_2}= 4.3 \times
10^{-11}(300\ {\rm K}/T)^{1/2} e^{-30\ {\rm K}/T}$ cm$^3$ s$^{-1}$).  
O$_2$, like H$_2$O and OH, is destroyed by 
photodissociation in the plateau region.  Therefore,

\begin{eqnarray}
x_{pl}(O_2) &\simeq& \left[{{1.7^2 \gamma _{O_2}}\over {\sigma _H v_O}}\right] \left({
R_{OH} \over R_{O_2}}\right) e^{0.1A_V}
[x_{pl}(H_2O)]^2\\
&\sim& 1 \times 10^{-7} e^{0.1A_V}\left({{2\times 10^{-21}\ {\rm cm^2}}
\over {\sigma _H}}\right)\left({30
\over{T}}\right)e^{-30/T} \left[{x_{pl}(H_2O) \over 10^{-7}}\right]^2,\nonumber
\end{eqnarray}
where $R_{O_2}=6.9\times 10^{-10}$ s$^{-1}$ is the unshielded rate of photodissociation
of O$_2$ in a $G_0=1$ field.
The $[x_{pl}(H_2O)]^2$ dependence, which translates to a $Y_{H_2O}^2$ dependence in
the plateau for low $Y_{H_2O} < 10^{-3}$, is clearly seen in Figure 8.  Note also 
that Eq. (33) explains
the lack of dependence of $x_{pl}(O_2)$ on $n$ and $G_0$, and also predicts
that  $x_{pl}(O_2) \propto \sigma _H$ as observed in the numerical models. 

In summary of the above, the approximate analytic equations can be used in
the cases with $G_0 \lta 500$ to 
understand and quantitatively estimate: (i) $x_{pl}(H_2O, OH, O\ {\rm and}\ O_2)$
and their dependence (or independence) on $n$, $G_0$, $Y$, and $\sigma _H$;
(ii) the depth $A_{Vf}$ of the onset of the plateau, the depth $A_{Vd}$ of the
termination of the plateau, the width $\Delta A_{V}= A_{Vd}-A_{Vf}$ of the plateau,
and their dependence on the above parameters.  At high $Y_{H_2O}> 10^{-3}$ the 
abundances of O, OH, O$_2$, and H$_2$O are overestimated by factors of order 2
in the above formulae because
we have ignored the presence of other ices, which reduce the fractional surface
coverage of water ice, and therefore the production of gas phase H$_2$O and
its photodissociation products.

Finally, the above analytic expressions can be used to predict the optimum conditions
for observing H$_2$O and O$_2$.  In general, for our model runs the intensity of the H$_2$O ground
state ortho line at 557 GHz and para line at 1113 GHz can be estimated by assuming that the emission
is ``effectively thin''.\footnote{We include the 1113 GHz line because it is accessible with the upcoming
Herschel Observatory.}  In other
words, although the optical depth in
the ground state may be larger than unity, the gas density is so far below the
critical density ($\sim 10^8$ cm$^{-3}$) that self absorbed photons are re-emitted and eventually escape
the cloud (see discussion in \S 5).  In this limit every collisional excitation results in an escaping
photon, and

\begin{equation}
I_{557} \simeq {{ h\nu [x_o n \gamma _{ex,o}+x_p n \gamma _{ex,p}] N(o-H_2O)}\over {4 \pi}},
\end{equation}
where $\gamma _{ex,o}\simeq 2.5 \times 10^{-10} e^{-26.7/T}$ and $\gamma _{ex,p}\simeq 
3.5 \times 10^{-11}e^{-26.7/T}$
 cm$^3$ s$^{-1}$ (valid for $T\lta 30$ K) are the rate coefficients for the  collisional
excitation of the 557 GHz transition by impact with ortho and para H$_2$ respectively 
(Faure et al 2007; M-L Dubernet, private communication).   Here,
 $x_o$($x_p$) is the abundance of ortho(para)-H$_2$ with respect to
H nuclei\footnote{Note that the maximum values of $x_o$ and $x_p$ are 0.5.}, 
$h\nu$ is the  energy
of the 557 GHz photon, and $N(o-H_2O)$ is the column of {\it ortho} water.  Similarly, the
para H$_2$O 1113 GHz line intensity is

\begin{equation}
I_{1113} \simeq {{ h\nu [x_o n \gamma _{ex,o}+x_p n \gamma _{ex,p}] N(p-H_2O)}\over {4 \pi}},
\end{equation}
where where $\gamma _{ex,o}\simeq  3 \times 10^{-10} e^{-53.4/T}$ 
and $\gamma _{ex,p}\simeq 9.0 \times 10^{-11} e^{-53.4/T}$
 cm$^3$ s$^{-1}$ (valid for $T\lta 30$ K) are the rate coefficients for the  collisional
excitation of the 1113 GHz transition by impact with ortho and para H$_2$.
From the
above equations, assuming most of the H$_2$O column arises in the plateau between
$A_{Vf}$ and $A_{Vd}$, the ortho(para) H$_2$O column is given

\begin{equation}
N(o(p)-H_2O) \simeq 2\times 10^{21}(A_{Vd}-A_{Vf})x_{pl}(H_2O) f_o(f_p)\ {\rm cm^{-2}},
\end{equation}
where  $f_o$($f_p$) is the fraction of H$_2$O in the ortho(para)
state.
 Substituting our analytic equations for these
parameters, we obtain
{\small
\begin{equation}
I_{557} \simeq  3 \times 10^{-7}f_o[x_o +0.14 x_p]\left({n \over{ 10^4 \ \rm
cm^{-3}}}\right)
\left({Y_{H_2O} \over{10^{-3}}}\right) \left({\sigma_H \over{2 \times 10^{-21} \rm cm^2}}\right)
\Delta A_V e^{-0.1A_V} e^{-26.7/T}  \ {\rm erg\ cm^{-2}\ s^{-1}\ sr^{-1}}
\end{equation}
}
and
\begin{equation}
I_{1113} \simeq 2.4 \left({f_p\over f_o}\right)\left({{x_o + 0.3x_p}\over {x_o + 0.14x_p}}\right)
e^{-26.7/T} I_{557},
\end{equation}
where $\Delta A_V = A_{Vd}-A_{vf}$.
Note that the column
of H$_2$O is relatively independent of $G_0$ and $n$.  However, the
intensity $I_{557}$ or $I_{1113}$ is proportional to the H$_2$O column times $\gamma _{ex}n
\propto ne^{-26.7/T}$ or $ne^{-53.4/T}$, and therefore is 
somewhat sensitive to  $G_0$ (which determines $T$)
but is more sensitive to $n$.  The 557 GHz line depends on $f_o$ which is a 
sensitive function of $T_{gr}$ for low $G_0$ (i.e., low $T_{gr}$), as will
be described below.
Therefore, the 557 or 1113 GHz H$_2$O line will be strongest in dense regions
illuminated by strong FUV fluxes (assuming sufficient column or $A_V$ to
encompass the plateau, and assuming the regions fill the beam).  We note
that the gas and grain temperatures vary somewhat through the water plateau
region.  We take the gas $T$ and the grain $T_{gr}$ to be
the average of the values at the freezeout point $A_{Vf}$ and at the $A_V$ where
the gas phase water abundance peaks.   Figure 11 shows
a contour plot of this average gas temperature $T_{ave}$ 
as a function of $G_0$ and $n$. Over nearly the entire range of $G_0$ and $n$,
$T_{ave}$ only varies from $\sim 15-40$ K.  Recall
that Eqs (37) and (38) break down for $G_0 \gta 500$, but note that from
our numerical results $I_{557}$ and $I_{1113}$ will increase further,
and the analytic equations will underestimate the intensities  with increasing
$G_0$ above $G_0 \sim 500$, since the water column rises here.  

Similarly, we can estimate the intensity of the O$_2$ transition.  Here,
the density is  above the critical density ($\sim 10^3$ cm$^{-3}$)
for the 487 GHz O$_2$ transition so the levels are in LTE.  In addition, because the transition has 
a low Einstein $A$ value, large columns are needed to make the transition
optically thick.  Thus, again assuming optically thin emission, the
intensity $I_{487} \propto N(O_2) e^{-25 K/T}/Z(T)$, where $Z(T)$ is the
partition function.   An approximation to Z(T) for 0 K$ < T < 25$ K is given as 
$Z(T) \simeq 3 + 2.1T + 0.1T^2$.
Using the equations from this subsection, we then estimate

\begin{equation}
I_{487} \simeq 6 \times 10^{-8}\left({{\Delta A_V} \over {Z(T)}}\right) \left({{30\ \rm K}\over T}\right) 
\left({Y_{H_2O}\over{10^{-3}}}\right)^2\left({\sigma _H \over{2\times 10^{-21} \rm cm^2}}\right)
e^{-55/T} e^{-0.05(A_{Vf}+A_{Vd})}\  \ {\rm erg\ cm^{-2}\
s^{-1}\ sr^{-1}}
\end{equation}
Note that again, this equation breaks down for $G_0 \gta 500$, and the
equation will underestimate the intensity for higher $G_0$ since the 
O$_2$ column rises.  In general, however, the equation predicts little
$n$ dependence (there is a small $n$ dependence in the $A_V$ terms).
At low $G_0 \lta 500$, the equation predicts only moderate $G_0$ dependence.
Although various terms depend on $T$, the temperature {\it of the
O$_2$ plateau region} is not sensitive to $G_0$ since the plateau retreats
further into the shielded region for higher $G_0$.  In addition, the $T$
dependence in the exp(-55/T) term is partially cancelled by the $TZ(T)$ term
in the denominator.

\subsection{The H$_2$O 557 and 1113 GHz and O$_2$ 487 GHz Line Intensities}

Using the numerical models, 
Figures 12 and 13 present H$_2$O $I_{557}$ and $I_{1113}$  and O$_2$ 
$I_{487}$ contours as functions of $n$
and $G_0$ for fixed standard values of the other parameters.  We have assumed
that the H$_2$O  ortho to para ratios are in LTE with the grain temperature
(which in the plateau is nearly the same as the gas temperature except for very
low ratios of $G_0/n$, which bring the plateau very close to the surface of
the cloud where $T > T_{gr}$).  The grain temperature is appropriate since a given H$_2$O
molecule forms on a grain and spends too little time in the gas to 
equilibrate to the gas temperature before it is photodissociated.
The grain temperature in the plateau is approximately (Hollenbach et al 1991)

\begin{equation}
T_{gr,pl} \simeq (3\times 10^4 + 2 \times 10^3 G_0^{1.2})^{0.2}.
\end{equation}
A fit to the LTE ortho to para ratio of H$_2$O valid to 10\% in
the region 5 K $< T_{gr} < 20$ K and good to 20\% for $T_{gr} > 20$ K
(e.g. Mumma et al 1987) is given:

\begin{equation}
{f_o\over f_p} = 0.8 e^{-{{22\,{\rm K}} \over T_{gr}}} +  2.2 e^{-\left({{15\,{\rm K}} 
\over T_{gr}}\right)^2}.
\end{equation}

The H$_2$ ortho to para ratio is calculated in steady state
using the rates and processes described by Burton et al (1992).
They are neither 3 nor LTE.  They are insensitive to $n$ in
the plateau, but sensitive to $G_0$, which affects $T$.
The H$_2$ ortho to para ratio $x_o/x_p \simeq 0.5$ at $G_0 =
3\times 10^4$, $0.3$ at $G_0 = 1.5 \times 10^4$, and 0.1
for $G_0 = 4 \times 10^3$.  For lower values of $G_0$, 
$x_o/x_p < 0.1$ so that para H$_2$ collisions dominate the
excitation of water.

The 557 GHz and 1113 GHz intensities behave as described by 
the analytic formulae (Eq. 37 and 38)
presented in \S 3.3 for $G_0 \lta 500$.  
The H$_2$O intensities rise roughly linearly with $n$. 
For fixed density $I_{557}$ and $I_{1113}$ tend to rise with $G_0$ (which raises $T_{gr}$
as shown in Eq.(40) and which tends to raise $T$ as can be seen in Fig. 11)
 partly due
the excitation factors $e^{-26.7/T}$ and $e^{-53.4/T}$, but also because
increasing $T$ increases the ortho/para ratio of H$_2$, and the ortho excitation
rates in collisions with H$_2$O are much higher. $I_{1113}$ does not rise as much
with increasing $G_0$ as one might naively expect given its excitation factor $e^{-53.4/T}$,
because the increasing $T$ reduces $f_p$ from a value of $\sim 1$ at low $G_0$ to
0.25 at high $G_0$.\footnote{Note also that because increasing $G_0$ pushes the water
and O$_2$ plateaus to higher $A_V$, the gas temperature $T_{ave}$ only rises weakly with $G_0$
However,
$I_{557}$ 
increases 
with increasing $f_o/f_p$, which is very sensitive to increasing $T_{gr}$ (or $G_0$)
for low $T_{gr}$ (or $G_0$). O$_2$ behaves somewhat differently, 
being less sensitive to $n$, as predicted
by the analytic models.}  

The following summarizes the accuracy of the analytic approximations for the
H$_2$O and O$_2$ intensities for  $Y_{H_2O} \sim 10^{-3}$.
For $G_0 < 500$ the analytic expression for $I_{557}$ agrees
with the numerical results to within a factor of 3 for $n= 10^3 - 10^5$ cm$^{-3}$.
The agreement for $I_{1113}$ is within a factor of 3 for $n= 10^4 - 10^5$ cm$^{-3}$,
but only to a factor of 5 (with the analytic underpredicting the intensity) for $n = 10^3$ cm$^{-3}$.
However, for this low value of $n$, the intensities are weak and not detectable.
For higher values of $G_0$ the analytic formula underpredicts the intensity
because of the larger columns of H$_2$O and O$_2$ (see Figure 10 and discussion).
For higher values of $n$ the water plateau is so close to the surface that the
gas temperature drops rapidly and significantly from $A_{Vf}$ to $A_{Vd}$; here
the analytic formula is not accurate since it assumes constant average temperatures
for both grains and gas.  Nevertheless, the analytic formulae are useful
in making estimates of line intensities, and also illustrate the dependences
of the intensities on the physical parameters.


\subsection{Model of Chemical Desorption}

Following O'Neill, Viti, \& Williams (2002), as discussed in \S 2.4, we have
run our standard case with the modification that all molecules formed on
grain surfaces instantly desorb.  In our standard case  only H$_2$ instantly desorbs.
The basic differences are that the freezeout depth $A_{Vf}$ moves inward (increases)
by about 2 magnitudes of visual extinction due to the chemical release of adsorbed OH and
H$_2$O, the
peak gas phase abundance of H$_2$O increases by about a factor of 10, 
and that of  O$_2$ by a factor of
roughly 100.  We do not pursue this type of model further here, but note
that it could warrant further study, although it appears to predict
excessive H$_2$O and O$_2$ intensities.  In the original O'Neill et al work,
photodesorption and cosmic ray desorption were not included.

\subsection{Model with No Grain Chemistry}

In order to test the importance of the formation of OH, H$_2$O and carbon molecules
on grain surfaces in our model, we have run the standard case ($G_0= 100$, $n= 10^4$)
and two other cases ($G_0= 100$, $n=10^3$ cm$^{-3}$ and $G_0=10$, $n= 10^4$ cm$^{-3}$)
 with only H$_2$
formed on grain surfaces and no other grain chemistry.  Like the model described
above in \S 3.5, one main effect is to push the peak in the H$_2$O and O$_2$
abundances to higher $A_V$, since grain formation of H$_2$O is suppressed. However,
in this case there is no enhancement of the peak abundances, because the formation
rates of OH and H$_2$O are suppressed even more than in the case studied in \S 3.5.
In addition, there can be significant depletion of gas phase elemental O at low $G_0$
by the formation of atomic O ice, which does not occur in \S 3.5 or in our models
with grain chemistry in \S 3.1 and \S 3.2.  In the ($G_0=100$, $n=10^4$ cm$^{-3}$) case
and the ($G_0= 10$, $n=10^4$ cm$^{-3}$) case, there is a reduction in both the
columns of H$_2$O and O$_2$ and in the gas and grain temperatures of the emitting region
compared to the cases with grain chemistry.  Thus, the predicted lines are significantly
weaker.  Because of the good quality of our grain chemistry model fits to the data
described in \S 4, and because of the laboratory and observational evidence for grain
chemistry described in \S 2.3 and \S 3.1, we do not consider this case further.

\subsection{Time Dependence}

The numerical results and analytic solutions presented so far are for
steady state solutions to the chemistry.  However, as noted in \S 2.3
and \S 2.4, in the opaque interior of molecular clouds at $A_V>A_{Vd}$,
the chemical timescales may exceed or be comparable to the cloud age,
so that time dependent solutions to the chemistry are needed.  The long
chemical timescales include the freezeout of gas species onto grains
for $n \lta 10^3$ cm$^{-3}$ (see Eq. 1), the cosmic ray desorption of
CO ice (see \S 2.4), and the conversion of gas phase CO to water ice
which is initiated by He$^+$ (see \S 2.3).  In the opaque "freezeout"
regions the attenuated FUV flux plays little role in the chemistry
so that the time dependent abundances do not vary with $A_V$.
Therefore, we present results at a representative $A_V=10$ for
the abundances of gas and ice species as a function of the age
of the cloud, the density of the cloud, and the cosmic ray
desorption rate of CO.

The time dependent code is that of Bergin et al (2000), updated to have the
same grain chemistry and desorption processes as described above for the
steady state code.  It includes more species than the PDR surface code,
and therefore tracks the carbon chemistry more accurately in the
inner regions.  It takes as initial conditions the chemistry of
translucent clouds: all the hydrogen is initially molecular H$_2$ with
the other species atomic (for ionization potentials $> 13.6$ eV) or singly
ionized otherwise.  

The
results are shown in Figures 14 and 15, where
we present  the abundances of  O$_2$, H$_2$O and CO plotted
as a function of the cosmic ray desorption rate of CO ice, for two times
representative of cloud ages ($2\times10^6$
years in Fig. 14 and $2\times 10^7$ years in Fig. 15), for the standard $G_0=100$,
and for 3 densities 
($10^3$, $10^4$, and $10^5$ cm$^{-3}$).  Of particular note is Figure 15,
where one sees the gas phase molecules increase in abundance as the
cosmic ray desorption rate increases, reach a peak when the timescale
for cosmic ray desorption of CO is about 0.1 - 1.0 times the cloud
evolution time of 2$\times 10^7$ years,
and then decreases as the cosmic ray rate increases further and the
desorption time becomes  less than $\sim  2\times 10^6$ years. 
If the rate is too low, then the CO ice remains on the grain and the
gas phases abundances are low.  If the rate is too high, then the
CO ice rapidly desorbs, but then the CO in the gas breaks down by
reaction with He$^+$ and the gas phase elemental oxygen freezes out as
water ice, thereby lowering the gas phase abundances of CO, H$_2$O, and
O$_2$. The main point is that in all cases for $t=2\times 10^7$ years and for
most cases with $t=2\times 10^6$ years, the abundances of
H$_2$O and O$_2$ at high $A_V$, even when time dependent chemistry is
taken into account (steady state chemistry is more extreme in this
respect as seen in Figure 3), are substantially
{\it lower} than the peak abundances of these molecules
at intermediate $A_V$.  Thus, most of the column of these molecules
in molecular clouds are not provided by the interior regions, where
they are frozen out, nor by the surface, where they are photodissociated.
Rather, most of the column arises from the intermediate zone.



\subsection{Chemistry: the Gas Phase C/O Ratio}

Figures 16 and 17 show the {\it gas phase elemental} abundance of C and the
{\it gas phase} abundance ratio of {\it elemental} C/O,
 as a function of $A_V$, for two different cloud ages, $t=2\times 10^6$ years and 
$t=2\times 10^7$ years, respectively.  In each figure, we plot the standard case,
the standard case but with the density changed to $n=10^3$ and $10^5$ cm$^{-3}$, and the
standard case but with the incident FUV field changed to $G_0=1$ and $10^3$.
  For these results, we have used our time dependent 
models, since the interesting effect of C/O exceeding unity occurs at high $A_V$,
where the time dependent models are more relevant and are more accurate in
treating the carbon chemistry.  At early times, $t\sim 2 \times 10^6$ years, the C/O
ratio generally stays below unity at high $A_V$. The exception is the high
density ($n= 10^5$ cm$^{-3}$) case, but here the gas phase elemental abundance
of carbon is extremely low, which implies low abundances of any organic molecules
produced in this carbon-rich environment.  At late times, 
$t\sim 2 \times10^7$ years, we see that all cases except the high $G_0=10^3$
case result in gas phase C/O$>1$ for $A_V \gta 8$, due to the preferential freezeout
of water ice which robs the gas of elemental O.  For those cases with C/O$>1$,
the gas phase elemental C abundance increases with increasing $G_0$ and decreasing $n$,
due to increased thermal desorption rates of carbon molecular ices
and decreased freezeout rates, respectively.\footnote{At $t=2 \times 10^7$ years
the time-dependent code shows that CO and CH$_4$ ice dominates the carbon ices for
$G_0 \lta 500$, and CO$_2$ and C$_3$H ice dominates at higher $G_0$, due to
thermal desorption of the more weakly bound CO and CH$_4$.}
Figure 17 shows that
 for densities $n\sim 10^3 - 10^4$ cm$^{-3}$ and $G_0=1-100$, strong chemical 
effects might be observed deep in
 clouds, with high abundances of gas phase hydrocarbons and other carbon
molecules rather than CO.
Unfortunately, these results can only be viewed as suggestive, because
 the gas phase C depends on the desorption
rate of CO, CH$_4$ and other carbon species frozen to grain surfaces, and these
are uncertain.  Nevertheless, it appears that regions with C/O $>1$ and high
gas phase elemental C abundance can appear
deep in the cloud. This is consistent with the chemistry of GMC cores
such as Orion A (e.g., Bergin et al 1997) and regions in the Taurus Molecular
Cloud (e.g., Pratap et al 1997).  In addition, in these deep
layers significant amounts of  CH$_4$ ice may form, and its
release to the gas stimulates the carbon chemistry.
In summary, the freezeout of the oxygen into water ice can have
 profound effects on the carbon chemistry
deep in the cloud and the ability of the cloud to produce carbon chain molecules
or hydrocarbons.  


\subsection{Limb Brightening and Surface to Volume Effects}

Because our model suggests a peak in the H$_2$O and O$_2$ abundance at intermediate depths
into a molecular cloud, one might expect clouds to be limb-brightened if viewed with
beams smaller than the cloud.  We make here two caveats to this prediction.  First,
GMCs are clumpy, and the clumpiness means that $A_{Vf}$ will occur at different physical
depths (i.e., radii) in the clouds, depending on the particular radial ray
through the cloud and where it intersects clumps.  This will
have the effect of smearing the limb-brightening.  In addition, although the 557 and 1113 
GHz H$_2$O transitions are "effectively thin", they do  suffer numerous scattering events
since
the optical depth in the lines are greater than unity.  This scattering will reduce the
limb-brightening effect for the H$_2$O transitions.  The O$_2$ does not suffer scattering,
so it may be more promising.  However, the sensitivity of {\it Herschel} may require large O$_2$ columns
$> 10^{16}$ cm$^{-2}$, and these may only be achieved for high $G_0$.  In such cases, the
peak O$_2$ abundance occurs deep in the cloud, $A_V \sim 8$, and makes limb brightening only
possible to detect in very high column but extended nearby clouds illuminated by high
fields.
Nevertheless, {\it Herschel} observations
of O$_2$ in clouds may find regions of limb-brightening.  

One still would like to
map the clouds, however, even if limb brightening is hard to detect. The regions
of high $G_0$ may be fairly limited in spatial extent, and our models predict higher
H$_2$O and especially O$_2$ columns in these regions which {\it Herschel} may spatially resolve.
In addition, the H$_2$O and O$_2$ plateaus extend for $\Delta A_V \sim 3$.  For gas
densities of $n\sim 10^4$ cm$^{-3}$ this corresponds to $\sim 0.2$ pc, which at 500 pc corresponds
to an angular size $\sim 1.5'$. Therefore, for a PDR slab viewed edge-on, {\it Herschel}
may resolve the plateau region.

Even for clouds too small in angular extent to map, our model can be tested
by correlation studies of the H$_2$O and O$_2$ intensities with the intensities
of rotational transitions of other molecules known to trace either the surface
or the volume of the cloud.  For example, $^{13}$CO and C$^{18}$O are thought to
generally\footnote{In some cases CO is frozen on grains in very high $A_V$
regions at cloud centers; see, for example Bergin \& Tafalla (2007) for a review
of the observations.} trace the
volume of the cloud, whereas PDR species such as CN trace the "surface" of the cloud
(Fuente et al 1995, Rodriguez-Franco et al 1998).  We are investigating such observational 
correlations in Melnick et al (in preparation).

\section{APPLICATIONS TO SPECIFIC SOURCES}

\subsection{Diffuse Clouds}

The diffuse clouds that we model, based on those observed by SWAS towards  
W51 and W49 where H$_2$O 557 GHz is seen in absorption, are characterized by
$G_0 \sim 1-10$, $n \sim 100$ cm$^{-3}$, and $A_V \sim 1-5$ [these are sometimes
called ``translucent clouds'' \citep{sm06}]. 
Diffuse clouds are illuminated on all sides by the diffuse interstellar FUV field, 
whereas our models are those of
1D slabs illuminated from one side.  To roughly construct a model of
a diffuse cloud of total thickness $A_{Vt}$, we take the results 
from $A_V=0$ to 0.5$A_{Vt}$ and assume a mirror image for the rest of the cloud from
0.5$A_{Vt}$ to $A_{Vt}$. 

We compare a single model fit of a diffuse cloud
with the 54 and 68 km s$^{-1}$ velocity components observed by SWAS
towards the background submillimeter source W49 \citep{p04} as well as in the single 
absorption component towards W51 \citep{n02} .  For all three of these components, 
both OH and H$_2$O column densities have been measured, with N(H$_2$O)$\sim 1-3\times 
10^{13}\,\rm cm^{-2}$ and N(OH)/N(H$_2$O) $\sim$ 3-4. Our diffuse cloud model with $n=10^2\,\rm cm^{-3}$
and $G_0=1$ gives N(H$_2$O)=$3\times 10^{13}\,\rm cm^{-2}$ and N(OH)/N(H$_2$O) =3.5
 for a total cloud extinction of $A_{Vt}$= 4.4 magnitudes. 
We find that assuming a photodesorption yield $Y_{OH,w} \simeq 2 \times Y_{H_2O}$ provides
the best fit to the observed N(OH)/N(H$_2$O) ratios.  
The same model also provides predicted $^{12}$CO and atomic C columns and line intensities of the $^{12}$CO J=1-0, J=3-2 and 
[CI] 609 $\mu$m lines, all of which were observed in the W49 components.  The models overpredict
the columns and the intensities  of the CO J=1-0 and [CI] lines by a factor $\sim 3$ which may be explained by
dilution
of the emission lines in the beam.  The J=3-2 line, however, is predicted to be a factor $\sim  2$ smaller
than is observed; a better fit for this line is found at higher densities, but  makes the fit of all the 
other observed lines poorer.   We note that Bensch et al (2003) had similar problems in
fitting a PDR model to the diffuse cloud CO and [CI] results. 

Comparing our H$_2$O column to the model $A_{Vt}$, we find a 
column-avearged H$_2$O abundance of $3 \times 10^{-9}$ with respect
to H nuclei. ÊThe observational papers cited above quote H$_2$O/H$_2$
abundance ratios $>10^{-7}$. ÊThese high abundances, however, were 
derived from the observed CO columns by multiplying the CO columns by
$10^4$ to obtain H$_2$ columns. ÊIn our models, we find H$_2$ columns
of $\sim 4 \times 10^{21}$ cm$^{-2}$ and H$_2$O/H$_2$ abundance ratios
of $\sim 10^{-8}$. ÊThe H$_2$/CO column-averaged abundance ratio in the
model is $1.3 \times 10^5$. ÊIn these translucent clouds, CO is 
photodissociated much more than H$2$, which self-shields much more
effectively. ÊConsiderable H$_2$ exists in regions where the gas phase
carbons is mostly C$^+$. These translucent clouds are examples of
"dark H$_2$" which is not traced by CO.

\subsection{Dense, Opaque Clouds}

\subsubsection{A$_V$ Thresholds for Ice Formation}

Whittet et al (2001) have presented plots of the water ice columns versus
cloud extinction in which a clear threshold $A_V$ for water ice is inferred.
In our model we identify that threshold $A_V$ with $A_{Vf}$, since the ice
very rapidly builds up with increasing $A_V$ once the grain has a monolayer of ice.  
Whittet et al (2001),
Tanaka et al (1990), and Williams, Hartquist \& Whittet (1992)
find that $A_{Vf}\sim 3$ for Taurus and $\sim 6$ for $\rho$ Oph. Assuming
the radiation field in Taurus is $G_0 \sim 1$ and that the gas density is $10^3$
cm$^{-3}$, we predict from our analytic model with $Y = 3\times 10^{-3}$ (i.e.,
$Y_{H_2O} = 10^{-3}$, recall that $Y$ is the total yield of both H$_2$O
and of OH+H being ejected from water ice) that
$A_{Vf}= 3$.  For $\rho$ Oph, \cite{lis99}  and Kamegai et al (2003) have estimated $G_0 \sim 100 - 1000$ 
as an average value in the SWAS beam.  Assuming $n = 10^3 - 10^4$ cm$^{-3}$ in
the ambient gas averaged over the beam, we predict $A_{Vf} \simeq 6$.
Thus, our model does a good job of matching these thresholds.
Note that the threshold $A_{Vf}$ in our model mainly depends on
$\ln (G_0Y/n)$ and so to the extent that we know $G_0/n$, we can constrain $Y$.
We therefore see that $Y\sim 3\times 10^{-3}$ is a reasonable choice for interstellar 
ice grains.\footnote{See also Nguyen et al (2002) for another model of the
ice distribution in Taurus}.

\subsubsection{B68}

B68 is one of the best-studied, nearby, isolated molecular clouds (Alves et al. 2001, Zucconi et al. 2001).
It lies at $\sim 160$ pc, with a mean density of $\overline{n}\sim 10^5$ cm$^{-3}$,
an $A_V$ through the cloud center of $\sim 30$, and
a central temperature of 7 K.  Extinction and dust continuum maps have recently provided a detailed
view of the density and temperature structure of the cloud (Alves et al. 2001, Zucconi et al.
2001). Previous
models of the CO lines and the exterior temperature of the cloud suggest $G_0 \sim 0.25
-1$ \citep{b02}.  The angular size of the cloud is $\sim 1.3\arcmin$, smaller than
the SWAS beamsize.

SWAS has performed long integrations on B68 searching for the 557 GHz H$_2$O line
and the 487 GHz O$_2$ line, setting stringent 3$\sigma$ upper limits on the line strengths
(assuming thermal linewidths) 
from these undetected transitions of $I_{557} \lta 4\times
10^{-9}$  erg cm$^{-2}$ s$^{-1}$ sr$^{-1}$ and $I_{487} \lta 6\times 10^{-9}$  
erg cm$^{-2}$ s$^{-1}$ sr$^{-1}$.
Utilizing the known distribution of density $n$ 
with $A_V$, and assuming $G_0 = 0.5$ and our standard values of $Y$ and $\sigma_H$, 
we have modeled the H$_2$O and O$_2$ abundance profiles and
the expected observed line strengths in B68.  To compare with the observed intensities, we 
calculate the intensity expected in the model and then dilute the intensity by the cloud area 
divided by the SWAS beam area. Averaged over the SWAS beam, the  predicted line
intensities of  $I_{557}
\sim 4\times 10^{-11}$  erg cm$^{-2}$ s$^{-1}$ sr$^{-1}$ and $I_{487} \sim 1\times 10^{-12}$
erg cm$^{-2}$ s$^{-1}$ sr$^{-1}$
lie  well below
the upper limits set by SWAS.
We predict a non-beam-diluted line intensity $I_{1113}\sim 10^{-8}$
erg cm$^{-2}$ s$^{-1}$ sr$^{-1}$, which is not detectable by {\it Herschel}. 
Because the 557 GHz and 487 GHz line intensities are  sensitive to $G_0$ (via the gas and dust 
temperatures), regions like B68 with low FUV fields are not promising targets
for H$_2$O and O$_2$ detections. 

Another interesting effect in our modeling of B68 arises because our models incorporate the
observed increase of density with depth $A_V$. From Figure 6 one sees that this reduces the width
$\Delta A_V$ of the H$_2$O plateau since the lower density surface gas means that $A_{Vf}$ is larger,
whereas the higher density gas in the interior makes $A_{Vd}$ smaller.  However, this effect is not
large, since our B68 model has $n= 1.3 \times 10^5$ cm$^{-3}$ at $A_V=1$, $n= 1.7 \times 10^5$ cm$^{-3}$ 
at $A_V=2$, and $n= 2.4 \times 10^5$ cm$^{-3}$ at $A_V=3$ (roughly the extent of the H$_2$O plateau).
We find that our predicted column of H$_2$O in our density-varying model differs by only 10\%
from a constant density, $n= 1.7 \times 10^5$ cm$^{-3}$, model.

\subsubsection{NGC2024}

NGC2024 is a dense, $n \sim 10^5$ cm$^{-3}$ star forming molecular cloud in Orion illuminated
by a relatively intense, $G_0 \sim 10^4$, FUV field \citep{gi00}.  Contrasted to
B68, it provides a test of our models for elevated radiation fields (i.e., $G_0 > 500$).  
SWAS has observed both the H$_2$O 557 GHz line and the O$_2$ 487 GHz line in this source.  We
have reanalyzed the SWAS data from the archives and obtain
 a beam-averaged intensity of $3 \times 10^{-7}$ erg cm$^{-2}$ s$^{-1}$ sr$^{-1}$ for the
557 GHz line and a 3 sigma upper limit on the 487 GHz line of $4 \times 10^{-8}$ erg cm$^{-2}$ s$^{-1}$
sr$^{-1}$  [see also Snell et al (2000) for the 557 GHz line and Goldsmith et al (2000) for
the 487 GHz upper limit].

Our models (see Figs. 12 and 13)
reproduce the H$_2$O 557 GHz line with beam
averaged gas densities
$n \sim 10^{4.8}$ cm$^{-3}$ and $G_0 \sim 10^{4}$, assuming that the ortho/para 
H$_2$O ratio is in LTE with the grain temperature and the ortho/para ratio of H$_2$
is the steady state value from the formulation by Burton et al (1992).
 Here, in our peak water plateau, the gas temperature
is $\sim$27~K, and the ortho/para ratio of H$_2$ is $\sim 1/6$. The grain temperature
is $\sim 42$ K and the ortho/para ratio of H$_2$O is 2.7. In the model the
O$_2$ 487 GHz line
intensity is predicted to be $4\times 10^{-9}$ erg cm$^{-2}$ s$^{-1}$ sr$^{-1}$, well
below SWAS detectability.  The peak (plateau) abundances for H$_2$O and O$_2$
in the model are (see Figs. 5 and 6) $10^{-7}$ and $3\times 10^{-6}$ respectively.  The column
averaged abundances, assuming $N \sim 4\times 10^{22}$ cm$^{-2}$
or $A_V \sim 20$ through the cloud \citep[based on CO column densities greater than few$\times 
10^{18}$ cm$^{-2}$ over a $\sim 4\arcmin$ region]{gi00} is $\overline{x}(H_2O) \equiv
N(H_2O)/N \simeq 1.1\times 10^{-8}$, and $\overline{x}(O_2) \simeq 1.7\times 10^{-7}$.  
We conclude that the models also provide good fits to NGC2024, using reasonable values of
$n$ and $G_0$, and predict peak abundances that are approximately 10 times greater
than the column averaged abundances.  We predict $I_{1113} \sim 3 \times 10^{-7}$
erg cm$^{-2}$ s$^{-1}$ sr$^{-1}$, which is detectable by the HIFI instrument on the
{\it Herschel Observatory}.





\subsubsection{O$_2$ Observed by Odin in $\rho$ Oph}

Larsson et al (2007) have reported the detection of the ground state
transition of O$_2$ at 119 GHz with the space submillimeter telescope {\it Odin}.
Note that this is a different transition than the O$_2$ transition observed by
SWAS, and is at a much longer wavelength.  Therefore, although the {\it Odin}
telescope has twice the diameter of SWAS, the {\it Odin} beam size at 119 GHz ($\sim 10'$)
is more than twice that of the SWAS beam at 487 GHz.  At the
distance of $\rho$ Oph, the {\it Odin} beam corresponds to 0.4 pc.  The beam
was centered on a 450 $\mu$m peak, and the submillimeter dust emission and
the CO isotopic rotational emission suggest an H$_2$ column of $2\times 10^{22}$
cm$^{-2}$ ($A_V \sim 20$).  A temperature of 30 K has been obtained for
both the gas (e.g., Loren et al 1990) and the dust (Ristorcelli et al, in preparation)
in this region.  Using these parameters, and assuming the O$_2$ to be optically
thin and in LTE, Larsson et al find a column $N(O_2) \sim 1 \times 10^{15}$ cm$^{-2}$ for
an abundance relative to H$_2$ of $\sim 5 \times 10^{-8}$ but an abundance in our
units (relative to H nuclei) of $x(O_2)= 2.5 \times 10^{-8}$.  The authors acknowledge
a possible uncertainty of at least a factor of 2 in their estimate of the O$_2$
abundance. 

The $\rho$ Oph region observed is illuminated on the back side by a B2 V star,
with the cool dense cloud core sitting in front of the photodissociation region.
The relatively high dust temperature, and the likely distance of the B star from
the cloud, suggest $G_0 \sim 100-1000$ shining on the cloud surface \citep{c76, kam03}.
The 0.4 pc size coupled with the observed total H$_2$ column suggest $n\sim 
10^5$ cm$^{-3}$ in this dense region.  Setting $G_0= 300$ and $n=10^5$ cm$^{-3}$,
and using standard parameters, we obtain a column $N(O_2) = 1.4\times 10^{15} $ cm$^{-2}$.  In our
model, the gas and dust temperatures at the O$_2$ abundance peak are $\sim 20$ K.
We predict (see Figs. 12 and 13) an H$_2$O 557 GHz intensity of 
$I_{557} = 10^{-7}$ erg cm$^{-2}$ s$^{-1}$ sr$^{-1}$,
and a O$_2$ 487 GHz intensity of  $I_{487} = 2\times 10^{-9}$ erg cm$^{-2}$ s$^{-1}$ sr$^{-1}$ from
this region.  We have analyzed SWAS archival data on this source and find a 
measured H$_2$O 557 GHz intensity of $1.3 \times 10^{-7}$ erg cm$^{-2}$ s$^{-1}$ sr$^{-1}$
and a 3$\sigma$ upper limit of $2\times 10^{-8}$ erg cm$^{-2}$ s$^{-1}$ sr$^{-1}$ for the
O$_2$ 487 GHz line.  We predict $I_{1113} \sim 3\times 10^{-7}$ erg cm$^{-2}$ s$^{-1}$ sr$^{-1}$,
which may be marginally detectable by the HIFI instrument on {\it Herschel}.
In summary, our model successfully fits the H$_2$O and O$_2$
observations of $\rho$ Oph, using reasonable parameters for this region.

\subsubsection{Upper Limits on O$_2$ Observed by Odin}

Pagani et al (2003) report 3$\sigma$ upper limits on $N(O_2)$ in a number of sources.
As examples of sources with low incident $G_0$, the upper limits for TMC1 and L183 (L134N)
are $7\times 10^{14}$ and $1.1\times 10^{15}$ cm$^{-2}$, respectively.  Although
the {\it Odin} beam is large, these sources are extended enough that beam dilution
should not be a significant factor.  In addition, although much of the Taurus cloud
has relatively low total $A_V$, these regions appear to have sufficiently high $A_V
\gta 10$ to encompass the O$_2$ plateau.  Using an estimated $G_0 \sim 1$ and $n \sim 10^4$
cm$^{-3}$, our models predict a column of $N(O_2) \sim 4\times 10^{14}$ cm$^{-2}$
(see Figure 10) on both sides of the cloud, for a total column of $\sim 8\times 10^{14}$
cm$^{-2}$.  This prediction lies very close to the observed upper limits.  If future
observations or analysis drive this upper limit down below our predicted value, then
either beam dilution or a lower value of $Y_{H_2O}$ need to be invoked to reconcile
observations with the models. Note that the upper limits on $N(O_2)$ set upper
limits on $Y_{H_2O} \lta 10^{-3}$.

More challenging  to our model are the upper limits quoted for sources with relatively
high $G_0$.  For example, Pagani et al report 3$\sigma$ upper limits of $N(O_2) <
1.9 \times 10^{15}$ cm$^{-2}$ for Orion A.   This column is averaged
over the $9'$ beam of {\it Odin}, which corresponds to a linear extent (diameter)
of about 1.4 pc at the distance of Orion.   If there were no extinction from
the source of the FUV (primarily $\Theta^1C$ in the Trapezium), the FUV flux
at a distance of 0.7 pc from the star corresponds to $G_0 \sim 4\times 10^3$.
Taking this as the average $G_0$ incident on the molecular cloud in the beam,
our model predicts $N(O_2) \sim 10^{16}$ cm$^{-2}$ (see Figure 10), significantly
higher than the observed upper limit.  We offer two possible explanations for
this discrepancy.  First, $\Theta^1C$ may be inside a cavity in the cloud carved
out by the champagne flow created by the Trapezium.  In this case, extinction by
the intervening cloud would diminish $G_0$ in the outer regions of the beam,
and thereby diminish the beam average column of O$_2$.  A second possibility arises
from the fact that the enhancement of the O$_2$ column at high $G_0$, as discussed in \S 3.2,
is caused by the thermal evaporation of O atoms from grain surfaces before they
can react to form water ice on the surface.  We have adopted a binding energy of
O atoms to the grain surface of $E_O/k= 800$ K (see Table 1).
However, this binding energy
is uncertain and in our view may be low, given that O is a radical.  A modest
increase in this binding energy would result in a much higher critical $G_0$
where the enhancement of $N(O_2)$ occurs, since the grain temperature is a weak
function of $G_0$ ($T_{gr} \propto G_0^{0.2}$) and thus the critical $G_0$ goes
as $E_O^5$.  In other words, increasing $E_O/k$ to 1200 K would increase the
critical $G_0$ from about 500 to about 4000, and provide a solution consistent
with the observations.

Sandqvist et al (2008) use {\it Odin} observations to place an upper limit
of $N(O_2) < 6 \times 10^{16}$ cm$^{-2}$ toward SgrA. As noted in that paper,
this is consistent with our model clouds for any incident $G_0$ and any density
$n$ for standard local gas phase elemental abundances of O and C.

\section{SUMMARY, DISCUSSION AND CONCLUSIONS}








We have constructed both a detailed numerical code and simple
analytical equations to follow the abundances of the main oxygen-bearing
species as a function of depth in molecular clouds.  We take
clouds of constant density $n$, illuminated by an external FUV
(6 eV$< h\nu < 13.6$ eV) field characterized by the unitless
parameter $G_0$ normalized to the local (Milky Way) interstellar
FUV field.  We explore the range 10$^3$ cm$^{-3} < n < 10^6$ cm$^{-3}$
and $1 < G_0 < 10^5$.  Our models incorporate thermal balance,
gas phase chemistry, a simple
grain surface chemistry that includes the formation of H$_2$, H$_2$O, and CH$_4$
on grain surfaces, the sticking of gas phase species to grain surfaces, and
various desorption processes such as photodesorption, thermal
desorption, and cosmic ray desorption.

The abundances of chemical species vary considerably as a function
of depth into a molecular cloud.  At the surface, molecules are
photodissociated and the gas is primarily atomic (e.g., O) or,
for species with ionization potentials less than 13.6 eV, 
singly ionized (e.g., C$^+$). Here, although dust grains may
be cold enough to prevent rapid {\it thermal} desorption, ices
do not form because {\it photodesorption} keeps the refractory
dust clear of ice mantles (the same is true of dust in the diffuse
ISM). Very deep in the cloud, again assuming grains cold enough
to prevent thermal desorption, photodesorption is negligible and
ices form, causing gas phase abundances of condensibles to plummet.
Here, time dependent effects (generally associated with cosmic rays) dominate the
gas phase chemistry due to the time required to deplete condensibles onto grains, 
to desorb  CO ice by cosmic rays, and to 
destroy gas-phase CO (by He$^+$ produced by cosmic rays).  These
processes are slow, and have timescales comparable to, or longer
than, typical molecular cloud lifetimes.  At intermediate depths,
photodesorption of ices supply gas phase elements, and the partial
shielding of the photodissociating external FUV flux allows gas phase molecular 
abundances to peak.

We have focused in this paper on oxygen chemistry, and followed 
in particular the gas phase species O, OH, O$_2$, and H$_2$O
as well as water ice.  Our answer to the longstanding 
astrochemical question, ``Where is elemental O in molecular clouds?'',
is that oxygen not in CO is primarily gas phase atomic O to a depth
$A_{Vf}$ (see Eq. 23), and is water ice at larger depths.  
The freezeout depth $A_{Vf}$ where significant water ice forms is
proportional to $\ln (G_0/n)$, and is typically 3-6.  
Refractory silicate grains contain a modest abundance of O ($\sim 10^{-4}$),
but the bulk is likely gas phase O and CO along with CO ice and H$_2$O ice.
For example, in our standard model, integrating to $A_V= 10$, the
O not in refractory grains is divided $56$\% H$_2$O-ice, $25$\% gas
phase O, $10$\% CO-ice, and $8$\% gas phase CO.

Our models also give an indication of the origin of water in
molecular clouds.  In the steady state standard model, at $A_{Vf}$
the formation rate of gas phase H$_2$O is 2\% via gas phase
reactions and 98\% by photodesorption of H$_2$O that has been formed
on grains.  At $A_{Vd}$ the percentages are 30\% and 70\%, respectively;
midway in the plateau, they are 8\% and 92\%, respectively.\footnote{A 
warning to future modelers: these percentages are
not trivial to compute since there is a great deal of cycling
of H$_2$O and OH to H$_3$O$^{+}$ (via reaction with H$_3^+$)
and then back to H$_2$O and OH (via dissociative electronic
recombination).  This cycling gives, of course, no net production
of these molecules.}  
These percentages depend on $G_0$ and $n$,
but as long as $G_0 \lta 500$, these numbers are representative.
At higher
$G_0>500$, where O atoms thermally desorb from grains before forming
OH and eventually water ice, most of the water production is via
gas phase reactions.  
 The production of water ice is about
3 times greater than the production of gas phase water by
photodesorption of water ice, because 2/3rds of the water ice
is photodesorbed as OH.  Thus, one needs to modify these
numbers if one wants to estimate the formation of water
in any form in a cloud: for example, in the standard
case midway in the plateau the percentages are 3\% by
gas phase reactions and 97\% by grain surface reactions.  

We plan a follow-up paper to examine carbon chemistry.  We note
that Figure 14 shows that for all densities considered and by 
$t> 2\times 10^6$ years, gas phase CO is largely absent in cloud interiors.
This is in rough agreement with observations that show CO depletions
in dense cloud interiors.  Thus, CO is not always a volume tracer.
However, the sequel paper will include more grain surface
carbon chemistry such as the formation of CO$_2$ and methanol
on grain surfaces. In addition, we will examine the effects of 
initial conditions on time-dependent chemistry, and compare observations
with a wide variety of models with varying cosmic ray desorption
rates and gas density profiles through the cloud.

For $G_0 \lta 500$, the abundances of gas phase OH, O$_2$ and H$_2$O peak
and plateau at values 
$\sim 10^{-7}$ between $A_{vf}$ and $A_{Vd}$ (see, Eqs. 26, 31, and 33).
These abundances are independent of $n$ and $G_0$ because
both their formation rate (ultimately from photodesorption
of H$_2$O ice) and their destruction rate (from photodissociation)
both depend on the product $nG_0$. However, since their formation
is linked with H$_2$O ice photodesorption, these plateau abundances
do increase with increasing ice photodesorption yield, $Y_{H_2O}$
and $Y_{OH,w}$, and also with the grain cross sectional area per
H nucleus $\sigma _H$ (see same equations).   

For higher values
of $G_0 \gtrsim 500$, the chemistry undergoes a change, as O atoms
no longer form H$_2$O ice on grain surfaces.  Here, the grains
are warm enough ($\gta 20$ K) that O atoms thermally desorb
before reacting with H atoms.  This pushes the freezeout to
greater depths and enhances the abundances of gas phase O, OH,
O$_2$ and H$_2$O by keeping more of the elemental oxygen in
the gas phase.  Because O$_2$ formation goes as the product
of O and OH abundances, O$_2$ is the most sensitive to this
change and its local abundance reaches $10^{-5}$, while its column climbs
to $\sim 2\times 10^{16}$ cm$^{-2}$.  We note that the observations
of Pagani et al (2003) may suggest a higher binding energy for O
on grains than adopted, and in this case the critical $G_0$ where
this occurs may be significantly higher than 500.

Deep in the cloud interior at $A_V >> A_{Vf}$,  the gas phase
abundances drop as ices become dominant, the exact values
of the abundances
depending on cloud lifetimes and the cosmic ray desorption rates
assumed (see Figures 14 and 15).  Here, time dependent chemistry dominates.

Poelman, Spaans, \& Tielens (2007) have suggested that the SWAS results do
not rule out substantial gas phase H$_2$O abundances and columns in the centers
of clouds, because of optical depth effects which ``hide'' the H$_2$O in
the centers.  However, as this saturation effect occurs, the line 
center antenna temperature approaches the gas temperature of the
emitting region.  In our model, this gas temperature is typically 10-20 K.
Even in the cold interior of clouds, one would expect temperatures of 5-10 K.
The SWAS observed antenna temperatures in regions of detected H$_2$O
are of order $T_A < 1$ K.  This result alone suggests the H$_2$O is
in the ``effectively optically thin'' regime.  Our model does include
an escape probability formalism for the H$_2$O lines.  We find, in
agreement with Poelman et al, that significant collisional de-excitation
and saturation of the line occurs when $\tau n/n_{cr} > 1$, where $\tau$
is the line center optical depth and $n_{cr}$ is the critical density
($\sim 10^8$ cm$^{-3}$ for the 557 GHz line).  Thus, the requirement
for 557 GHz line saturation is roughly

\begin{equation}
\left[{N(H_2O) \over {6\times 10^{12}\ {\rm cm^{-2}}}}\right]{n\over n_{cr}} \gta 1.
\end{equation}
Since our models have $N(H_2O) \sim 10^{15}$ cm$^{-2}$ and $n \lta 10^6$ cm$^{-3}$,
we are almost always within the effectively optically thin regime, except perhaps
for the highest densities $n\sim 10^6$ cm$^{-3}$.  Furthermore, if
the H$_2$O columns and densities were high enough that line saturation occurred, the
line fluxes would exceed those observed by SWAS.  We argue that SWAS observations
indicate effectively thin 557 GHz emission, and that there is no evidence for hidden
H$_2$O in the centers of clouds.

Bergin et al (2000) listed ingredients for a successful 
astrochemical model of a molecular cloud.  Slightly updated,
they included: (1) low column-averaged abundances ($\lta 10^{-7}$)
of gas phase O$_2$, (2) low column-averaged abundances ($\lta 3 \times 10^{-8}$)
of gas phase H$_2$O, (3) an explanation of the apparently higher abundances
of H$_2$O when it is observed in absorption in translucent clouds, (4) column-averaged water
ice abundances of $\sim 10^{-4}$, and (5) existence of complex carbon-bearing
species in the gas phase in some regions.  As noted above, the model presented
in this paper fulfills all of these criteria.

In addition, we have successfully applied, in a simple way, these
models to observations of the $A_V$ threshold for the onset of water
ice, the strict upper limits for the gas phase H$_2$O abundance in
B68 made by SWAS, the possible detection of O$_2$ in $\rho$ Oph made
by {\it Odin}, the detection of H$_2$O and upper limits for O$_2$
made by SWAS for NGC 2024, and upper limits of the O$_2$ abundance
set by {\it Odin}.

These successes imply that the proper modeling of the interstellar
chemistry of molecular clouds must include as ingredients: photodissociation,
freezeout on grains, grain surface chemistry, desorption processes,
and, in the very opaque central regions, time-dependent chemistry.  At the surface
photodissociation and photodesorption dominate.  Deep in the cloud, dust
vacuums the condensibles and slow cosmic ray processes are important, requiring
time-dependent chemistry.  At intermediate depths, gas phase molecular
species such as O$_2$ and H$_2$O peak in abundance, and provide the
source of their submillimeter emission.

\acknowledgements

We would like to thank M. Baragiola, U. Gorti,  R. Liseau, L. Pagani, 
and E. van Dishoeck  for many useful
discussions. We also thank the referee for a thorough and constructive report.
We gratefully acknowledge the financial support of NASA grant
NNG06GB30G from the Long Term Space Astrophysics (LTSA) Research Program.

\appendix
\section{Appendix}
\placetable{symtab}

\clearpage
\input{tab1.tex}
\clearpage
\input{tab2.tex}
\clearpage
\input{tab3.tex}
\clearpage

\begin{figure}[ht!]
\label{fig:gasphase}
\plotone{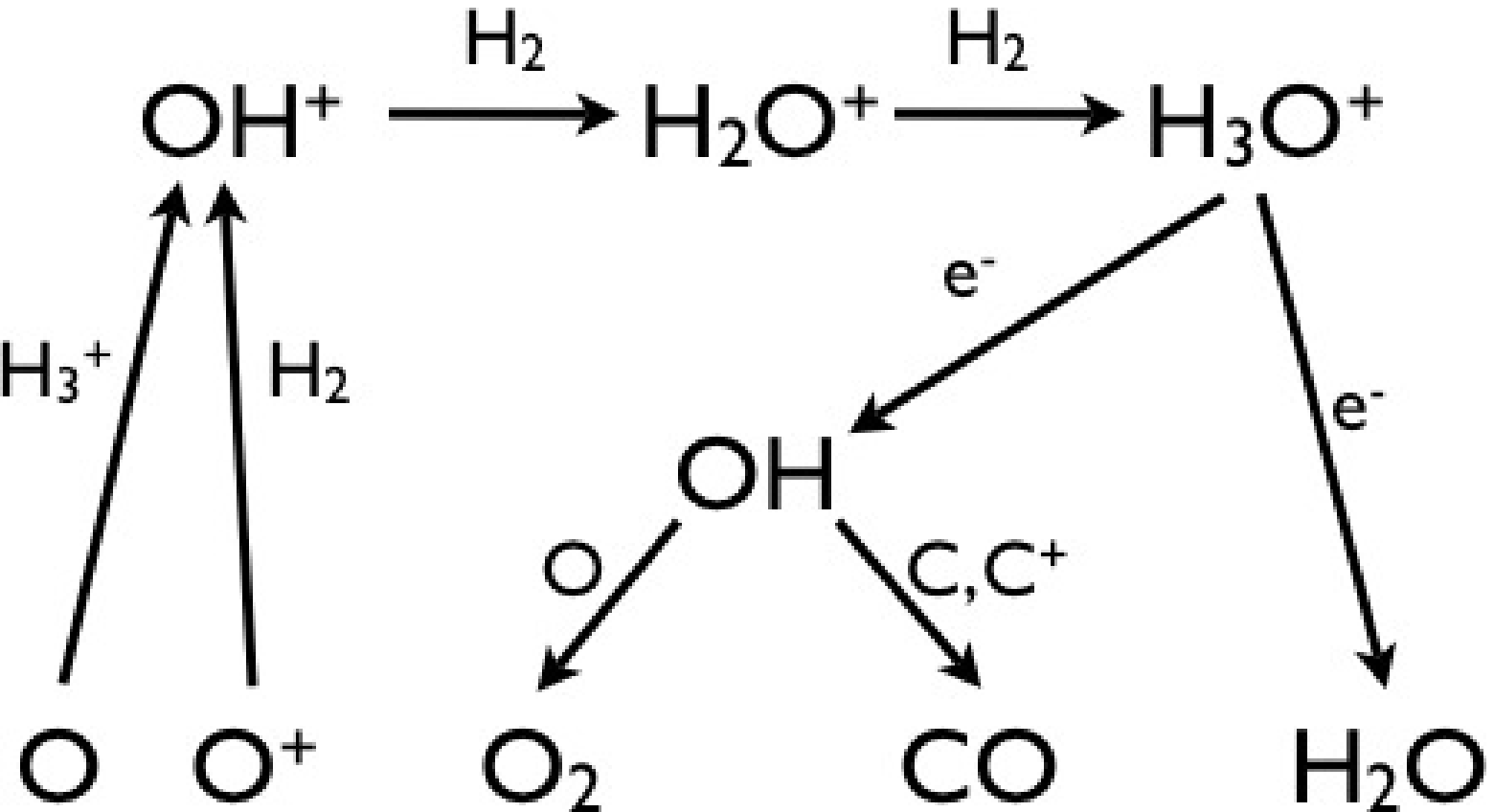}
\caption{Standard ion-neutral chemistry leading to the formation of H$_2$O
and O$_2$ in shielded regions of the ISM. The chemistry is often initiated by
cosmic ray ionization of H$_2$.  The resultant H$_2^+$ reacts with H$_2$
to form H$_3^+$. An 
alternate route starts with the cosmic ray ionization of hydrogen and
the resultant change exchange of H$^+$ with O to form O$^+$.}
\end{figure}
\clearpage
\begin{figure}[hb!]
\plotone{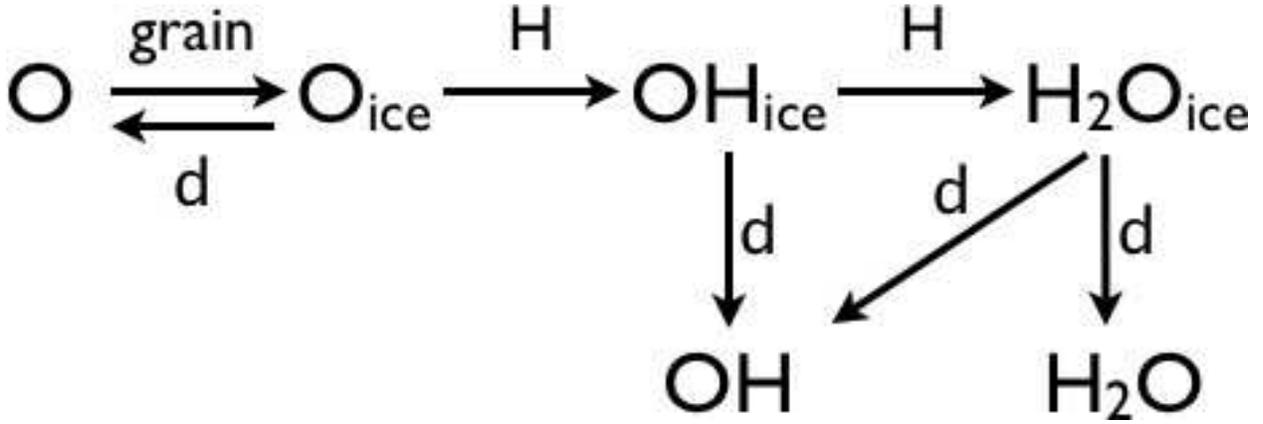}
\caption{Grain-surface oxygen chemical network included in this modeling effort.
Note that in addition we include grain surface carbon chemistry (see text) and
allow the sticking of all species onto grain
surfaces and various desorption mechanisms.   The grain surface oxygen
reaction network 
begins with O sticking to grains, followed by accretion of H to form 
OH$_{ice}$ and then
H$_2$O$_{ice}$. Here, ``$d$'' stands for the three desorption
processes: thermal, photo-, and cosmic ray. 
Near the cloud surface, FUV photons photodesorb ices, leading to 
some gas-phase H$_2$O and O$_2$. In addition, for high FUV fields which
warm the grains, thermal desorption of the weakly bound O atoms can be
important both near the surface and deep in the cloud. Finally, deep in 
the cloud, cosmic ray desorption is important. }
\label{fig:grainsurface}
\end{figure}

\clearpage
\begin{figure}[ht!]
\plotone{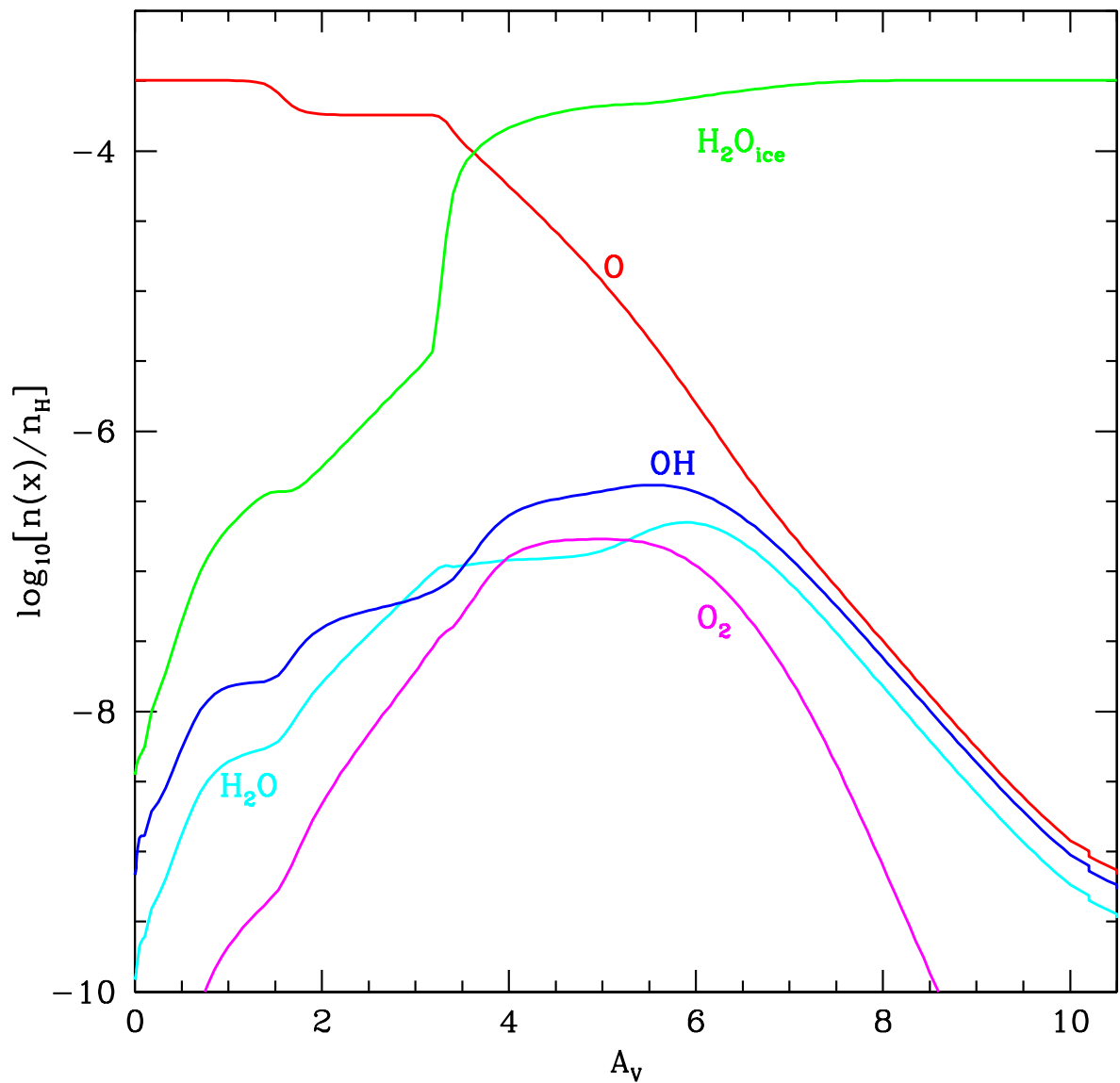}

\caption{Oxygen chemistry for the standard ($n= 10^4$ cm$^{-3}$, $G_0= 100$)
steady-state PDR model, including freeze-
out, grain-surface reactions, thermal
desorption, photodesorption, and cosmic ray desorption. 
Model parameters are those given in Table 1. The
abundances of major O-bearing species are shown as a function of visual 
extinction 
from the cloud surface, $A_V$.}
\label{fig:std}\end{figure}

\clearpage
\begin{figure}[ht!]
\plotone{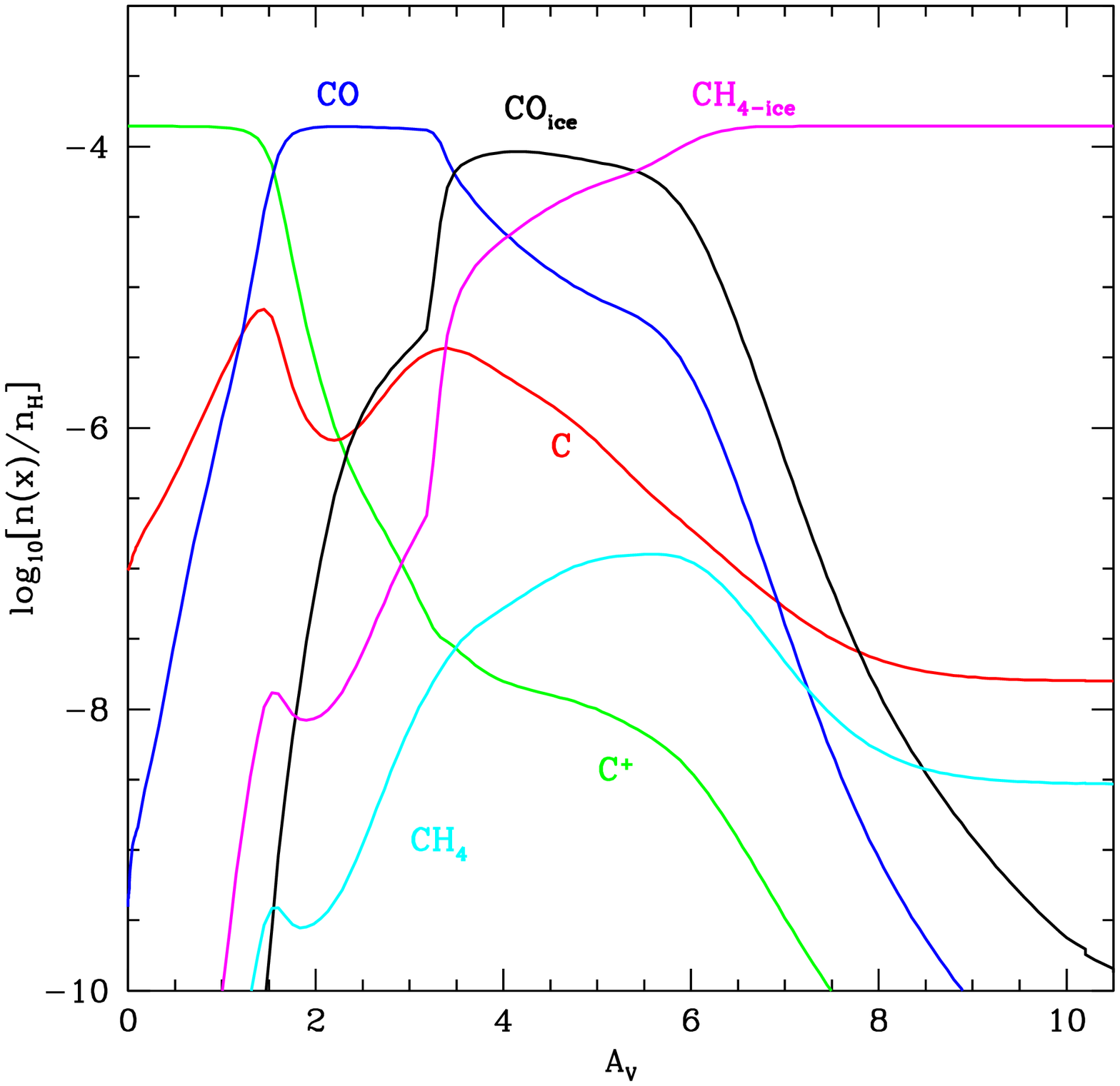}

\caption{Carbon chemistry for the standard  ($n= 10^4$ cm$^{-3}$, $G_0= 100$)
steady-state PDR model, including freeze-
out, grain-surface reactions, thermal
desorption, photodesorption and cosmic ray desorption. 
Model parameters are those given in Table 1. The
abundances of major C-bearing species are shown as a function of visual 
extinction 
from the cloud surface, $A_V$. For $A_V>5$, the gas phase CO abundance wlll not be
as low as is shown here for the steady-state solution; time-dependent effects will 
keep the CO abundance elevated (see \S3.7).  In addition, the limited chemistry in
the steady state PDR code drives the high abundance of CH$_4$ ice at high $A_V$; the
time dependent code, which has more extensive chemistry, finds a mixture of CO
and CH$_4$ ice for the carbon-bearing ices in this case.}
\label{fig:stdC}\end{figure} 
\clearpage
\begin{figure}[ht!]
\plotone{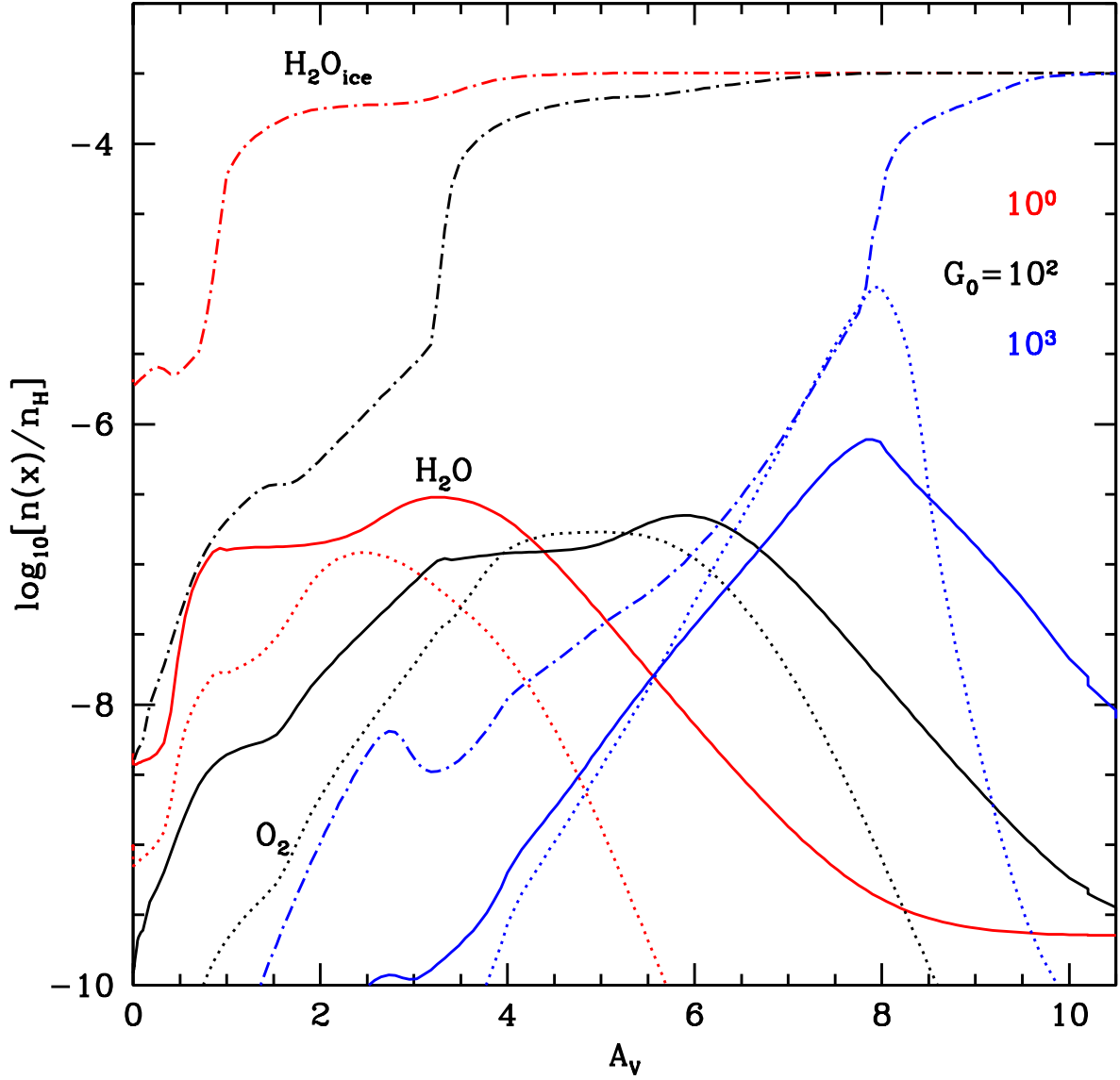}

\caption{H$_2$O, O$_2$ and H$_2$O$_{ice}$ abundances for a cloud with 
$n=10^4\,\rm cm^{-3}$ but with a variety of FUV field strengths incident on 
the cloud surface. Results 
are shown for FUV fields G$_0$=1, 10$^2$ and 10$^3$ times the average interstellar
field. H$_2$O$_{ice}$ is dot-dash, H$_2$O is solid, and O$_2$ is dotted lines.
Higher $G_0$ drives curves to the right.
Although the depth at which freezeout occurs is affected by $G_0$, the 
total H$_2$O column is not. The increase in the peak abundance of O$_2$ seen for
$G_0 = 1000$ is caused by thermal desorption of atomic O from the
warm grains, which suppresses H$_2$O$_{ice}$ formation and keeps more elemental
O in the gas phase. }
\label{fig:varyg}\end{figure} 

\clearpage
\begin{figure}[ht!]

\plotone{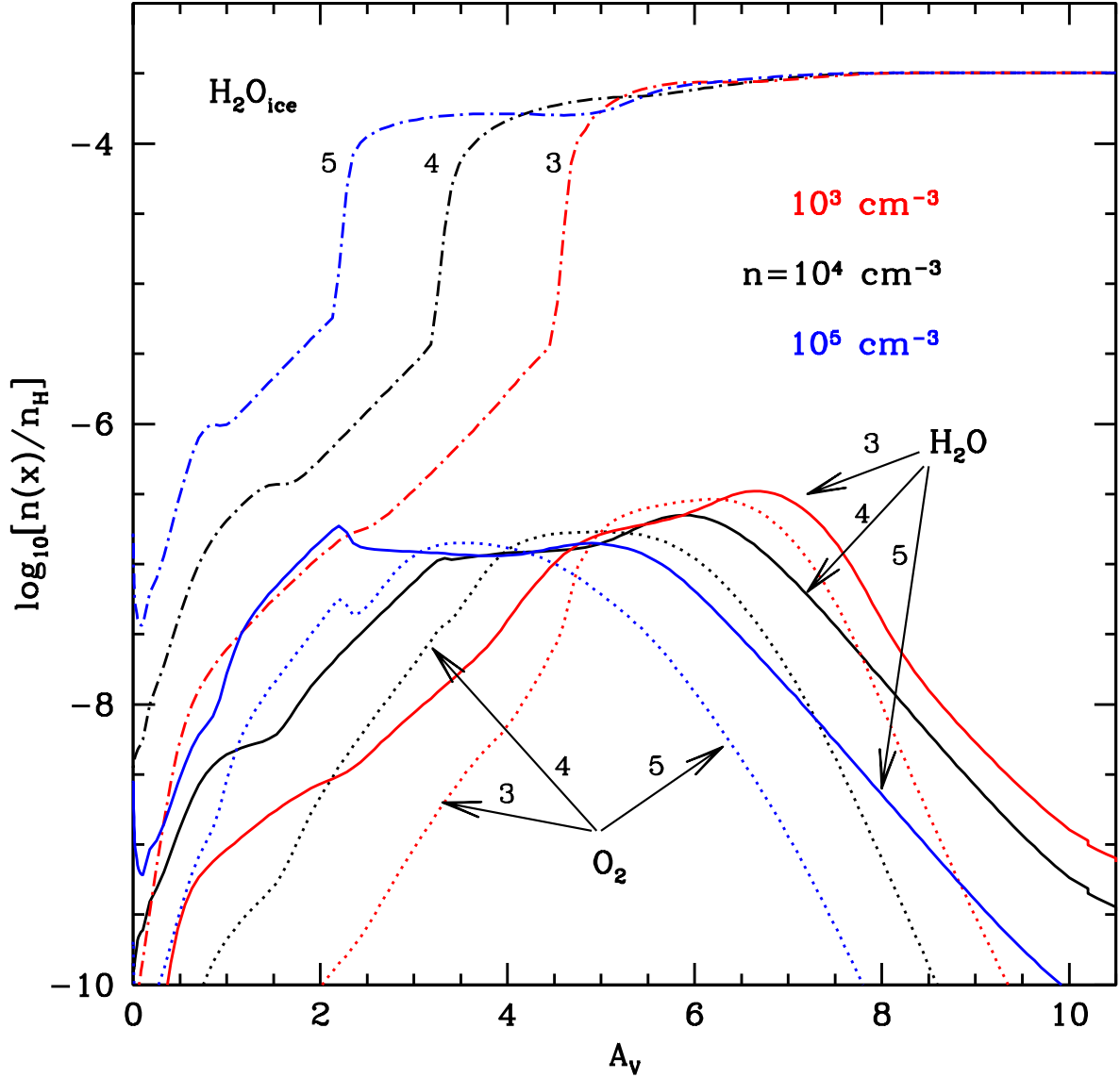}
\caption{Effect of changing the gas density. Results 
are shown for $n=10^3$, $10^4$ and $10^5\,\rm cm^{-3}$ and for
the standard FUV field $G_0=10^2$. 
The threshold $A_V$ for water ice formation and the $A_V$ where
H$_2$O and O$_2$ peak increase for increasing $G_0/n$.  The peak
abundances do not change with $n$.  The labels
on the arrrows refer to the log of the density $n$.}
\label{fig:varyn}\end{figure}
\clearpage 
\begin{figure}

\plotone{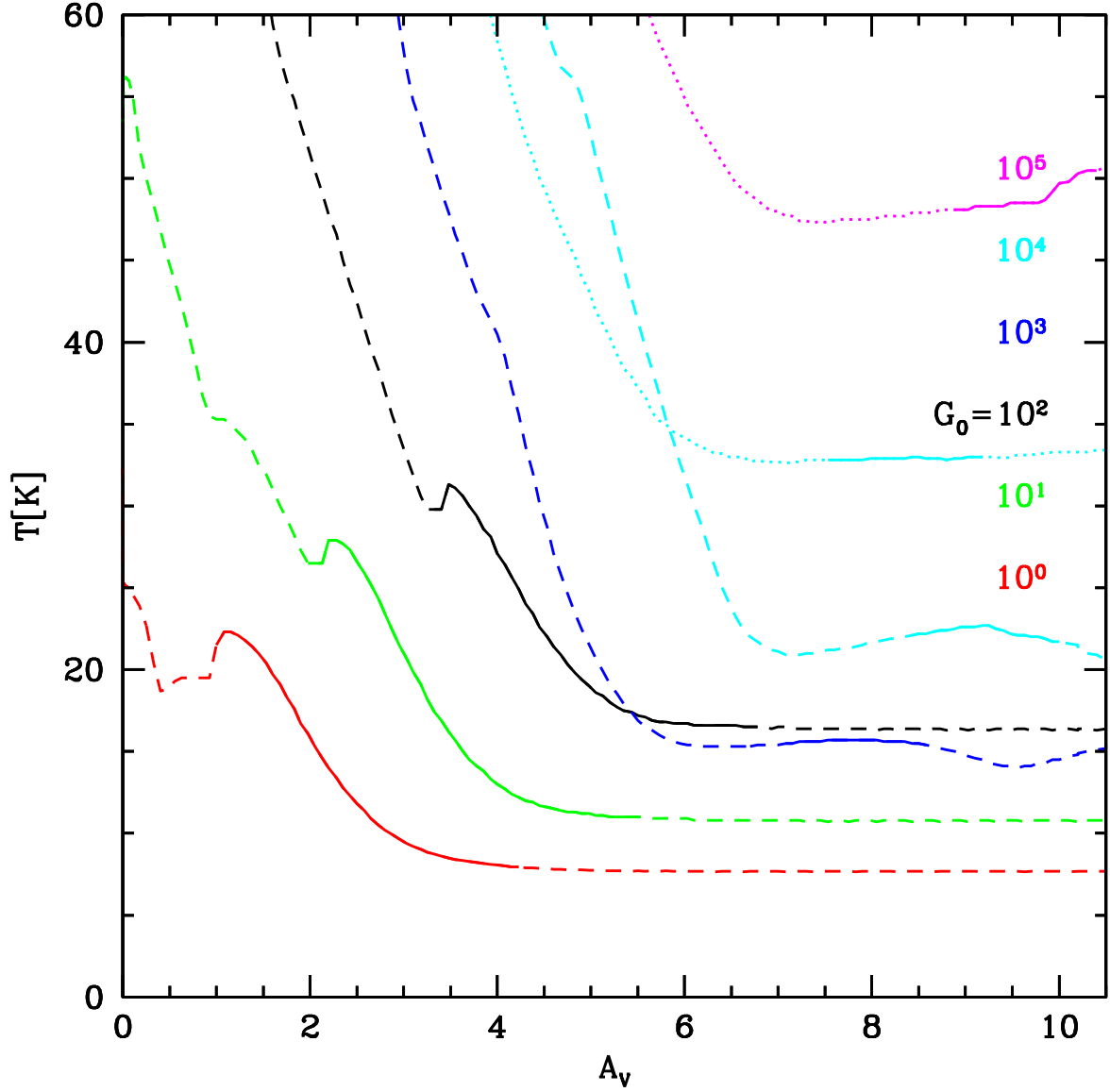}
\caption{The gas temperature $T$ as a function of $A_V$ for a
variety of $n$ and $G_0$ with all other parameters standard.
Dashed lines refer to gas density $n= 10^4$ cm$^{-3}$, and in
order of increasing $T$ at low $A_V$ correspond to $G_0 = 1$,
10, 100, 10$^3$, and 10$^4$.  Dotted lines refer to gas density
$n= 10^5$ cm$^{-3}$, and in order of increasing $T$ correspond
to $G_0 = 10^4$ and $10^5$ (Orion PDR conditions). The solid
portion of each curve refers to the region of the ``water 
plateau'' for $G_0 < 500$, or to regions where the water abundance
is at least 0.33 times its peak value for $G_0 > 500$.}
\label{fig:TvsAV}\end{figure}
\clearpage 
\begin{figure}

\plotone{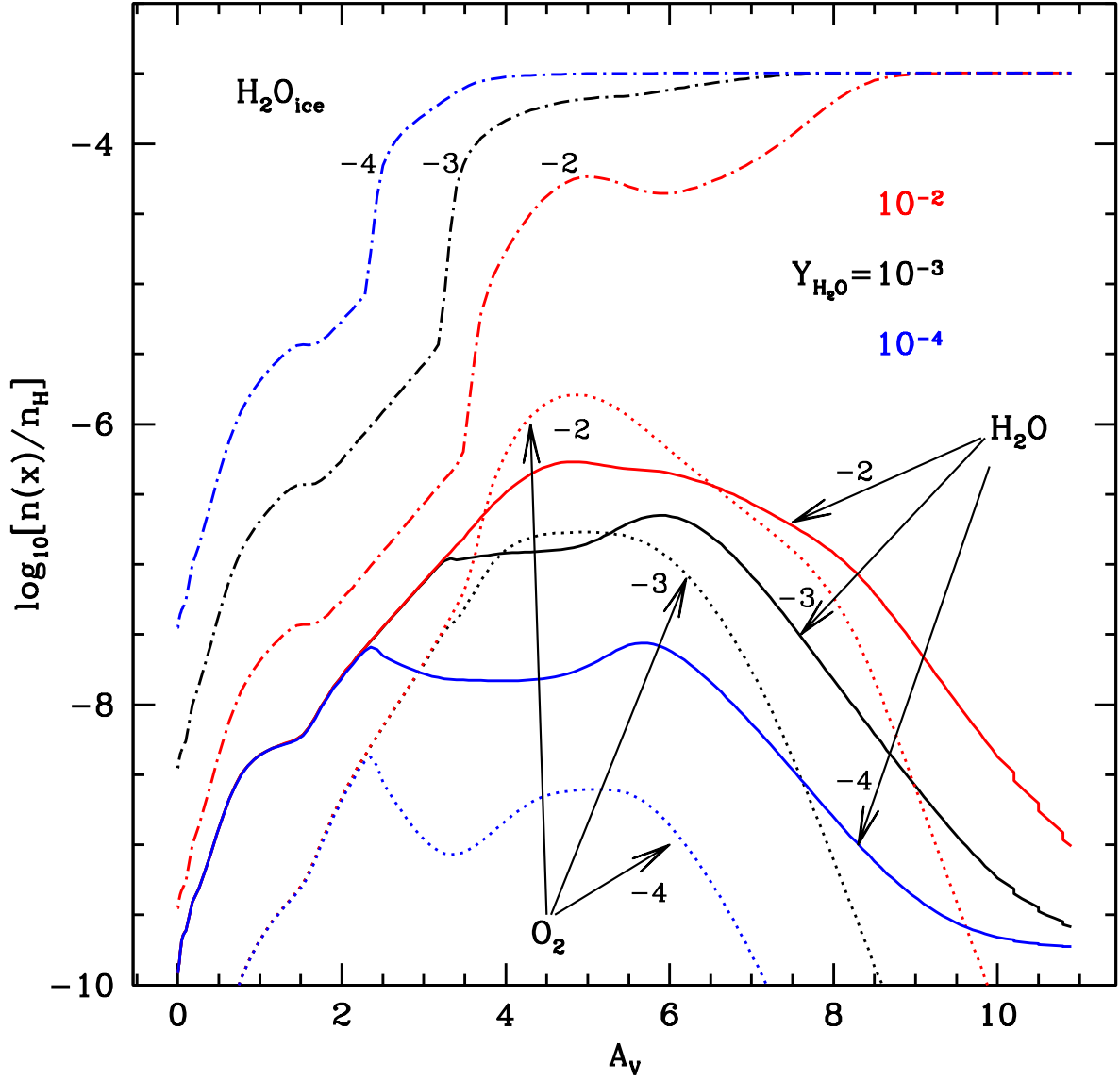}
\caption{Abundances of H$_2$O, H$_2$O$_{ice}$ and O$_2$ for the
standard case ($n= 10^4$ cm$^{-3}$, $G_0= 100$) as a function
of $A_V$ and of the 
photodesorption yield of water ice to form gas-phase H$_2$O, $Y_{\rm H_2O}$. 
Results are shown for 
the standard yield of 10$^{-3}$ and for yields of 10$^{-2}$ and
10$^{-4}$. H$_2$O$_{\rm ice}$ also photodesorbs OH with a rate two times
$Y_{\rm H_2O}$ (see text).  The labels refer to the log
of the yield. }
\label{fig:varyy}\end{figure}
\clearpage
\begin{figure}

\plotone{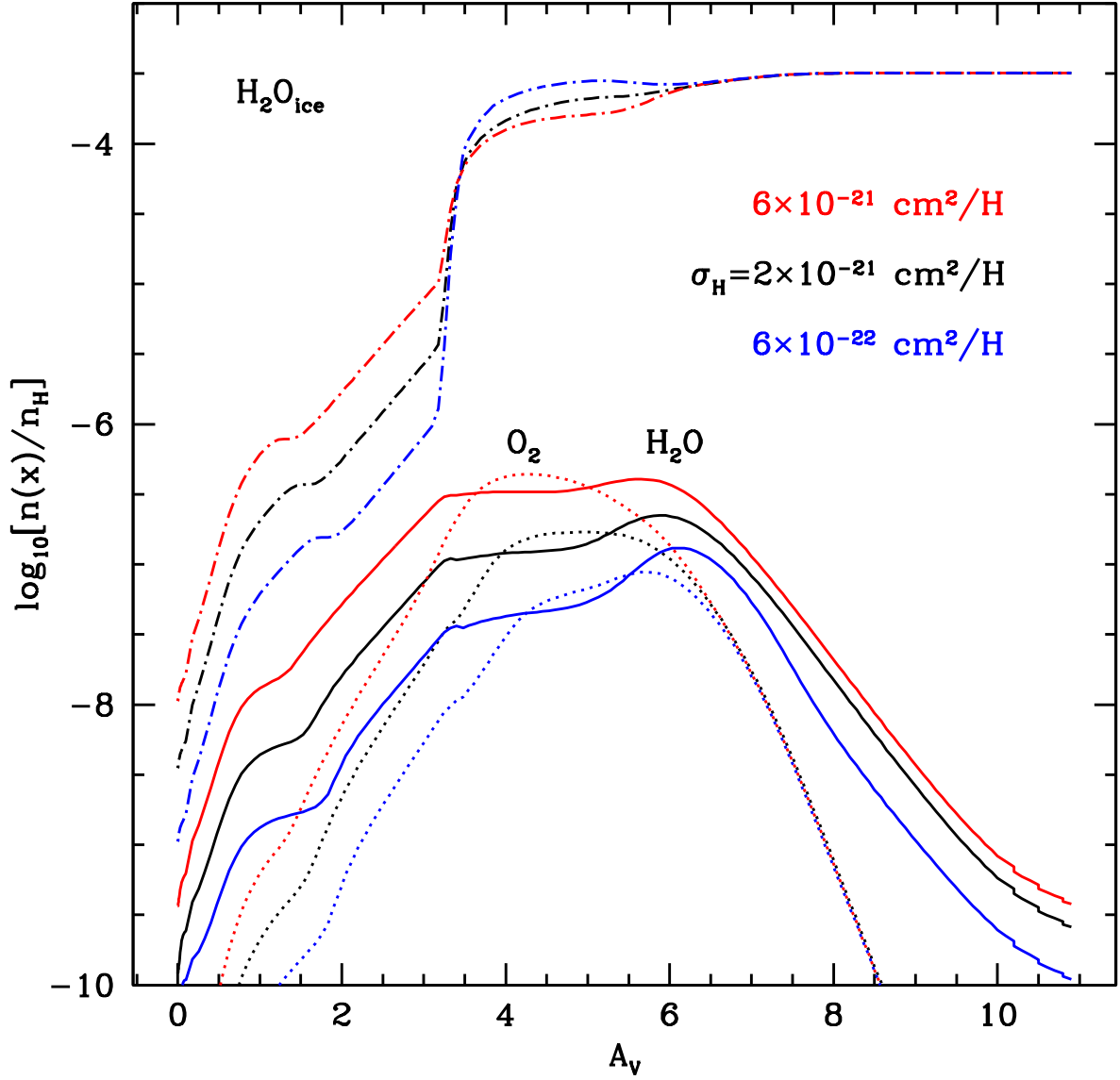}
\caption{Abundances of H$_2$O, O$_2$ and H$_2$O$_{ice}$ as a function
of $A_V$ and of 
the total grain cross section per H nucleus, 
$\sigma_{H}$.  The higher curves for O$_2$ and H$_2$O and the lower curves
for H$_2$O$_{ice}$ refer to higher
values of $\sigma _H$.}
\label{fig:varygrain}\end{figure} 

\clearpage
\begin{figure}[ht!]
\plottwo{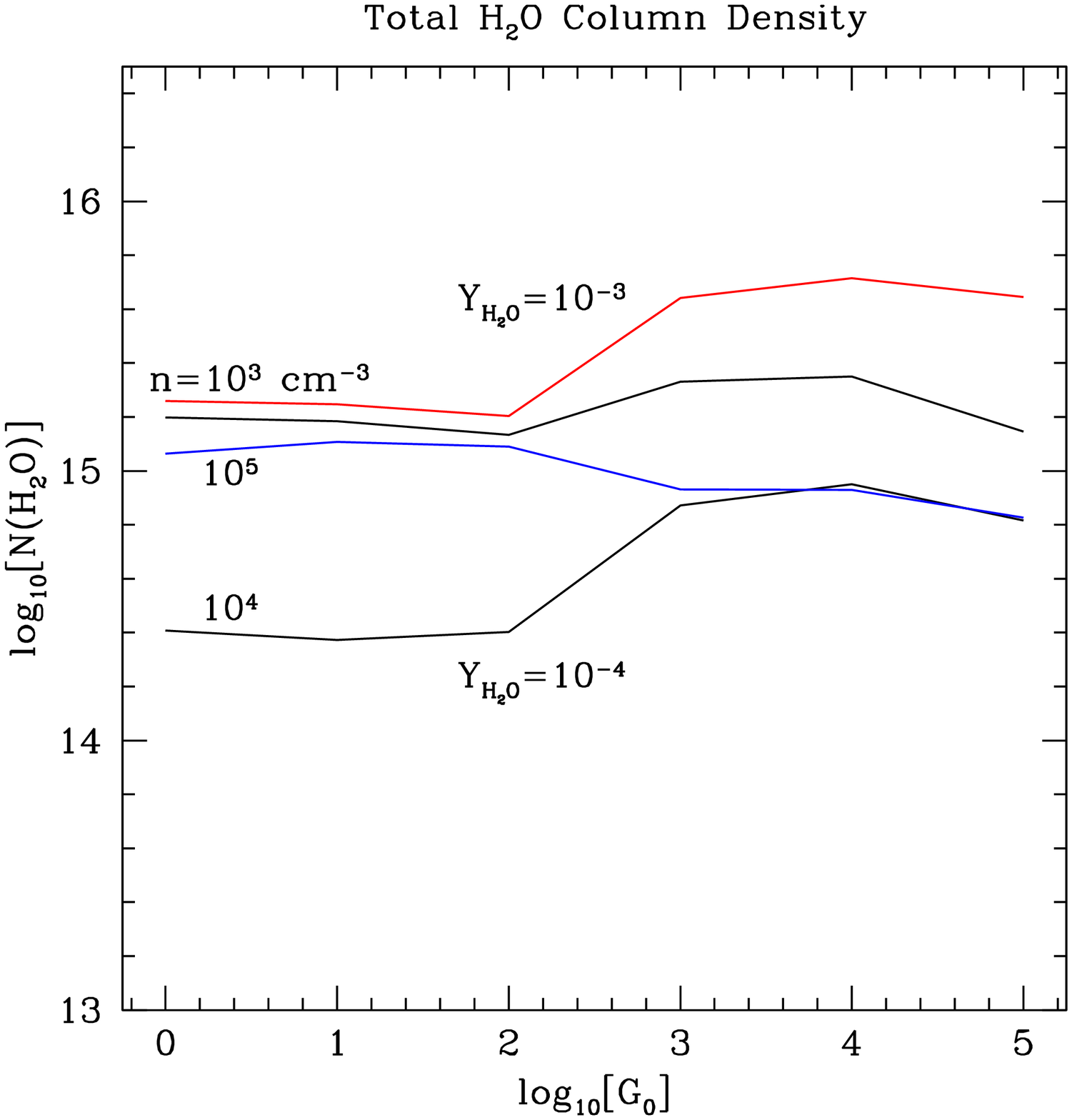}{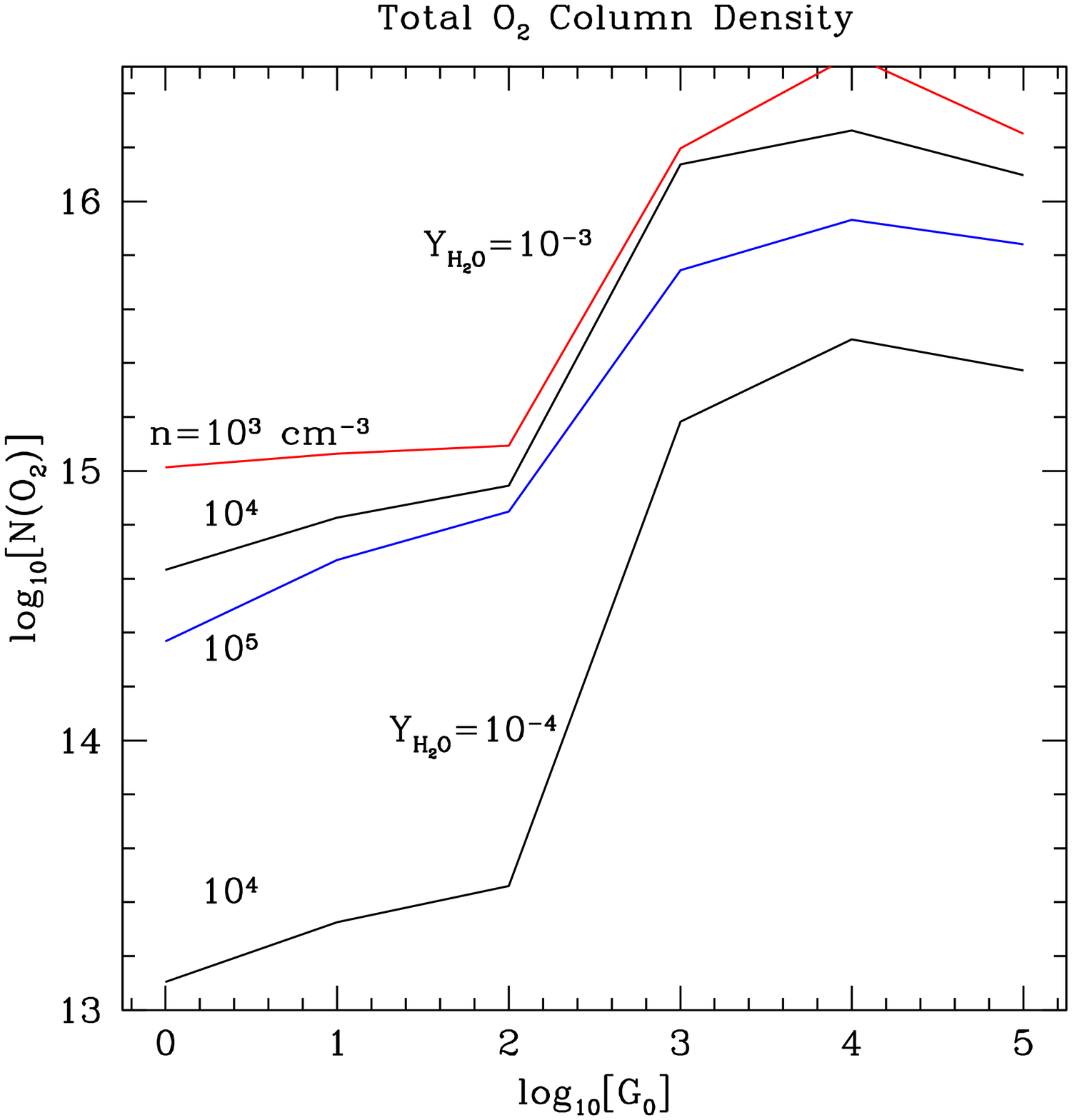}
\caption{The total columns of H$_2$O (left panel) and O$_2$ (right panel)
as a function of $G_0$ and $n$,
assuming clouds with  $A_V$ greater than the plateau values, or roughly $A_V \gta 3-6$.
The increase of the column of H$_2$O and O$_2$ at $G_0 \sim 500$ is caused by
the thermal desorption of O atoms from grain surfaces, which suppresses the formation
of water ice and enhances the gas phase abundance of elemental O.  Note that since
the column is produced at intermediate $A_V$, the column average abundance of H$_2$O
and O$_2$ will fall as $A_V^{-1}$ for high $A_V$. Results are shown for the standard 
photoyield $Y_{H_2O}=10^{-3}$ for densities of $10^3$, $10^4$, and $10^5$ cm$^{-3}$,
and for $Y_{H_2O}=10^{-4}$ for a density of $10^4$ cm$^{-3}$.}
\label{fig:nave}\end{figure} 

\clearpage
\begin{figure}[ht!]
\plotone{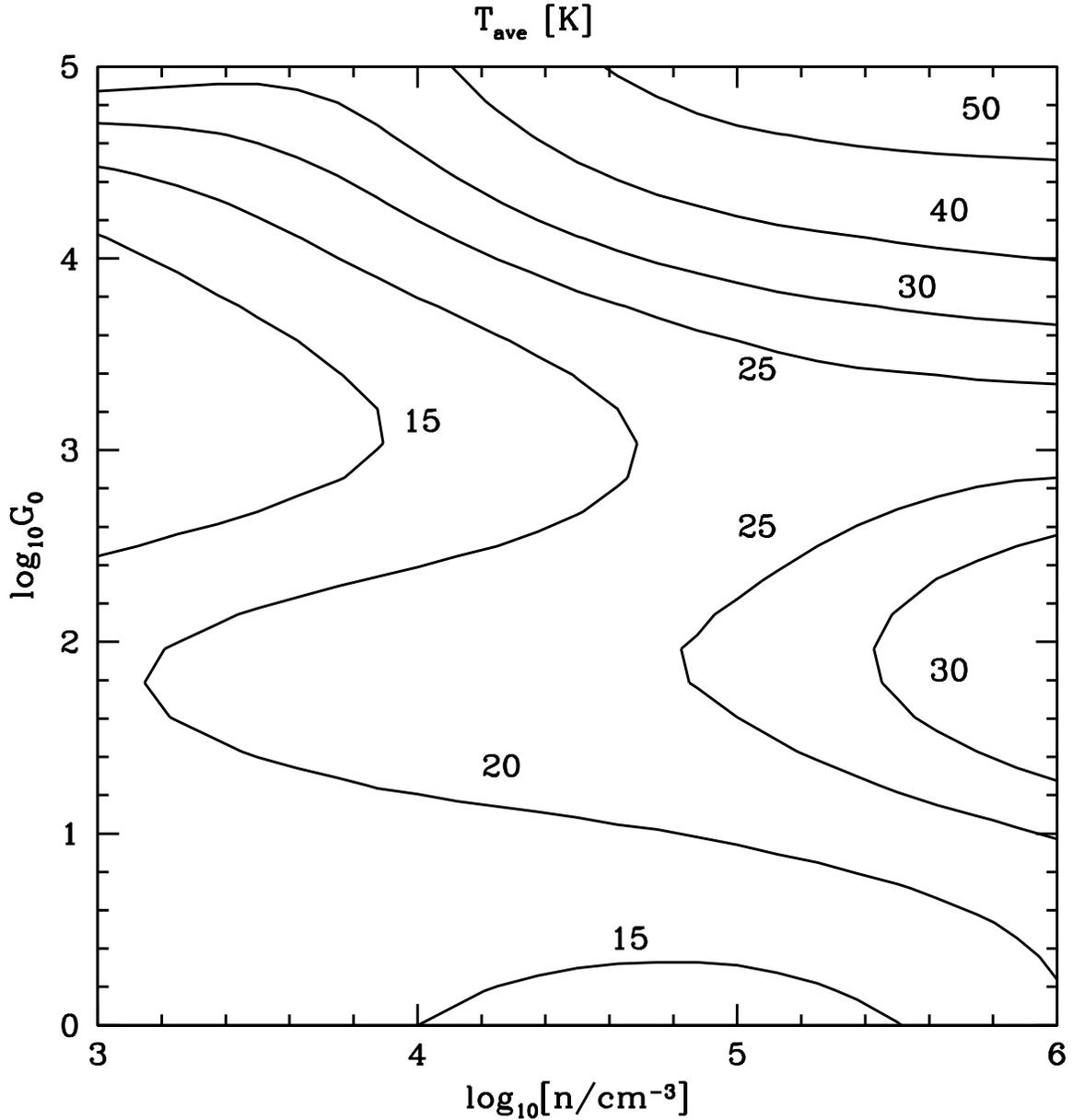}
\caption{The average (see text) gas temperature in the water plateau region or the peak
water abundance region for a range of  
densities and FUV field strengths incident on 
the cloud surface.  Note that the temperature only increases by a factor of $\sim 3$ while
the incident FUV flux $G_0$ increases by $10^5$.  The plateau retreats deeper into the cloud
as $G_0$ increases, which suppresses a more dramatic increase in $T$.}
\label{fig:Tave}\end{figure} 

\clearpage
\begin{figure}[ht!]
\plottwo{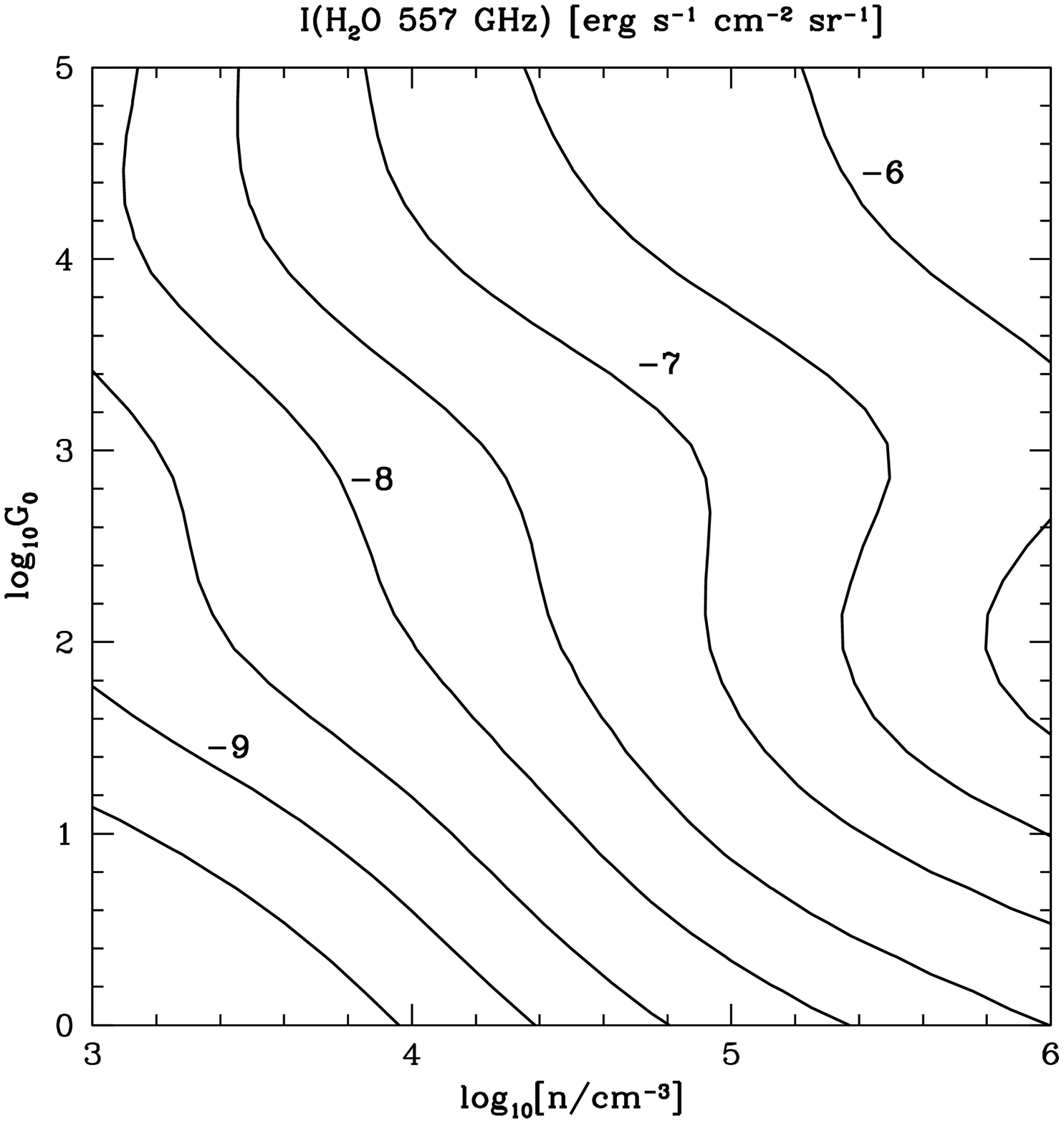}{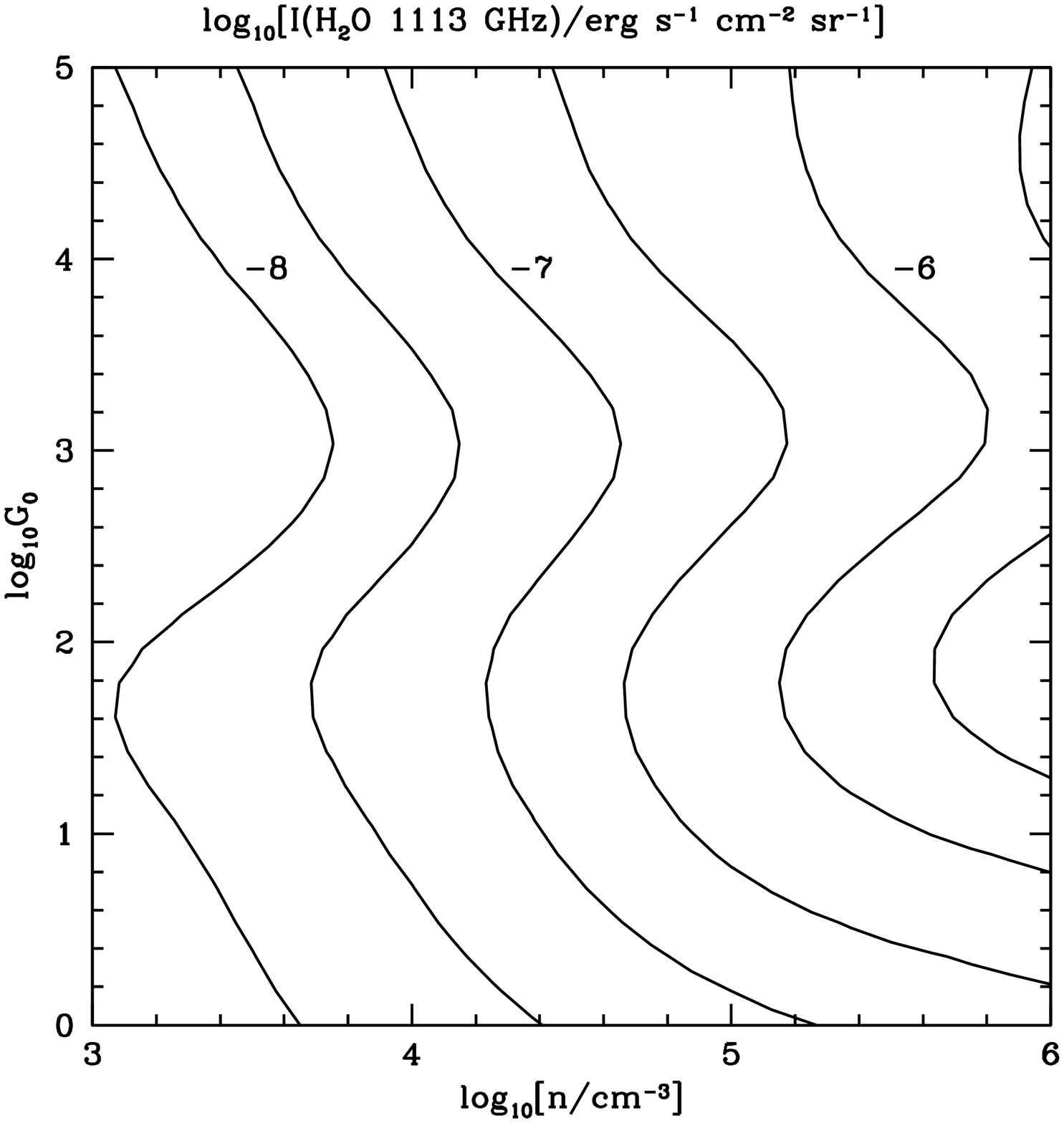}
\caption{H$_2$O intensities for a range of  
densities and FUV field strengths incident on 
the cloud surface. Contours are labeled with the logarithm of 
intensity, measured in units of  
${\rm erg\ s^{-1}\ cm^{-2}\ sr^{-1}}$.
{\it Left Panel:} Intensity of the ortho-H$_2$O groud state transition $1_{1,0}-1_{0,1}$ at 557 GHz.  To convert to units of
${\rm K\,km\,s^{-1}}$, multiply contour levels by $5.6\times 10^6$; {\it Right Panel:} Intensities of the para-H$_2$O
ground
state transition $1_{1,1}-0_{0,0}$ at 1113 GHz.  To convert to units of
${\rm K\,km\,s^{-1}}$, multiply contour levels by $7.0\times 10^5$.}
\end{figure}
\clearpage
\begin{figure}[ht!]
\plotone{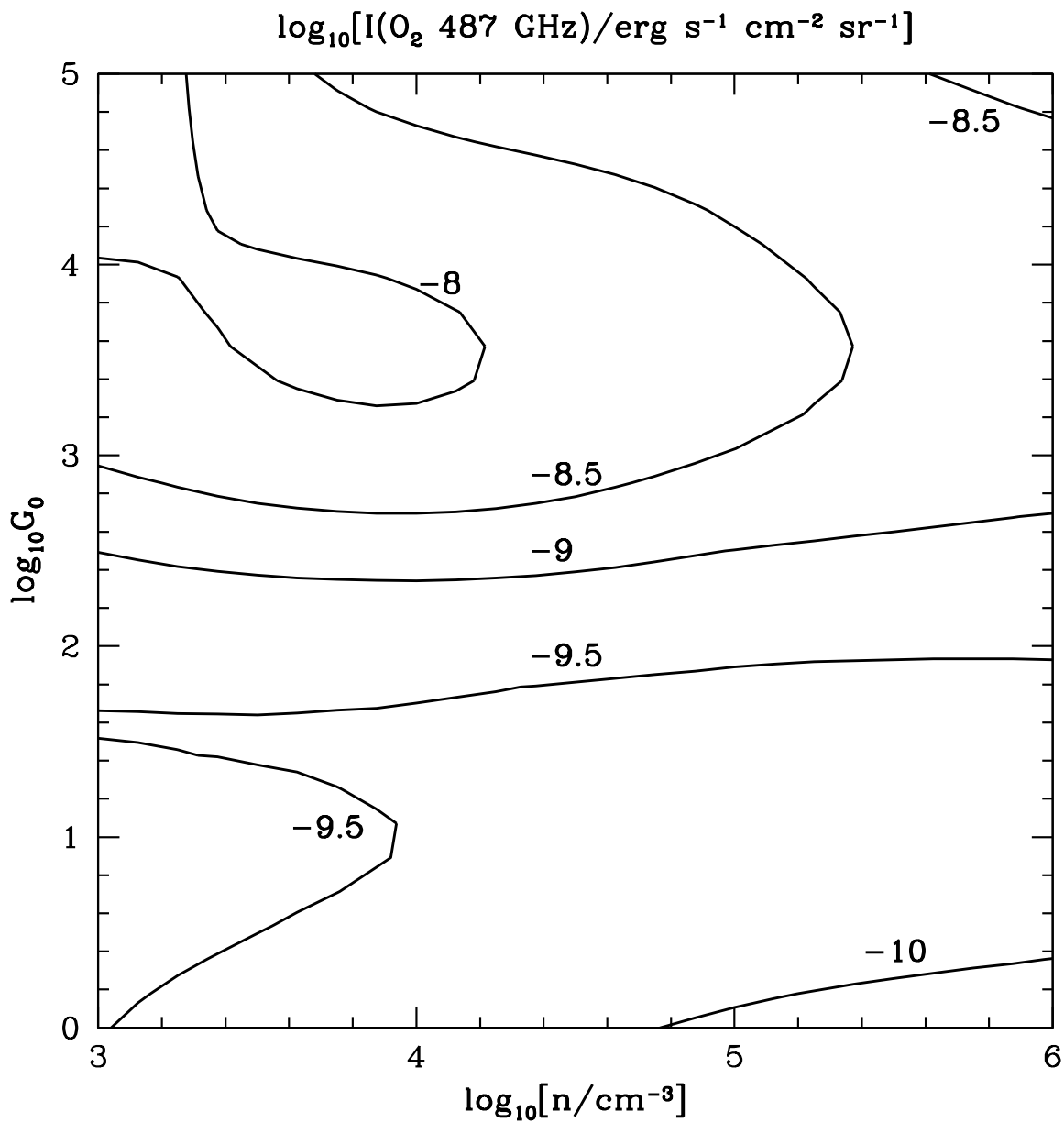}
\caption{The O$_2$ 487 GHz intensity for a variety of  
densities and FUV field strengths incident on 
the cloud surface. Contours are labeled with the logarithm of 
intensity, measured in units of  
${\rm erg\ s^{-1}\ cm^{-2}\ sr^{-1}}$. To convert to units of
${\rm K\,km\,s^{-1}}$, multiply contour levels by $8.4\times 10^6$.
Note the relative constancy of the O$_2$ intensity for $G_0 < 100$, since
the column and temperature of the O$_2$ plateau region stay quite constant.}
\end{figure}
\clearpage

\begin{figure}[ht!]
\plotone{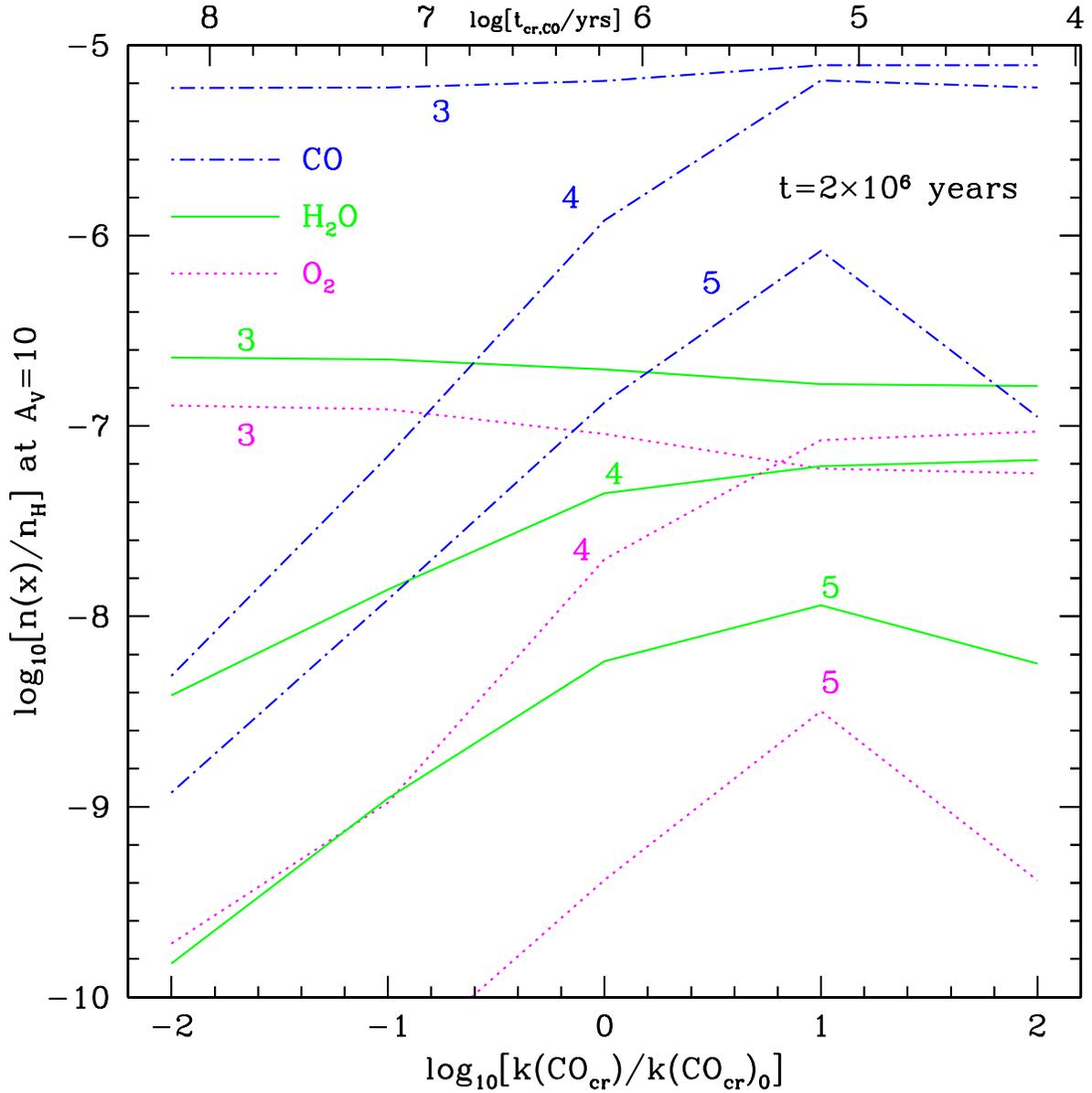}
\caption{Abundances of H$_2$O (solid), CO (dot-dash) and O$_2$ (dotted)
at A$_V=10$ at a cloud age
of $2\times 10^6$ years, for a range of CO cosmic-ray
desorption rates. Our standard CO cosmic-ray rate is given
by $k(CO_{cr})_0$ and is given in Table 1. Results are shown for rates from $10^{-2}$ to
$10^2$ times the 
standard rate, for G$_0=100$ and densities $n_{\rm H}= 10^3$, $10^4$, and 
$10^5 {\rm cm^{-3}}$.  The higher densities have lower abundances due
to freezeout of gas species. Note that the H$_2$O and O$_2$ abundances are only
significant (i.e., comparable to the plateau value), for low densities.  The
timescale $t_{cr,CO}$ for cosmic ray desorption of CO ice (see Eq. 8) is given at
the top.}  
\end{figure}
\clearpage

\begin{figure}
\plotone{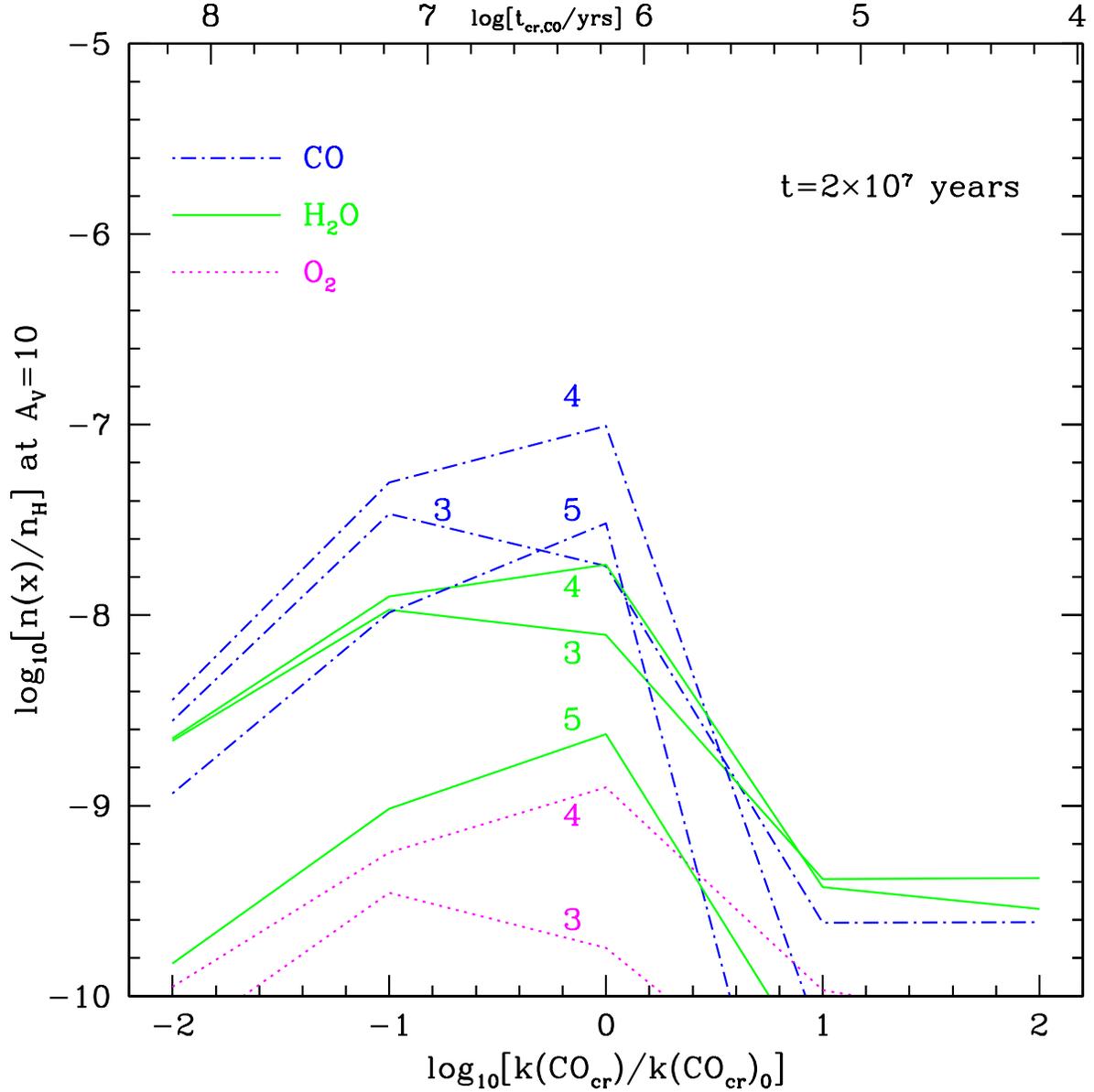}
\caption{Same as Fig. 14 but for an age of 
of $2 \times 10^7$ years.  Note that the H$_2$O and O$_2$ abundances
are never significant compared to their plateau values.
If the cosmic ray desorption timescale for CO ice, $t_{cr,CO}$ is long, 
the  H$_2$O and O$_2$ abundances are low because
the frozen CO is not liberated to the gas to provide elemental gas phase O to form them.
If the timescale is very short, the CO ice is desorbed quickly and the gas has time
to deplete the elemental gas phase O as water ice on the grain surfaces. }
\end{figure}
\clearpage

\begin{figure}
\plotone{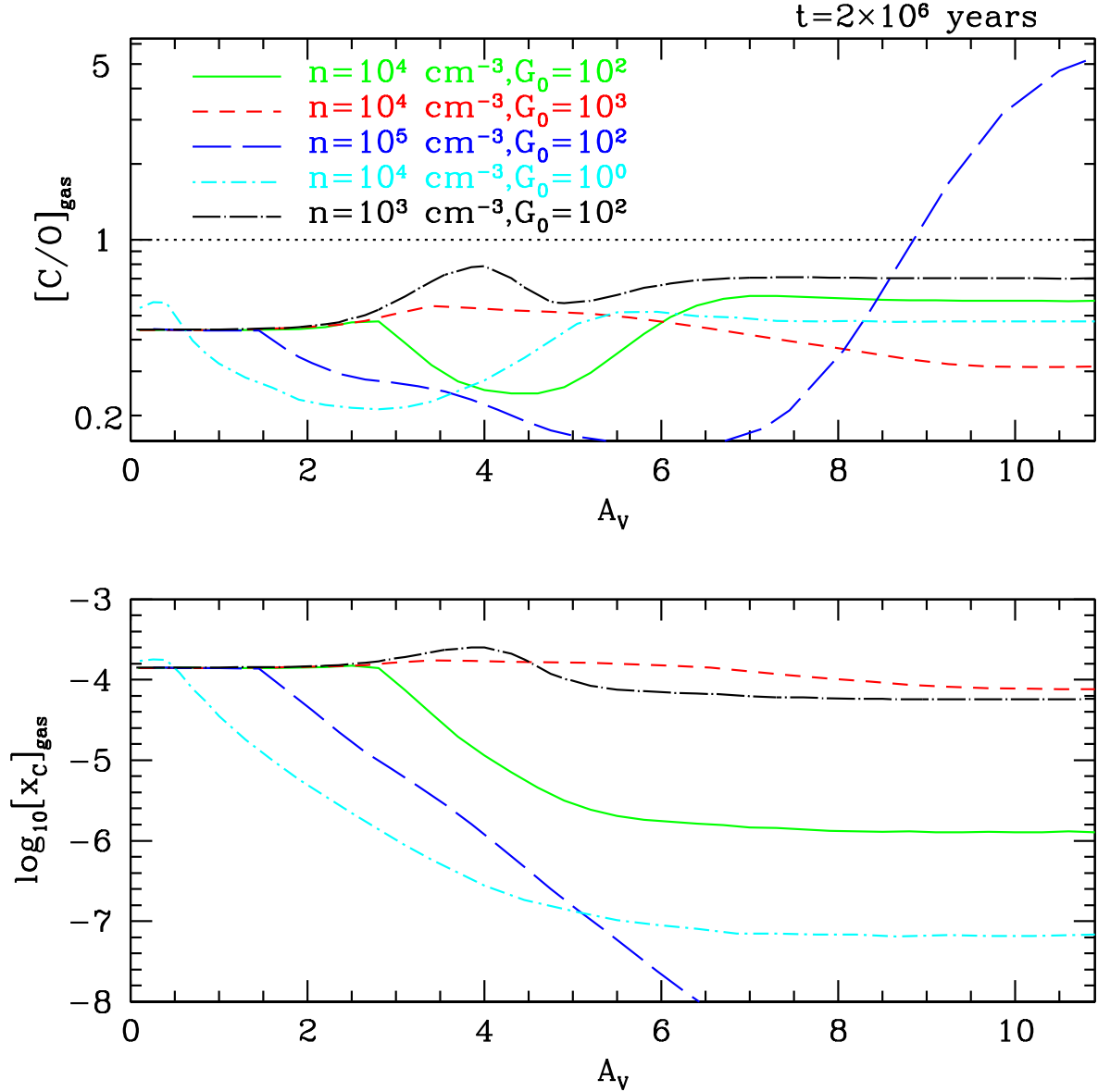}
\caption{Ratio of carbon nuclei to oxygen nuclei in the gas phase
 for clouds with a variety of densities and FUV field strengths. Results 
are shown for a cloud 
age of 2$\times 10^6$ years.  If the C/O ratio rises above unity, then
interesting organic chemistry ensues (see text). Note that the C/O ratio
only rises above unity for one case (the high density, $n=10^5$ cm$^{-3}$ case).
However, in this case the time to deplete the gas phase carbon is very short
so that the gas phase elemental C abundance is very low (bottom panel).  Thus,
the abundance of organic molecules will be very low.}
\end{figure} 

\clearpage
\begin{figure}[ht!]

\plotone{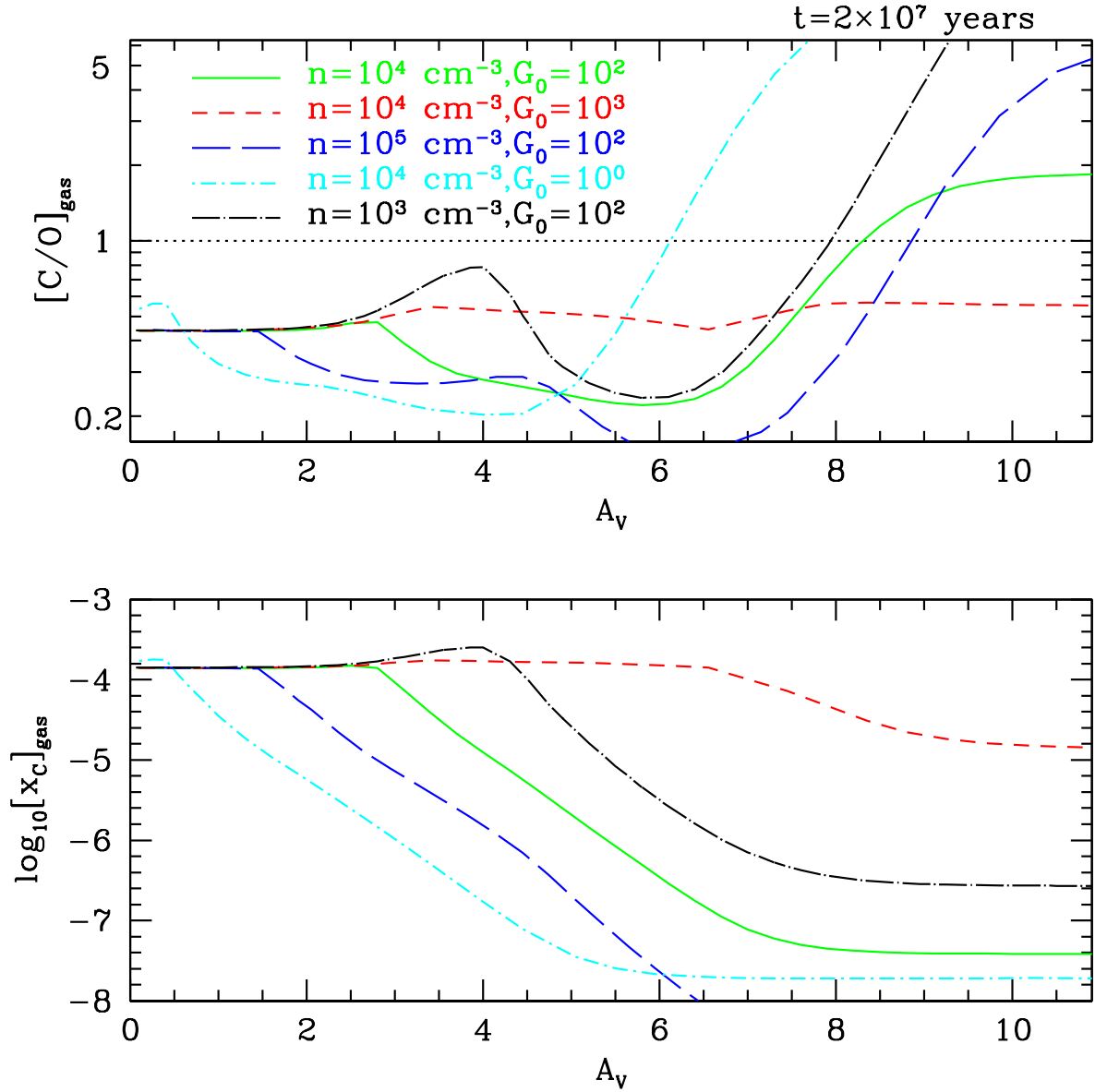}
\caption{Same as Figure 16 but for cloud
age of 2$\times 10^7$ years. Note that C/O rises 
above unity for all cases except the case with high $G_0=10^3$.
Thus, in all cases but this one, complex carbon chemistry is initiated, since not
all C will be tied up in CO.  The bottom panel shows which clouds will have
the highest abundance of these molecules (the cases with C/O$>1$ and high elemental
C abundance).  The elemental C abundance rises with increasing $G_0/n$.}
\end{figure} 
\clearpage


\end{document}

%% file: tab1.tex
\begin{deluxetable}{lccc}
\tablecaption{Adsorption Energies, Photodesorption Yields, and Cosmic-Ray Desorption Rates of Surface
Monolayer Atoms and Molecules}
\tablehead{\colhead{Species} &\colhead{Adsorption Energy\tablenotemark{a}}&
\colhead{Photodesorption Yield}&\colhead{Cosmic-Ray Desorption Rate}\\
\colhead{}&\colhead{[K]}&\colhead{[UV Photon]$^{-1}$}&\colhead{[mol$^{-1}$ s$^{-1}$]}}
\startdata
O&800&$10^{-4}$&---\\
OH&1300&$10^{-3}$&---\\
H$_2$O&4800\tablenotemark{b}&$10^{-3}$ as H$_2$O&$4.4\times 10^{-17}$ \tablenotemark{c}\\
&&$2\times 10^{-3}$ as OH&\\
O$_2$&1200&$10^{-3}$&$9.1\times 10^{-15}$ \tablenotemark{d}\\
C&800&$10^{-3}$&---\\
CH&650&$10^{-3}$&---\\
CH$_2$&960&$10^{-3}$&---\\
CH$_3$&1200&$10^{-3}$&---\\
CH$_4$&1100\tablenotemark{b}&$10^{-3}$&$2.7\times 10^{-15}$ \tablenotemark{d}\\
CO&960\tablenotemark{b}&$10^{-3}$&$6.0\times 10^{-13}$ \tablenotemark{c}\\
S&1100&$10^{-3}$&---\\
Si&2700&$10^{-3}$&---\\
Fe&4200&$10^{-3}$&---\\

\enddata
\tablenotetext{a} {All adsorption energies are from \cite{hh93} unless otherwise 
noted.}
\tablenotetext{b} {\cite{aik96}}
\tablenotetext{c} {\cite{hc06}, Bringa \& Johnson (2004)}
\tablenotetext{d} {\cite{hh93}}
\end{deluxetable}

%% file: tab2.tex
\begin{deluxetable}{lll}
\tablecaption{Standard Model Parameters}
\tablehead{\colhead{Parameter}&\colhead{Symbol}&\colhead{Value}}
\startdata
H nucleus density&$n$&$10^4\,\rm cm^{-3}$\\
FUV field&$G_0$&100\\
Oxygen abundance\tablenotemark{a}&$x_O$&$3.2\times 10^{-4}$\\
Carbon abundance\tablenotemark{a}&$x_C$&$1.4\times 10^{-4}$\\
Sulfur abundance\tablenotemark{a}&$x_{S}$&$2.8\times 10^{-5}$\\
Silicon abundance\tablenotemark{a}&$x_{Si}$&$1.7\times 10^{-6}$\\
Magnesium abundance\tablenotemark{a}&$x_{Mg}$&$1.1\times 10^{-6}$\\
Iron abundance\tablenotemark{a}&$x_{Fe}$&$1.7\times 10^{-7}$\\
Total grain cross section per H nucleus&$\sigma_{H}$&$2\times 10^{-21}\, \rm cm^{2}$\\
\enddata
\tablenotetext{a} {Abundances of nuclei in the gas phase at the cloud surface,
relative to H nuclei; Savage \& Sembach (1996)}
\end{deluxetable}

%% file: tab3.tex
\begin{deluxetable}{ll}
\tablecaption{Symbols Used in the Text\label{symtab}}
\tablehead{\colhead{Symbol}&\colhead{Definition}}
\startdata
$a$&grain radius\\
$A_V$&visual extinction from surface of the cloud\\
$A_{Vf}$&visual extinction where there is onset of water ice freezeout, monolayer of ice forms\\
$A_{Vd}$&visual extinction where the gas phase water abundance drops\\
$\Delta A_V$&width of the plateau of gas phase H$_2$O or O$_2$, or $(A_{Vd}-A_{Vf})$\\
$E_i$&binding energy of species $i$ to grain surface\\
$f_{\rm o},(f_{\rm p})$& fraction of H$_2$O in the ortho(para) state\\
$f_{s,i}$& fraction of surface sites occupied by species $i$\\
$F_{\rm FUV}$&flux of FUV photons at the cloud surface\\
$F_0$&$10^8$ photons cm$^{-2}$ s$^{-1}$, the flux of FUV photons corresponding to $G_0=1$\\
$F_{pd,i}$&flux of molecules of species $i$ photodesorbing from ice surface\\
$F_{td,i}$&flux of molecules of species $i$ thermally desorbing from ice surface\\ 
$G_0$&unitless parameter measuring incident FUV flux normalized to $1.6\times 10^{-3}$ erg cm$^{-2}$ s$^{-1}$ \\
&\hfil  or to $10^8$ photons cm$^{-2}$ s$^{-1}$, roughly the local interstellar value\hfil\\
$\gamma_{\rm He^+}$& net rate coefficient for destruction of H$_2$O by He$^+$\\
$\gamma_{\rm H_3^+}$& net rate coefficient for destruction of H$_2$O by H$_3^+$\\
$\gamma_{\rm O_2}$& rate coefficient for gas phase reaction of OH with O to form O$_2$\\
$I_{487}$& intensity of the O$_2$ 487 GHz transition, in erg cm$^{-2}$ s$^{-1}$ sr$^{-1}$\\
$I_{557}$& intensity of the H$_2$O 557 GHz transition, in erg cm$^{-2}$ s$^{-1}$ sr$^{-1}$\\
$I_{1113}$& intensity of the H$_2$O 1113 GHz transition, in erg cm$^{-2}$ s$^{-1}$ sr$^{-1}$\\
$n$& gas phase hydrogen nucleus number density [$\sim n({\rm H}) + 2n({\rm H_2} + n({\rm H^+})$]\\
$n_{gr}$&number density of grains\\
$\nu_i$&vibrational frequency of species $i$ bound to a grain surface\\
$N$&gas-phase column density of hydrogen nuclei\\
$N_i$&gas-phase column density of species $i$\\
$N_f$&column density of hydrogen nuclei for onset of water ice freezeout\\
$N_{s,i}$&number of adsorption sites per cm$^2$ on a grain surface ($\simeq 10^{15}\,{\rm cm^{-2}}$)\\
$R_{td,i}$&thermal desorption rate of per atom or molecule of species $i$ from a surface\\
$R_{cr,i}$&cosmic ray desorption rate per surface atom or molecule, $i$\\
$R_{\rm H_2O}$&$5.9\times 10^{-10}$ s$^{-1}$, the unshielded photodissociation rate per molecule of H$_2$O for G$_0$=1\\
$R_{\rm O_2}$&$6.9\times 10^{-10}$ s$^{-1}$, the unshielded photodissociation rate per molecule of O$_2$ for G$_0$=1\\
$R_{\rm OH}$&$3.5\times 10^{-10}$ s$^{-1}$, the unshielded photodissociation rate per molecule of OH for G$_0$=1\\
$\sigma_{\rm H}$&grain cross sectional area per H nucleus\\
$\sigma_{gr}$&cross sectional area of a single grain\\
$t_{cr,i}$&timescale for cosmic ray desorption of adsorbed ice species $i$ with abundance $x(i)$\\
$t_{evap,{\rm H_2O}}$&time for an adsorbed H$_2$O molecule to thermally evaporate from a grain surface\\
$t_{evap,{\rm O}}$&time for an adsorbed O atom to thermally evaporate from a grain surface\\
$t_{f,i}$&timescale for species $i$ to hit (and stick to) a grain\\
$t_{\gamma}$&timescale for a grain of size $a>100\,$\AA\, to absorb an FUV photon at the cloud surface\\
$t_{\gamma,f}$&timescale for a grain of size $a>100\,$\AA\, to absorb an FUV photon at $A_{V,f}$\\
$t_{\gamma -des}$&timescale to photodesorb an O atom from a grain with covering factor $f_{s,{\rm O}}$\\
$t_{gr,i}$&time for a given grain to be struck by a gas phase atom or molecule of species $i$\\
$t_r$&timescale for a grain to radiate much of its thermal energy\\
$T_{ave}$&the average of the gas $T$ at $A_{Vf}$ and at the $A_V$ where the H$_2$O abundance peaks\\
$T_{max}$& peak temperature of a grain after absorbing one FUV photon\\
$T_{300}$& $T/300$~K, where $T$ is the gas temperature\\
$x(i)$&abundance of species $i$ relative to H nuclei\\
$x_o$& abundance of ortho-H$_2$\\
$x_p$& abundance of para-H$_2$\\
$x_{pl,i}$&abundance of species $i$ in the plateau region between $A_{Vf}$ and $A_{Vd}$\\

$Y_i$&photodesorption yield of species $i$ from a grain surface\\
$Y$&$Y_{\rm H_2O}+Y_{{\rm OH},w}$, the total yield of H$_2$O and OH being desorbed from water ice surface\\
$Z({\rm T})$& partition function for O$_2$\\
\enddata
\end{deluxetable}